\documentclass[
reprint,
superscriptaddress,
preprintnumbers,
 amsmath,
 amssymb,
 prc,
]{revtex4-1}

\usepackage{graphicx}
\usepackage{dcolumn}
\usepackage{bm}
\usepackage[colorlinks,
linkcolor=blue,
anchorcolor=blue,
citecolor=blue,
urlcolor=blue]{hyperref}

\usepackage{xcolor}
\usepackage{soul}

\newcommand{\pT}{p_{\rm T}}
\begin{document}

\preprint{LA-UR-22-29012}
\title{Global constraint on the jet transport coefficient from single hadron, dihadron and $\gamma$-hadron spectra in high-energy heavy-ion collisions}

\author{Man Xie}
\affiliation{Key Laboratory of Quark and Lepton Physics (MOE) \& Institute of Particle Physics, Central China Normal University, Wuhan 430079, China}
\affiliation{Key Laboratory of Atomic and Subatomic Structure and Quantum Control (MOE), Guangdong Basic Research Center of Excellence for Structure and Fundamental Interactions of Matter, Institute of Quantum Matter, South China Normal University, Guangzhou 510006, China}
\affiliation{Guangdong-Hong Kong Joint Laboratory of Quantum Matter, Guangdong Provincial Key Laboratory of Nuclear Science, Southern Nuclear Science Computing Center, South China Normal University, Guangzhou 510006, China}

\author{Weiyao Ke}
\email[]{weiyaoke@lanl.gov}
\affiliation{Department of Physics, University of California, Berkeley, California 94720, USA}
\affiliation{Nuclear Science Division MS 70R0319, Lawrence Berkeley National Laboratory, Berkeley, California 94720, USA}
\affiliation{Theoretical Division, Los Alamos National Laboratory, Los Alamos NM 87545, United States}

\author{Hanzhong Zhang}
\email[]{zhanghz@mail.ccnu.edu.cn}
\affiliation{Key Laboratory of Quark and Lepton Physics (MOE) \& Institute of Particle Physics, Central China Normal University, Wuhan 430079, China}
\affiliation{Key Laboratory of Atomic and Subatomic Structure and Quantum Control (MOE), Guangdong Basic Research Center of Excellence for Structure and Fundamental Interactions of Matter, Institute of Quantum Matter, South China Normal University, Guangzhou 510006, China}
\affiliation{Guangdong-Hong Kong Joint Laboratory of Quantum Matter, Guangdong Provincial Key Laboratory of Nuclear Science, Southern Nuclear Science Computing Center, South China Normal University, Guangzhou 510006, China}

\author{Xin-Nian Wang}
\email[]{xnwang@lbl.gov}
\affiliation{Department of Physics, University of California, Berkeley, California 94720, USA}
\affiliation{Nuclear Science Division MS 70R0319, Lawrence Berkeley National Laboratory, Berkeley, California 94720, USA}

\date{\today}
\begin{abstract}
Modifications of large transverse momentum single hadron, dihadron, and $\gamma$-hadron spectra in relativistic heavy-ion collisions are direct consequences of parton-medium interactions in the quark-gluon plasma (QGP). The interaction strength and underlying dynamics can be quantified by the jet transport coefficient $\hat{q}$. We carry out the first global constraint on $\hat{q}$ using a next-to-leading order pQCD parton model with higher-twist parton energy loss and combining world experimental data on single hadron, dihadron, and $\gamma$-hadron suppression at both RHIC and LHC energies with a wide range of centralities. The global Bayesian analysis using the information field (IF) priors provides the most stringent constraint on $\hat q(T)$. We demonstrate in particular the progressive constraining power of the IF Bayesian analysis on the strong temperature dependence of $\hat{q}$ using data from different centralities and colliding energies. We also discuss the advantage of using both inclusive and correlation observables with different geometric biases. As a verification, the obtained $\hat{q}(T)$ is shown to describe data on single hadron anisotropy at high transverse momentum well. Predictions for future jet quenching measurements in oxygen-oxygen collisions are also provided.
\end{abstract}

\maketitle


\section{Introduction}
\label{sec:intro}

Relativistic heavy-ion collisions produce a high density of partons with strong final-state interactions and lead to the formation of the quark-gluon plasma (QGP) \cite{PHENIX:2004vcz,STAR:2010vob}. 
Experimental evidence at the Relativistic Heavy-Ion Collider (RHIC) and the Large Hadron Collider (LHC) indicates that QGP is a strongly-coupled system and can be described well by the relativistic hydrodynamics with a surprisingly small specific shear viscosity \cite{Romatschke:2007mq,Dusling:2007gi,Song:2007ux,Gale:2012rq}. In the early stage of these collisions, rare high transverse momentum ($\pT$) partons are produced through hard scatterings of beam partons. These high-$\pT$ (hard) partons will traverse the QGP and interact with the hot and dense medium along their propagation paths. Multiple scatterings between hard partons and the medium then lead to parton energy loss and the suppression of high-$\pT$ jet and hadron spectra~\cite{Gyulassy:1990ye,Wang:1991xy,Qin:2015srf}. Apart from the strong suppression of single inclusive hadrons \cite{PHENIX:2008saf,Aamodt:2010jd,Adare:2012wg,CMS:2012aa,Abelev:2012hxa,Aad:2015wga,Khachatryan:2016odn,Acharya:2018qsh}, medium modifications of dihadron and $\gamma$/Z-hadron spectra have also been observed in experiments at both RHIC and LHC\cite{Adams:2006yt,Abelev:2009gu,Aamodt:2011vg,Conway:2013xaa,STAR:2016jdz,STAR:2016jdz,Adam:2016xbp}. These experimental data will provide important information on the properties of QGP.

Phenomenological models that have successfully explained the observed jet quenching phenomena are usually based on a modified factorized parton model. In these models, the production of initial hard jets is described using the perturbative QCD parton model with nuclear-modified parton distributions (nPDF), while the final-state interactions are treated either in the energy loss or quenching weight approach \cite{Baier:2001yt,Salgado:2003gb,Feal:2019xfl,PhysRevLett.89.252301}, transport models \cite{Jeon:2003gi,Zapp:2008gi,Ghiglieri:2015ala,He:2015pra,Cao:2017hhk,Ke:2020clc}, and the modified fragmentation approach \cite{Zhang:2003wk,Wang:2009qb,Chen:2010te,Chen:2011vt,Chien:2015vja,Chien:2015ctp}. In recent years, a multi-stage approach that combines jet transport and modified DGLAP evolution equations has also been developed within the JETSCAPE Collaboration\cite{Cao:2017zih,Putschke:2019yrg,JETSCAPE:2021ehl}.

In all of these models, the jet transport coefficient $\hat{q}$ can be used to quantify the interaction strength between jet partons and the medium \cite{Baier:1996kr, Baier:1996sk, Baier:1998kq,Guo:2000nz, Wang:2001ifa, Majumder:2009ge}. Microscopically, $\hat{q}$ can be related to the gluon field strength correlation or the local gluon number density of the QCD medium along the light-like trajectory of the hard parton \cite{Casalderrey-Solana:2007xns,PhysRevD.77.125010} in the QGP medium. Phenomenological values of $\hat{q}$ have been extracted through model-to-data comparisons using hadron suppression data  \cite{Chen:2010te,Qin:2007rn,Schenke:2009gb,Chen:2011vt,JET:2013cls,Liu:2015vna,Das:2015ana,Andres:2016iys,Cao:2017umt,Feal:2019xfl,Xie:2019oxg,Xie:2020zdb}. See Ref. \cite{Apolinario:2022vzg} for a compilation of the extracted values of $\hat q$ from different phenomenological studies. Among these studies, the JET Collaboration \cite{JET:2013cls} surveyed five different parton energy loss approaches with a realistic hydrodynamic description of the bulk evolution to determine $\hat{q}$ from single inclusive hadron suppression in the most central nuclear collisions at both RHIC and LHC. In the more advanced analysis recently performed by the JETSCAPE Collaboration \cite{JETSCAPE:2021ehl}, not only the newly-developed multi-stage approach of in-medium parton shower evolution was applied but the Bayesian inference technique was also employed for uncertainty quantification. Meanwhile, analyses that combine more than one type of observables have become available, for example within a modified partonic transport model, Ref. \cite{Ke:2020clc} used both inclusive hadron and jet suppression to calibrate $\hat{q}$ in central nuclear collisions at RHIC and LHC.

In this study, we use the next-to-leading order (NLO) pQCD parton model with medium-modified fragmentation functions obtained from the high-twist calculation to study the quenching of inclusive hadron, dihadron, and $\gamma$-hadron production.
We will carry out the first combined global analysis using existing world experimental data on all three channels of nuclear modifications: single inclusive hadron spectra, back-to-back dihadron yield, and $\gamma$-hadron yield in heavy-ion collisions at both RHIC and LHC energies with a wide range of centralities. 
We will focus in particular  on constraining the temperature dependence of $\hat{q}(T)$.
Since the QGP fireballs reach different initial temperatures in central and non-central collisions and in collisions at different beam energies, it is, therefore, important to include datasets from both RHIC and LHC with a wide coverage of centrality classes in order to maximize the sensitivity to the temperature dependence of $\hat{q}$.

Simultaneous use of the modifications of inclusive hadron, dihadron, and $\gamma$-hadron also utilizes the different geometric biases to probe the temperature dependence of $\hat{q}$.
Inclusive hadron spectra has a steep power-law shape in $p_{\rm T}$, $dN^h \sim 1/p_{\rm T}^N$ with $N\gg 1$, leading to an $R_{AA} \approx \int f(\Delta E) e^{-N \Delta E/p_{\rm T}} d\Delta E$ with $f(\Delta E)$ being the energy loss distribution. This makes inclusive hadron $R_{AA}$ mostly sensitive to the behavior of $f(\Delta E)$ at small energy loss.
Because $\Delta E$ strongly correlates with the path length of the hard parton traveling inside the QGP, this leads to the so-called surface bias that hadron yields at fixed $p_{\rm T}$ are dominated by the fragmentation of partons produced near the surface of the QGP with a smaller path length and smaller energy loss, limiting the sensitivity of the single inclusive hadron $R_{AA}$ (especially at intermediate $p_{\rm T}$) to the jet transport coefficient in the high-temperature region at the center of the fireball.
In the case of dihadron spectra, trigger bias is different from inclusive hadron. The initial parton production locations of the high-$p_{\rm T}$ trigger hadron are biased to the surface, while the lower $p_{\rm T}$ associated hadron on the away side will have to come from partons that traverse the whole length of the QGP and probe the hotter core of the QGP.
Nevertheless, if both the trigger and associate hadrons have large $p_{\rm T}$, the events are biased to the ``tangential'' geometry where both partons are produced at the surface and moves back-to-back parallel to the QGP surface.
Finally, in the $\gamma$-hadron process, since the trigger photon does not interact strongly with the medium, the photon $p_{\rm T}$ selection is free from any geometric bias and the modifications of the associated hadron distribution carry more information of the high-temperature inner region of the QGP fireball.
A quantitative study of the geometric bias in different observables is discussed in Refs. \cite{Zhang:2007ja,Zhang:2009rn}.
In this work, we will take advantage of these different biases to probe different regions of the QGP to extract the temperature-dependent jet transport coefficient.

We will not include data on full-jet suppression in this work, since the mechanism of full-jet suppression is more complicated and depends more directly on collective dynamics of QGP. Single-hadron suppression is only controlled by energy loss of hard-partons or their modified fragmentation functions, while jet modifications depend not only on the in-medium dynamics of hard shower partons but also jet-induced medium response, which transport the energy-momentum lost by jet shower partons to very large angles $\mathcal{O}(1)$.
Both contributions are needed to compute the full-jet energy loss, and parton transport model that couples 3+1D hydrodynamic evolution \cite{Chen:2017zte,PhysRevC.95.044909,Tachibana:2020mtb,JETSCAPE:2020uew} has been developed.
These simulations are computationally intense and may cause additional model-dependent uncertainty in modeling the jet-induced medium response. 

To propagate experimental constraints and uncertainties to a functional object--the temperature-dependent jet transport parameter $\hat{q}(T)$, we employ the newly developed information field Bayesian inference method that does not require an explicit parametrization of the functional form of $\hat q(T)$.
We have discussed its advantage in a companion short paper \cite{Xie:2022ght}.
The information field approach has been proposed to infer probability distribution functions (e.g. the parton distribution function) decades ago \cite{Bialek:1996kd,Periwal:1998ny,https://doi.org/10.48550/arxiv.physics/9912005} and is actively applied in astrophysics
\cite{Ensslin:2013ji,PhysRevD.80.105005}. The information field approach treats the prior distribution of the unknown function ($\hat{q}$(T)) as a random field over the input variable (temperature $T$), which greatly generalizes the prior functional distribution to reduce unnecessary bias when choosing an explicit functional parametrization. 
Specifically to our application to extract $\hat{q}(T)$, the random field prior strongly suppresses long-range correlations between prior values of $\hat{q}(T)$ in different temperature regions; therefore,
the experimental data for collisions with different centralities and at different colliding energies can provide robust, independent, and progressive constraints on the temperature dependence of $\hat q(T)$. 
In principle, with the inclusion of dihadron and $\gamma$-hadron data, one would expect improved constraining power on the jet transport coefficient than past efforts with only single inclusive hadron spectra, if the large experimental uncertainties in the correlation measurements were under control.

The remainder of the paper is organized as follows. In Sec. \ref{sec:phy}, we will give a brief introduction to the NLO pQCD parton model for single inclusive hadron, dihadron, and $\gamma$-hadron spectra in both high-energy proton-proton ($p+p$) and nucleus-nucleus ($A+A$) collisions with medium modified fragmentation functions. In Sec. \ref{sec:stat}, we will describe the machine learning assisted Bayesian inference method with the information field approach, the selection of prior of an unknown function, and a discussion of the sensitivity of the observables to the temperature dependence of $\hat q(T)$ (Sec. \ref{sec:stat:sensitivity}).  
Main results are presented in Sec. \ref{sec:discussion}, where we will discuss how $\hat{q}(T)$ are progressively constrained in different regions of temperature by combining datasets from different beam energies and centralities (Sec. \ref{sec:discussion:qhat_vs_T}), the momentum dependence of $\hat{q}$ (Sec. \ref{sec:discussion:qhat_vs_p}), and the impact of dihadron and $\gamma$-hadron measurements (Sec. \ref{sec:discussion:Raa_vs_IAA_qhat}).
In Sec. \ref{sec:validation}, we will validate the constrained $\hat{q}(T)$ by computing azimuthal anisotropy of high-$p_T$ hadrons and also make predictions to jet quenching observables for future measurements in oxygen-oxygen collisions.
Before summarizing in Sec. \ref{sec:summary}, we also consider in Sec. \ref{sec:new-data} the possibility of adding two more data sets, including the modified fragmentation function of $\gamma$-jet events, into the main analysis.

\begin{figure*}[htb!]
\begin{center}
\includegraphics[width=1.0\textwidth]{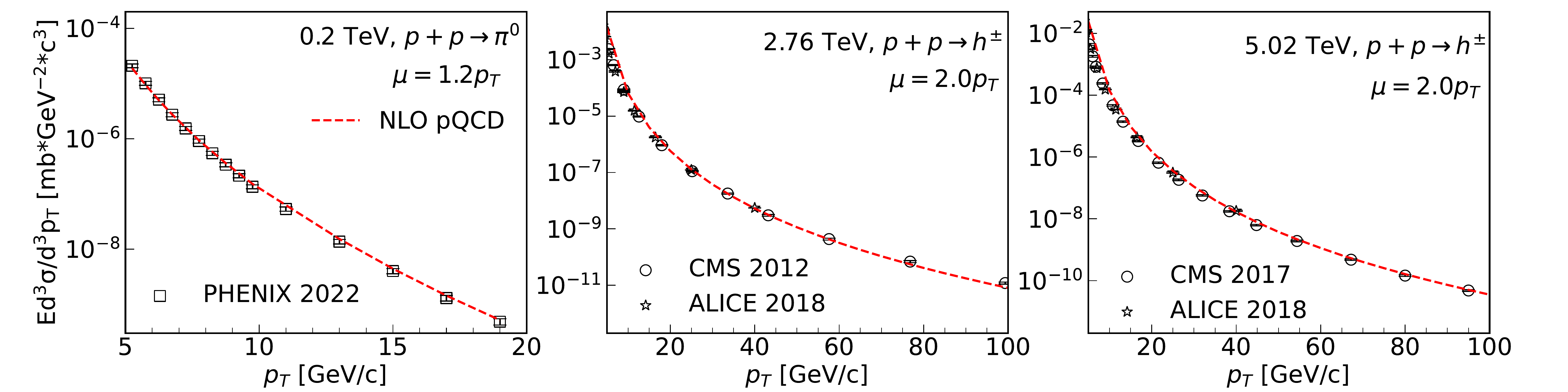}
\end{center}
\vspace{-5mm}
\caption{Cross sections of single hadron production in $p+p$ collisions compared with experimental data \cite{PHENIX:2021dod,PHENIX:2008saf,CMS:2012aa,Acharya:2018qsh, Khachatryan:2016odn}. From left to right corresponds to colliding energies $\sqrt{s_{\rm NN}}=0.2$, 2.76 and 5.02 TeV.}
\label{fig:pp-h}
\end{figure*}

\begin{figure*}
\begin{center}
\includegraphics[width=0.33\textwidth]{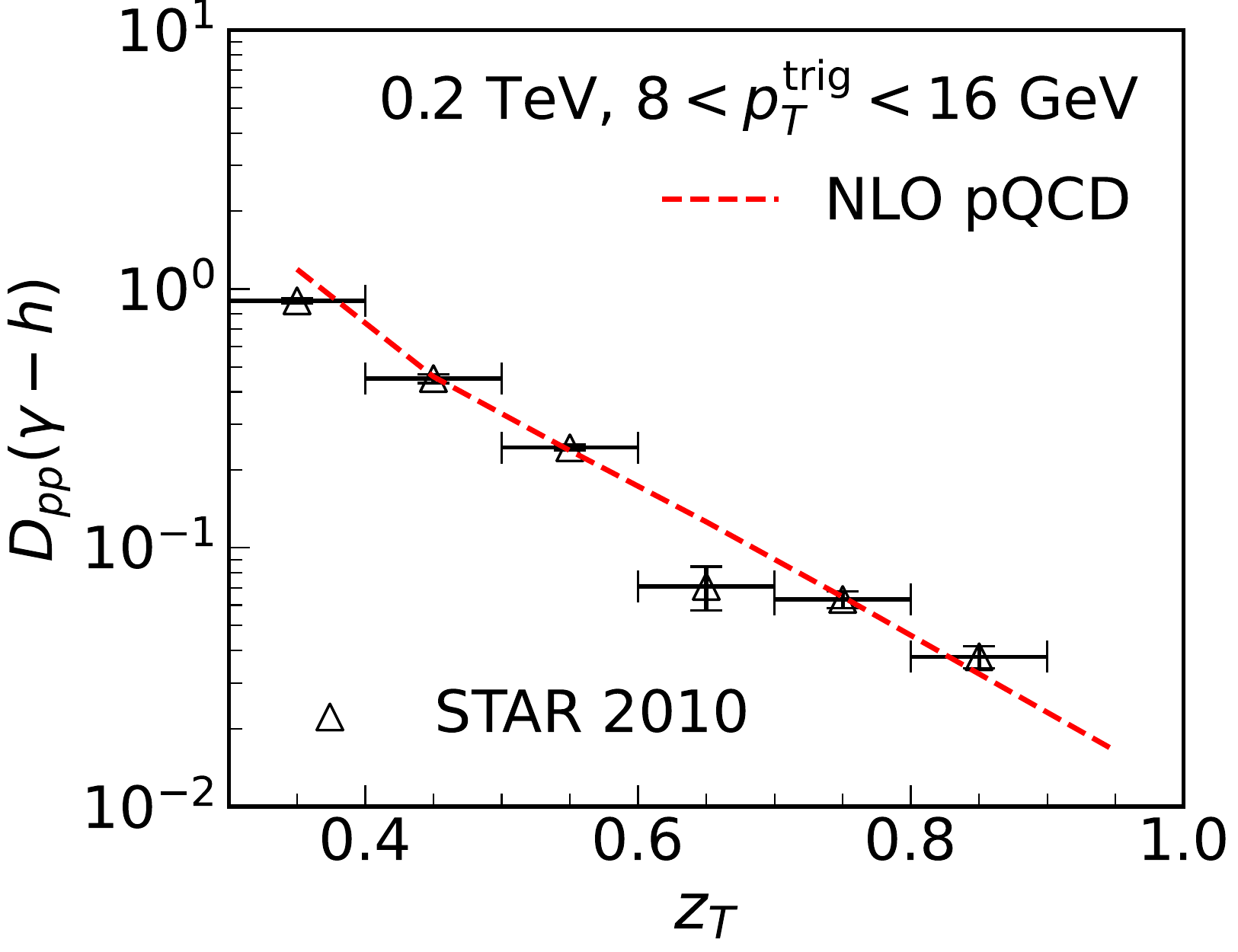}
\includegraphics[width=0.32\textwidth]{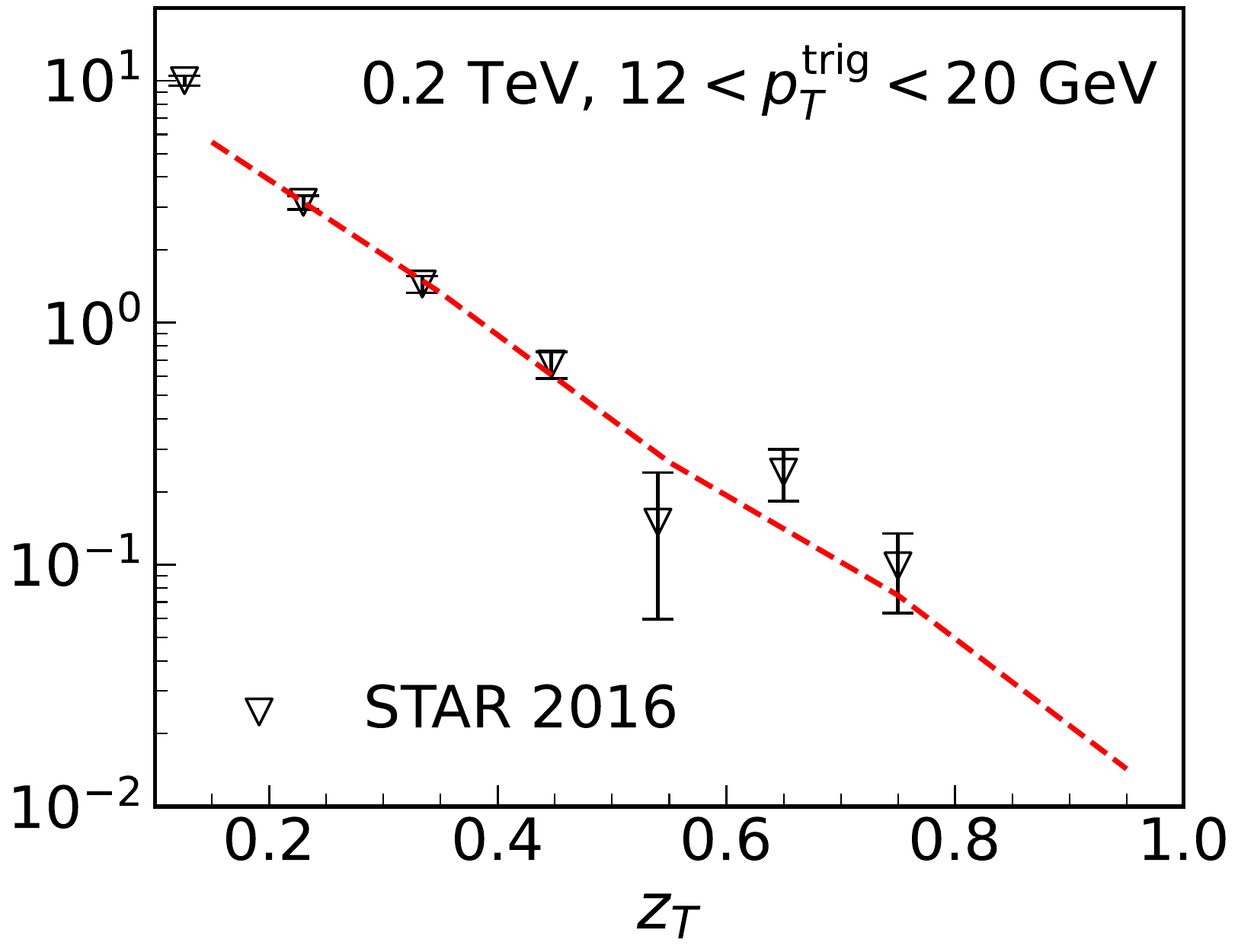}
\end{center}
\vspace{-5mm}
\caption{$\gamma$-triggered fragmentation functions in $p+p$ collisions at $\sqrt{s_{\rm NN}}=0.2$ TeV with ($8<p_{\rm T}^{\rm trig}<16$ GeV$/c$, $3<p_{\rm T}^{\rm assoc}<16$ GeV$/c$) (left plot) and ($12<p_{\rm T}^{\rm trig}<20$ GeV$/c$, $1.2<p_{\rm T}^{\rm assoc}<p_{\rm T}^{\rm trig}$) (right plot) and compared with STAR data \cite{Abelev:2009gu,STAR:2016jdz}.}
\label{fig:Dpp-gam-had}
\end{figure*}

\begin{figure*}[htb!]
\begin{center}
\includegraphics[width=1.0\textwidth]{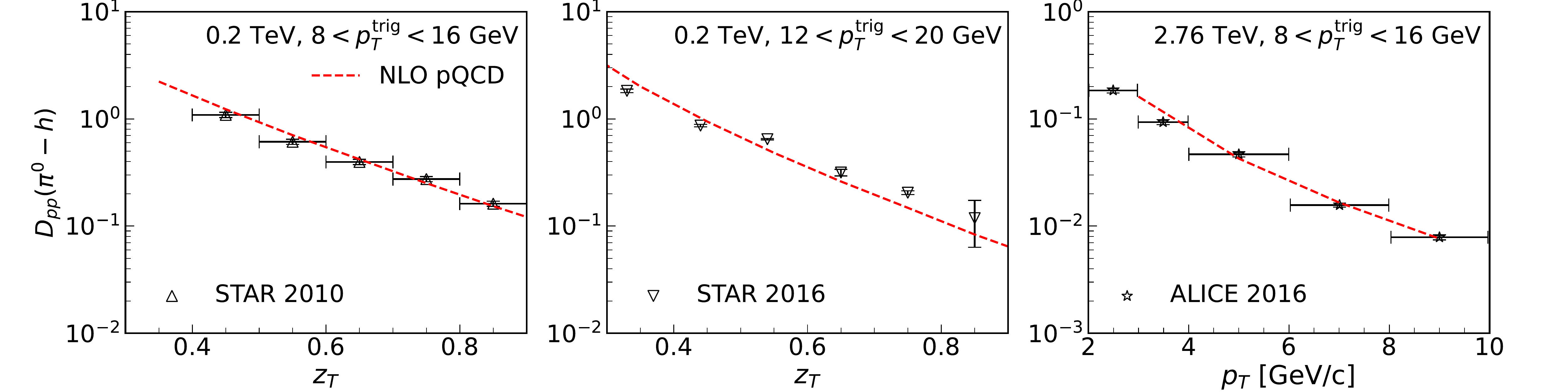}
\end{center}
\vspace{-5mm}
\caption{$\pi^{0}$-triggered fragmentation functions in $p+p$ collisions at $\sqrt{s_{\rm NN}}=0.2$ TeV ($8<p_{\rm T}^{\rm trig}<16$ GeV$/c$, $3<p_{\rm T}^{\rm assoc}<16$ GeV$/c$) (left plot) and ($12<p_{\rm T}^{\rm trig}<20$ GeV$/c$, $1.2<p_{\rm T}^{\rm assoc}<p_{\rm T}^{\rm trig}$) (middle plot) and at $\sqrt{s_{\rm NN}}=2.76$ TeV with ($8<p_{\rm T}^{\rm trig}<16$ GeV$/c$, $2<p_{\rm T}^{\rm assoc}<10$ GeV$/c$) (right plot), as compared with STAR data \cite{Abelev:2009gu,STAR:2016jdz} and ALICE preliminary data.}
\label{fig:Dpp-dihad}
\end{figure*}

\section{\label{sec:phy}NLO pQCD parton Model} 

In this section, we will briefly review the next-to-leading-order (NLO) perturbative QCD (pQCD) calculations in the collinear factorized parton model for single inclusive hadron, $\gamma$-hadron and dihadron production cross sections at large $p_{\rm T}$ in proton-proton ($p+p$) and nucleus-nucleus ($A+A$) collisions.

\subsection{pQCD parton model for $p+p$ collisions}
In high-energy $p+p$ collisions, the invariant cross section for single inclusive hadron production at large transverse momentum $p_{\rm T}$ within the parton model can be factorized into the convolution of collinear parton distribution functions (PDFs), short-distance partonic cross-sections and the collinear fragmentation functions (FFs) \cite{Owens:1986mp,CTEQ:1993hwr}.
The hadron transverse-momentum $p_{\rm T}^h$ and rapidity $y_h$ differential cross-section is
\begin{eqnarray}
	\frac{d\sigma_{pp}^h}{dy_{h}d^2p_{\rm T}^{h}}&&=\sum_{abcd}\int dx_a dx_b f_{a/p}(x_a,\mu^2) f_{b/p}(x_b,\mu^2)  \nonumber \\
	&&\times \frac{1}{\pi z_c}\frac{d\sigma_{ab\rightarrow cd}}{d\hat{t}}D_{h/c}(z_c,\mu^2)+ \Delta\sigma_{pp}^h(\alpha_s^3),
\label{eq:pp-sin-spec}
\end{eqnarray}
where the tree-level $2\rightarrow2$ partonic scattering cross-sections $d\sigma_{ab\rightarrow cd}/d\hat{t}$ depend on the partonic Mandelstam variables $\hat s=x_ax_bs$, $\hat t=-p_{\rm T}^hx_a\sqrt{s}e^{-y_h}/z_c$, $\hat u=-p_{\rm T}^hx_a\sqrt{s}e^{y_h}/z_c$. The final hadron momentum fraction $z_c$ is given by the identity for massless two-body scattering, $\hat s+ \hat t+\hat u=0$. 
The NLO correction $\Delta\sigma_{pp}^h$ at order $\mathcal {O}(\alpha_s^3)$ contains $2 \to 3$ processes at the tree level and virtual corrections to the $2 \to 2$ processes. 
Finally, we take the CT14 parameterization \cite{Hou:2016nqm} of the parton distribution functions of a free nucleon $f_{a/p}(x_a,\mu^2)$ and the Kniehl-Kramer-Potter parametrization \cite{Kniehl:2000fe} of parton fragmentation functions $D_{h/c}(z_c,\mu^2)$ in the vacuum. 
The renormalization scales are given by $\mu=1.2 p_{\rm T}^h$ at the RHIC energy and $\mu=2.0 p_{\rm T}^h$ at the LHC energies to describe large $p_T$ hadron production in $p+p$ collisions.

The dihadron cross-section in pQCD collinear factorized parton model can be expressed as \cite{Owens:1986mp},
\begin{eqnarray}
\frac{d\sigma_{pp}^{hh}}{dy_{h_c} d^2p_{{\rm T}}^{h_c} dy_{h_d} d^2p_{{\rm T}}^{h_d}}
	&&= \sum_{abcd}\int\hspace{-4pt} dz_c dz_df_{a/p}(x_a, \mu^2) f_{b/p}(x_b, \mu^2) \nonumber\\
	&&\hspace{-0.5 in}\times  \frac{x_a x_b}{\pi z_c^2 z_d^2} \frac{d\sigma_{ab\rightarrow cd}}{d\hat{t}} D_{h/c}(z_c, \mu^2) D_{h/d}(z_d, \mu^2) \nonumber\\
	&&\hspace{-0.5 in}\times \delta^2 (\vec{p}_{\rm T}^{~c} + \vec{p}_{\rm T}^{~d}) + \Delta\sigma_{pp}^{hh}(\alpha_s^3),
\label{eq:pp-dihadron}
\end{eqnarray}
where $\vec p_{{\rm T}}^{~c}=\vec p_{{\rm T}}^{~h_c}/z_c$ and $\vec p_{{\rm T}}^{~d}=\vec p_{{\rm T}}^{~h_d}/z_d$ are the transverse momenta of parton $c$ and $d$ from the hard processes that fragment into hadron $h_c$ with transverse momentum $\vec p_{{\rm T}}^{~h_c}$ and hadron $h_d$ with
transverse momentum $\vec p_{{\rm T}}^{~h_d}$, respectively. The renormalization scales are given by $\mu=1.2 M$ at the RHIC energy and $\mu=2.0 M$ at the LHC energies, where $M$ is the invariant mass of parton $c$ and $d$, $M = \sqrt{(p_c +p_d)^2}$ \cite{Zhang:2007ja}. The NLO correction $\Delta\sigma_{pp}^{hh}(\alpha_s^3)$ also contains $2 \to 3$ real corrections and virtual corrections to $ 2\to 2$ processes.

To compute the per-trigger $\gamma$-hadron yield, we will need the cross-section for direct photon production in $p+p$ collisions \cite{Owens:1986mp, Zhou:2010zzm},
\begin{eqnarray}
	\frac{d\sigma_{pp}^{\gamma}}{dy_{\gamma}d^2p_{\rm T}^{\gamma}}&&=\sum_{abd}\int_{x_{a{\rm min}}}^1 dx_a f_{a/p}(x_a,\mu^2) f_{b/p}(x_b,\mu^2)\nonumber \\
&&\times
	\frac{2}{\pi} \frac{x_a x_b}{2x_a-x_{\rm T} e^{y_\gamma}}\frac{d\sigma_{ab\rightarrow {\gamma}d}}{d\hat{t}}+\Delta\sigma_{pp}^{\gamma}(\alpha_e \alpha_s^2),
\label{eq:pp-pho-spec}
\end{eqnarray}
where $x_{\rm T}=2p_{\rm T}^{\gamma}/\sqrt{s}$, $x_b=x_ax_{\rm T}e^{-y_\gamma}/(2x_a-x_{\rm T}e^{y_\gamma})$, $x_{a{\rm min}}=x_{\rm T}e^{y_\gamma}/(2-x_{\rm T}e^{-y_\gamma})$. 
The partonic cross sections $d\sigma_{ab\rightarrow {\gamma}d}/d\hat{t}$ in the LO $2 \to 2$ processes only include Compton scattering $qg\rightarrow q\gamma$ and annihilation $q\bar{q}\rightarrow g\gamma$ sub-processes.  Fragmentation photons are ignored here due to their small contribution to cross sections with isolation cuts \cite{Zhang:2009rn,Xie:2020zdb}. 
Similarly, the cross-section for $\gamma$-hadron production is \cite{Owens:1986mp},
\begin{eqnarray}
	\frac{d\sigma_{pp}^{\gamma h}}{dy_\gamma d^2p_{\rm T}^{\gamma} dy_{h} d^2p_{\rm T}^{h}}&&=\sum_{abd}\int dz_d f_{a/p}(x_a,\mu^2) f_{b/p}(x_b,\mu^2) \nonumber \\
	&& \times \frac{x_ax_b}{\pi z_d^2} \frac{d\sigma_{ab\rightarrow {\gamma}d}}{d\hat{t}} D_{h/d}(z_d,\mu^2) \nonumber \\
	&& \times \delta^2(\vec{p}_{\rm T}^{~\gamma}+\frac{\vec{p}_{\rm T}^{~h}}{z_d})\ + \Delta\sigma_{pp}^{\gamma h}(\alpha_e \alpha_s^2),
\label{eq:pp-pho-h}
\end{eqnarray}
where $z_d=p_{\rm T}^h/p_{\rm T}^d$, and $\mu=1.2M$ at the RHIC energy.

In both dihadron and $\gamma$-hadron production, the associated hadron spectra per trigger are often called triggered fragmentation functions $D_{pp}^{\rm trig}(z_{\rm T})$ with $z_{\rm T}=p_{\rm T}^{\rm assoc}/p_{\rm T}^{\rm trig}$ and the trigger being either a high-$\pT$ hadron or a direct photon. It is defined in $p+p$ collisions as \cite{Wang:2003aw},
\begin{eqnarray}
	D_{pp}^{\rm trig}(z_{\rm T}) =  p_{\rm T}^{\rm trig}
	\frac{  \frac{d\sigma_{pp}}{dy^{\rm trig}dp_{\rm T}^{\rm trig}dy^{\rm assoc}dp_{\rm T}^{\rm assoc}} }{\frac{d{\sigma}_{pp}}{dy^{\rm trig}dp_{\rm T}^{\rm trig}}   },
\label{eq:D_pp}
\end{eqnarray}
where the numerator is the cross section of triggered-hadron production, the denominator is the cross section of single inclusive trigger particle cross section.
The triggered fragmentation functions are sometimes also expressed as a function of the associated hadron momentum  $D_{pp}^{\rm trig}(\pT^{\rm assoc})=D_{pp}^{\rm trig}(z_{\rm T})/p_{\rm T}^{\rm trig}$.

We have verified that the NLO pQCD parton model provides a good description of the baseline cross-section measured in $p+p$ collisions.
Shown in Fig.~\ref{fig:pp-h} are the numerical results for the single inclusive hadron cross-sections in $p+p$ collisions as compared to the experimental data at both RHIC and LHC \cite{PHENIX:2008saf,CMS:2012aa,Acharya:2018qsh, Khachatryan:2016odn} with good agreement. 
From left to right, each plot corresponds to the cross-section at colliding energies $\sqrt{s_{\rm NN}}=0.2$ TeV, 2.76 TeV, and 5.02 TeV, respectively.

The direct photon spectra from the above NLO pQCD parton model have been compared to the experimental data systematically in Ref. \cite{Xie:2020zdb} at different colliding energies. 
Contributions from the fragmentation photons were also found to be less than 10\% \cite{Zhang:2009rn}. 
Therefore, direct photon-triggered fragmentation functions are sufficient to describe the experimental data with the isolation cut \cite{Xie:2020zdb}. The direct photon-triggered fragmentation functions from the NLO pQCD parton model are shown in Fig. \ref{fig:Dpp-gam-had} with the experimental data for two different $p_{\rm T}^{\rm trig}$ ranges in $p+p$ collisions at $\sqrt{s_{\rm NN}}=0.2$ TeV. 

Finally, the hadron-triggered fragmentation functions from the NLO pQCD parton model are shown in Fig.~\ref{fig:Dpp-dihad} as compared to the experimental data \cite{Abelev:2009gu,STAR:2016jdz}. The left and middle plots are for colliding energy $\sqrt{s_{\rm NN}}=0.2$ TeV with two different $p_{\rm T}^{\rm trig}$ ranges, and the right plot is for $\sqrt{s_{\rm NN}}=2.76$ TeV. 

These comparisons between numerical results and experimental data give us confidence that the NLO pQCD parton model can give a good description of all the experimental data of single hadron, $\gamma$-hadron, and dihadron production at large $p_{\rm T}$ in baseline $p+p$ collisions.

\subsection{pQCD parton model for $A+A$ collisions}

The pQCD parton model for particle production can be extended to nucleus-nucleus collisions with the vacuum fragmentation functions replaced by medium-modified ones and the proton's PDF's by the nuclear PDF's. At a fixed impact parameter $\vec{b}$ for $A+B$ collisions, the single inclusive hadron spectra at large $p_{\rm T}$, similar to Eq. (\ref{eq:pp-sin-spec}), can be expressed as \cite{Chen:2010te,Liu:2015vna},
\begin{eqnarray}
	\frac{dN_{AB}^h}{dy_{h}d^2p_{\rm T}^h} &&=\sum_{abcd} \int d^2r dx_a dx_b t_A(\vec{r}) t_B(\vec{r}+\vec{b}) \nonumber \\
	&& \hspace{-0.5in}\times f_{a/A}(x_a,\mu^2,\vec{r})f_{b/B}(x_b,\mu^2,\vec{r}+\vec{b}) \nonumber \\
	&& \hspace{-0.5in} \times \frac{1}{\pi z_c}\frac{d\sigma_{ab\rightarrow cd}}{d\hat{t}}\tilde{D}_{h/c}(z_c,\mu^2,\Delta{E_c}) + 
	\Delta N_{AB}^h(\alpha_s^3).
\label{eq:AA-sin-spec}
\end{eqnarray}

Similarly, the diharon spectra in $A+B$ collisions, according to Eq.(\ref{eq:pp-dihadron}), can be written as \cite{Zhang:2007ja},
\begin{eqnarray}
\frac{dN_{AB}^{hh}}{dy_{h_c} d^2p_{{\rm T}}^{h_c} dy_{h_d} d^2p_{{\rm T}}^{h_d}}
	&&= \sum_{abcd}\int d^2 r dz_c dz_d t_A(\vec{r}) t_B(\vec{r}+\vec{b}) \nonumber \\
	&&\hspace{-0.8 in} \times f_{a/A}(x_a,\mu^2,\vec{r})f_{b/B}(x_b,\mu^2,\vec{r}+\vec{b}) 
	\frac{x_a x_b}{\pi z_c^2 z_d^2} \nonumber \\
	&& \hspace{-0.8 in} \times \frac{d\sigma_{ab\rightarrow cd}}{d\hat{t}} \tilde{D}_{h/c}(z_c,\mu^2,\Delta{E_c}) \tilde{D}_{h/d}(z_d,\mu^2,\Delta{E_d}) \nonumber \\
	&& \hspace{-0.8 in} \times \delta^2 (\vec{p}_{\rm T}^{~c} + \vec{p}_{\rm T}^{~d}) + \Delta N_{A B}^{hh}(\alpha_s^3).
\label{eq:AA-dihadron}
\end{eqnarray}

For direct $\gamma$ spectrum in $A+B$ collisions, one only needs to consider the cold nuclear matter effect in the nuclear PDF's in the initial state \cite{Zhou:2010zzm},
\begin{eqnarray}
	\frac{dN_{A B}^{\gamma}}{dy_{\gamma}d^2p_{\rm T}^{\gamma}}&&=\sum_{abd} \int d^2 r\int_{x_{a{\rm min}}}^1 dx_a t_A(\vec{r}) t_B(\vec{r}+\vec{b}) \nonumber \\
	&&\hspace{-0.3 in}\times f_{a/A}(x_a,\mu^2,\vec{r})  f_{b/B}(x_a,\mu^2,\vec{r}+\vec{b})\nonumber \\
     &&\hspace{-0.3 in}\times \frac{2}{\pi} \frac{x_a x_b}{2x_a-x_{\rm T} e^{y_\gamma}} \frac{d\sigma_{ab\rightarrow {\gamma}d}}{d\hat{t}} +\Delta N_{AB}^\gamma(\alpha_e \alpha_s^2).
\label{eq:AA-pho-spec}
\end{eqnarray}

Similarly to Eq.(\ref{eq:pp-pho-h}), the spectrum for $\gamma$-hadron production in $A+B$ collisions can be given as \cite{Zhang:2009rn},
\begin{eqnarray}
	\frac{dN_{A B}^{\gamma h}}{dy_\gamma d^2p_{\rm T}^{\gamma} dy_h d^2p_{\rm T}^{h}}&&=\sum_{abd}\int d^2 r dz_d t_A(\vec{r}) t_B(\vec{r}+\vec{b}) \nonumber \\
	&& \hspace{- 1 in}\times f_{a/A}(x_a,\mu^2,\vec{r})f_{b/B}(x_b,\mu^2,\vec{r}+\vec{b}) \frac{x_a x_b}{\pi z_d^2} \frac{d\sigma_{ab\rightarrow {\gamma}d}}{d\hat{t}} \nonumber \\
	&& \hspace{-1 in} \times  \tilde{D}_{h/d}(z_d,\mu^2,\Delta{E_d}) \delta^2(\vec{p}_{\rm T}^{~\gamma}+\frac{\vec{p}_{\rm T}^{~h}}{z_d}) + \Delta N_{AB}^{\gamma h}(\alpha_e \alpha_s^2).
\label{eq:AA-pho-h}
\end{eqnarray}

Correspondingly, the triggered fragmentation functions in $A+B$ collisions are defined as,
\begin{eqnarray}
	D_{AB}^{\rm trig}(z_{\rm T}) =  p_{\rm T}^{\rm trig}
	\frac{  \frac{d\sigma_{AB}}{dy^{\rm trig}dp_{\rm T}^{trig}dy^{\rm assoc}dp_{\rm T}^{\rm assoc}} }{  \frac{d{\sigma}_{AB}}{dy^{\rm trig}dp_{\rm T}^{\rm trig}}   }.
\label{eq:D_AA}
\end{eqnarray}

In the above equations for single hadron, dihadron, direct photon and $\gamma$-hadron spectra in $A+B$ collisions, $t_A$ and $t_B$ are the nuclear thickness functions of the projectile and target nucleus, shifted by the impact parameter $\vec{b}$. They are given by integrating the Woods-Saxon nuclear density distributions along the beam direction and are normalized to the mass number $A$ of each nucleus, e.g., $\int d^2 r t_A(\vec{r})=A$.

We take into account both the cold nuclear matter (CNM) effects through the nuclear modification of the PDF's in the initial state and parton energy loss in hot QGP medium in the final state. We do not consider initial-state parton energy loss in the nucleus, which is shown to be a small effect for hadron production in the central rapidity region \cite{Vitev:2007ve,Kang:2015mta,Ke:2022gkq}.
The nuclear parton distribution functions (nPDFs) $f_{a/A}(x_a,\mu^2,\vec{r})$ are factorized into iso-spin dependent nucleon PDF's and spatial-dependent nuclear modification factor $S_{a/A}(x_a,\mu^2,\vec{r})$ \cite{Wang:1996yf,Wang:1998ww,Li:2001xa,Chen:2008vha},
\begin{eqnarray}
f_{a/A}(x_a,\mu^2,\vec{r}) &&= S_{a/A}(x_a,\mu^2,\vec{r})\left[\frac{Z}{A}f_{a/p}(x_a,\mu^2)\right. \nonumber\\
&&+\left.\left(1-\frac{Z}{A}\right)f_{a/n}(x_a,\mu^2)\right],
\label{eq:fa/A}
\end{eqnarray}
where $Z$ is the charge number of the nucleus. The spatial dependence of the nuclear modification factor $S_{a/A}(x_a,\mu^2,\vec{r})$ for nPDF's is modeled as \cite{Emelyanov:1999pkc,Hirano:2003pw,Zhou:2010zzm},
\begin{eqnarray}
S_{a/A}(x_a,\mu^2,\vec{r})=1+ A \frac{t_A(\vec{r})[S_{a/A}(x_a,\mu^2)-1]}{\int{d^2}r [t_A(\vec{r})]^2},
\label{eq:Sa/A}
\end{eqnarray}
where the spatial-averaged nuclear modification factor $S_{a/A}(x_a,\mu^2)$ is taken from the EPPS16 \cite{Eskola:2016oht} parametrization.

The medium-modified fragmentation functions (mFFs) $\tilde{D}_{h/d}(z_d,\mu^2,\Delta{E_d})$ as a result of medium-induced emissions and parton energy loss inside the hot QGP medium are given by \cite{Wang:2004yv,Zhang:2007ja,Zhang:2009rn},
\begin{eqnarray}
	&& \tilde{D}_{h/d}(z_d,\mu^2,\Delta{E_d}) = (1-e^{-\langle{N_g^d}\rangle})\left[\frac{{z_d}'}{z_d}D_{h/d}({z_d}',\mu^2)\right. \nonumber\\
&& ~~+\left.{\langle{N_g^d}\rangle}\frac{{z_g}'}{z_d}D_{h/g}({z_g}',\mu^2)\right]+e^{-\langle{N_g^d}\rangle}D_{h/d}({z_d},\mu^2),
\label{eq:mFF}
\end{eqnarray}
where ${z_d}'=p_{\rm T}^{h}/(p_{\rm T}^d-\Delta{E_d}$) is the momentum fraction of a hadron from the fragmentation of parton $d$ with initial transverse momentum $p_{\rm T}^d$ after losing a total amount of energy $\Delta{E_d}$ in the medium,
$z_d$ is the momentum fraction of a hadron from a parton fragmentation in vacuum, ${z_g}'=\langle{N_g^d}\rangle p_{\rm T}^{h}/\Delta{E_d}$ is the momentum fraction of a hadron from the fragmentation of a radiated gluon that carries an average energy $\Delta{E_d}/\langle{N_g^d}\rangle$. The number of radiated gluons is assumed to follow the Poisson distribution with the average number $\langle{N_g^d}\rangle$. In the above equation, the first weighting factor $1-e^{-\langle{N_g^d}\rangle}$ is the probability for a parton to radiate at least one gluon induced by multiple scattering. The second factor $e^{-\langle{N_g^d}\rangle}$ is the probability of no induced gluon radiation. Considering that the elastic energy loss of the jet is much smaller than the inelastic energy loss \cite{Qin:2007rn,Cao:2016gvr},  we only take the radiative parton energy loss into account in this work. Nevertheless, we should keep in mind that collisional processes can be important at low and intermediate $p_{\rm T}$.

The radiative parton energy loss $\Delta{E_d}$ for a light quark $d$ with initial energy $E$ can be calculated within the high-twist approach \cite{Wang:2004yv,Zhang:2007ja,Zhang:2009rn},
\begin{eqnarray}
\frac{\Delta{E}_d}{E} &&= \frac{2C_A\alpha_s}{\pi} \int_{\tau_0}^{\infty} d\tau \int_{0}^{E^2} \frac{dl_{\rm T}^2}{l_{\rm T}^2\left(l_{\rm T}^2+{\mu_D}^2\right)} \nonumber \\
	&&\times  \int_{\epsilon}^{1-\epsilon} dz  \left[1+(1-z)^2\right] \nonumber \\
&& \times \hat{q}_d(\tau, \vec r+(\tau-\tau_0)\vec n) \sin^2\left[\frac{l_{\rm T}^2(\tau-\tau_0)}{4z(1-z)E}\right],
\label{eq:deltaE}
\end{eqnarray}
which is integrated over the quark propagation path starting from the initial transverse position $\vec r$ at an initial time $\tau_0=0.6$ fm/$c$ along the direction $\vec n$.
Here $C_A=3$, $\alpha_s$ is the strong coupling constant, $l_{\rm T}$ is the transverse momentum of the radiated gluon, and $z$ is its longitudinal (along the jet direction) momentum fraction.
$\mu_D=\sqrt{(1+n_f/6)}gT$ is the plasma Debye screening mass with $n_f=3$ the number of active flavors. $\epsilon=\frac{1}{2}[1-\sqrt{1-(2l_{\rm T}^2/E\omega)}]$ and $l_{\rm T}^2 \leqslant 2E\omega$ are the constraints imposed on the integral limit of $z$ and $l_{\rm T}^2$, where $E$ is the energy of the hard parton and $\omega$ is the average medium parton energy. Two kinematic constraints $l_{\rm T}^2/E^2 \leqslant z^2$ and $l_{\rm T}^2/E^2 \leqslant (1-z)^2$ are also imposed in the calculations.
The parton energy loss for a propagating gluon is assumed to be the same, except that the jet transport coefficient of a gluon $\hat{q}_A$ differs from that of quark $\hat{q}_F$ by a color factor $C_A/C_F$. Therefore, the radiative energy loss of a gluon is $9/4$ times that of a quark \cite{Wang:2009qb,Deng:2009ncl}. The average number of radiated gluons $\langle{N_g^d}\rangle$ from the propagating parton $d$ is \cite{Chang:2014fba},
\begin{eqnarray}
	\langle N_g^d \rangle &&= \frac{2C_A\alpha_s}{\pi} \int_{\tau_0}^{\infty} d\tau \int_{0}^{E^2} \frac{dl_{\rm T}^2}{l_{\rm T}^2\left(l_{\rm T}^2+{\mu_D}^2\right)} \nonumber \\
	&&\times  \int_{\epsilon}^{1-\epsilon} \frac{dz}{z}  \left[1+(1-z)^2\right]  \nonumber \\
&& \times \hat{q}_d(\tau, \vec r+(\tau-\tau_0)\vec n) \sin^2\left[\frac{l_{\rm T}^2(\tau-\tau_0)}{4z(1-z)E}\right].
\label{eq:deltaNg}
\end{eqnarray}
Note that the Debye screening mass is necessary to regulate the average number of gluon emissions $\langle N_g^d \rangle$.

The jet transport coefficient in the hydrodynamic QGP fluid $\hat q=\hat q(T)p^\mu\cdot u_\mu/p^0$ depends both on the fluid velocity $u^\mu$ and the temperature $T$ in the local co-moving frame, where $p^{\mu}=(p_0,\vec{p})$ is the four-momentum of the parton. In this work, we only consider parton energy loss in the QGP phase with a pseduocritical temperature $T_c=0.165$ GeV.  Effectively, we assume the jet transport coefficient $\hat q(T)$ in the local co-moving frame vanishes below $T<T_c$.  The dynamical evolution of the QGP medium created in $A+A$ collisions is provided by the CLVisc 3+1 D hydrodynamics model simulations \cite{Pang:2012he,Pang:2014ipa,Pang:2018zzo}. 
The initial conditions for the hydrodynamic simulations are given by averaging over 200 events from the TRENTo initial-condition model \cite{Moreland:2014oya} for each range of centrality of collisions. An overall envelop function in the spatial rapidity is used to generalize the 2D TRENTo initial condition at middle rapidity to a 3D distribution. 
The parameters of the envelop function are fitted to the final charged hadron rapidity distribution \cite{Pang:2018zzo}. 
The overall normalization factors in the TRENTo initial conditions are also adjusted to fit the final total charged hadron multiplicity at each colliding energy. 

\subsection{Nuclear modification factors}

With the calculations of hadron spectra in $p+p$ and $A+A$ collisions explained above, we compute the nuclear modification factors for the single inclusive hadron spectra in nuclear collisions,
\begin{eqnarray}
R_{AB}=\frac{\int_{b_{\rm min}}^{b_{\rm max}}  \frac{dN_{AB}}{dyd^2p_{\rm T}} b db}{\int_{b_{\rm min}}^{b_{\rm max}} T_{AB}(\vec{b}) \frac{d{\sigma}_{pp}}{dyd^2p_{\rm T} } b db},
\label{eq:R_AA}
\end{eqnarray}
where $T_{AB}(\vec{b})=\int d^2 rt_A(\vec{r})t_B(\vec{r}+\vec{b})$ is the nuclear overlapping function of two colliding nuclei $A$ and $B$ at a given impact parameter $\vec{b}$ and $[b_{\rm min},b_{\rm max}]$ is the range of the impact-parameter corresponding to the given range of centrality selections in the experimental data.

For hadron-triggered and $\gamma$-triggered hadron spectra at large transverse momentum in $A+B$ collisions, nuclear modification factors are defined as the ratios of the hadron spectra per trigger or the triggered fragmentation functions in $A+B$ and $p+p$ collisions,
\begin{eqnarray}
I_{AB}(z_{\rm T})=\frac{D_{AB}(z_{\rm T})}{D_{pp}(z_{\rm T})}.
\label{eq:I_AA-zt}
\end{eqnarray}
Note that in the calculations of the triggered fragmentation functions according to Eq.~(\ref{eq:D_AA}), both the cross sections of the trigger (denominator) and trigger-hadron (numerator) are integrated over the range of the impact-parameter $[b_{\rm min},b_{\rm max}]$ for each given selection of centrality of $A+A$ collisions. 
The above nuclear modification factors are often also expressed as a function of the associated hadron momentum $I_{AB}(p_{\rm T}^{\rm assoc})$ with $D(p_{\rm T}^{\rm assoc})=D(z_{\rm T})/p_{\rm T}^{\rm trig}$ for a given value of $p_{\rm T}^{\rm trig}$.
These are the observables that we will rely on for the Bayesian inference of the temperature dependence of the jet transport coefficient $\hat q(T)$ in this paper.

\section{\label{sec:stat}Bayesian inference of $\hat{q}(T)$ with information field method} 

The NLO parton model with medium-induced parton energy loss, as outlined in the last section, has been successfully used to describe the experimental data on suppression of high-$p_{\rm T}$ single inclusive and triggered hadron spectra in high-energy heavy-ion collisions \cite{Zhang:2007ja,Zhang:2009rn,Chen:2011vt}. 
Using this method, the extracted effective jet transport coefficient (averaged over the range of initial temperatures in the corresponding collision systems), the only input of the model, is consistent with the values extracted by the JET collaboration \cite{JET:2013cls}. 
In this work, we will generalize the previous studies and perform a global Bayesian inference of the temperature dependence of $\hat q(T)$ using single hadron, dihadron and $\gamma$-hadron correlations. Since dihadron and $\gamma$-hadron correlations are known to be more sensitive to the parton energy loss, such a global analysis should produce a much more stringent constraint on the temperature dependence of the jet transport coefficient.

Furthermore, unlike earlier Bayesian analyses \cite{JETSCAPE:2021ehl,Ke:2020clc} where $\hat{q}(T)$ is parametrized explicitly, we will introduce the method of inferring unknown functions ($\hat{q}(T)$ in this case) from a non-parametric approaching using the information field method. This method is extremely flexible and avoids some unnecessary complications of using an explicit parametrization as explained below. It also greatly simplifies the analysis of temperature-dependence sensitivity of $\hat{q}(T)$ on different observables. 

\subsection{The prior distribution of an unknown function as a random field}
Bayesian inference of the jet transport coefficient has been performed in the past for heavy quarks \cite{Xu:2017obm,PhysRevC.98.064901,Liu:2021dpm}, for inclusive hadron and jet spectra \cite{Ke:2020clc}, and recently within the JETSCAPE Collaboration \cite{JETSCAPE:2021ehl}. 
However, in all existing statistical inferences in the literature of both the jet transport coefficient and the bulk transport coefficients \cite{PhysRevC.94.024907,JETSCAPE:2020mzn,JETSCAPE:2020shq}, one normally first parametrizes the temperature dependence of these transport coefficients as unknown functions with several parameters and then applies the Bayesian statistical method to infer these parameters. 
One then marginalizes over the posterior of these parameters to define the posterior distribution of the temperature-dependent $\hat{q}(T)$ and the bulk transport coefficients. 
Such an approach to extracting unknown functions is intuitive, and yet brings several problems to the statistical inference:
\begin{itemize}
\item First, a parametrization of $\hat{q}(T)$ is a strong (and often biased) form of prior assumption. Often, an explicit parametrization introduces long-range correlations among values of $\hat{q}(T)$ at different temperatures. The consequence is that data that should only probe the low-temperature region (for example, measurements in peripheral collisions or at lower beam energy) will also constrain $\hat{q}$ in the high-temperature region through correlations in the prior distributions.
Such a parametrization method not only restricts the generality of the final results but also brings the risk of introducing tension when fitting to low-temperature-sensitive and high-temperature-sensitive data sets simultaneously.
\item Second, modern Bayesian inference of the parameters of computationally intensive models heavily relies on machine learning acceleration, where ``model emulators'' are trained on a finite set of training data to fast infer the full-model predictions.
Although an explicit parametric form is intuitive and meaningful to humans, it is unnecessarily complicated for machine learning as it often contains strong non-linear correlations.  
For example, in the parametrization of a previous study \cite{Ke:2019jbh} $\hat{q}/T^3 = (1+(a T/T_c)^p)^{-1}$, a linear variation in $a$ or $p$ causes a strong non-linear change in the effective value of $\hat{q}$. 
Even though the physical model maps values of $\hat{q}$ to observables in a rather well-behaved fashion, the performance of the emulator can still be impaired by the non-linear response to those individual input parameters.
\item Third, generalization to higher-dimensional functions becomes increasingly complicated with explicit parametrization. For example, in Ref.~\cite{JETSCAPE:2021ehl}, the energy, temperature and virtuality ($Q$) dependence of $\hat{q}$ are considered. Unless one has a very clear physical guidance, the number of parameters and complexity increases exponentially with the number of control variables (e.g. $T$, $E$, and $Q$).
\end{itemize}

In an attempt to overcome these issues, we switch to describe the prior of an unknown function using methods from the information field theory (IFT) \cite{Bialek:1996kd,Ensslin:2013ji,https://doi.org/10.48550/arxiv.physics/9912005}, which provides a non-parametric way of representing an unknown function and propagates constraints and uncertainties from model-to-data comparison. We will first show the construction of the functional prior $\hat{q}(T)$ and then discuss how it avoids the above-mentioned problems.

We describe the unconstrained functional $F(x)$ by an unconditioned Gaussian random field that is specified by the one-point (mean) and two-point (correlation) functions,
\begin{eqnarray}
  && \langle F(x) \rangle = \mu(x),\\
  && \langle [F(x)-\mu(x)][F(x')-\mu(x')]\rangle = C(x, x'). 
\end{eqnarray}
To relate $\hat{q}(T)$ to $F(x)$, we make the following transformation,
\begin{eqnarray}
F &=& \ln \left( \hat{q}/T^3 \right),\\
x &=& \ln \left( T/{\rm GeV} \right).
\end{eqnarray}
The first logarithmic transformation is used to guarantee $\hat{q}>0$. 
The second transform comes from a physical consideration that there are no other relevant scales above the transition temperature. So it is reasonable to assume $\ln T$ is the natural measure of the variation of $\hat{q}(T)$.

\begin{figure}
    \centering
    \includegraphics[width=\columnwidth]{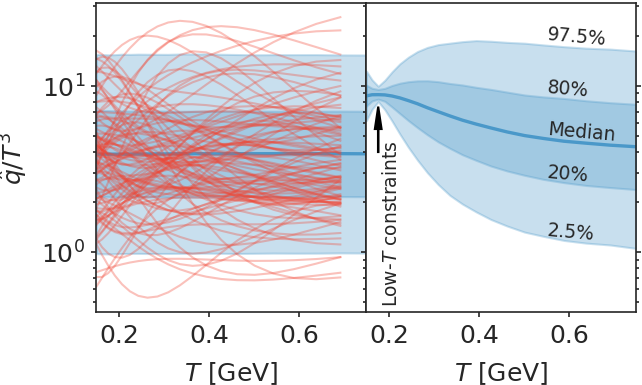}
    \caption{Left panel: the horizontal blue line indicates the median value of the prior distribution of $\hat{q}/T^3$. The darker and lighter bands show the 60\% and 95\% percentile region of the prior distribution. Red lines are the 100 random realizations of $\hat{q}/T^3$ pulled from the information field to train, which will be used to train the model emulator. Right panel: a conditional information field with $8<\hat{q}/T^3<10$ at $T=0.176$ GeV. The lines and bands show the median and percentile region of the conditional distribution.}
    \label{fig:design}
\end{figure}

We take the mean of the function to be a constant $\mu(x) \equiv \mu$ since we do not expect the dimensionless quantity $\hat{q}/T^3$ to change by orders of magnitude in the QGP phase within $T_c<T<4T_c$.
The correlation function $C(x, x')$ is assumed to be Gaussian with variance $\sigma^2$ and correlation length $L$,
\begin{eqnarray}
C(x, x') = \sigma^2 e^{-\frac{(x-x')^2}{2L^2}}.
\end{eqnarray} 
With the above assumptions, $\mu, \sigma$ and  $L$ control what type of functions are favored in the prior distribution, and in principle, they should have a certain degree of arbitrariness.
However, as a first application of this approach to a problem in heavy-ion collisions, we fix the prior values
$\mu=\left\langle\ln \hat{q} / T^3\right\rangle=1.36, \sigma=0.7$, and $L=\ln (2)$ with the following considerations,
\begin{itemize}
    \item With $\mu=1.36$ and $\sigma=0.7$, the resulting range of high-probability variation already covers most of the past analysis ranges $0.8<\hat{q}/T^3<15$ as can be seen in left panel\footnote{There are models that have preferred very large values of $\hat{q}$ near the transition temperature \cite{Xu:2014tda}. But as a post check in the current analysis, large values of $\hat{q}/T^3 > 10$ are already strongly disfavored, so we consider this range of prior to being adequate.} of Fig.~\ref{fig:design}.
    \item The choice of $L$ is much more subtle. 
    From the discussion in Ref.~\cite{Bialek:1996kd}, we argue that an optimal length scale should match the ``temperature resolution'' of the observables. Because the model always involves a temperature integration of $\hat{q}(T)$, we choose $L$ to be the smallest $\ln(T_{\rm max}/T_{\rm min})$ in the hydrodynamic simulations of all colliding systems and centralities. In our analysis, this is the 40-50\% centrality class of Au+Au collisions at $\sqrt{s}=200$ GeV with $T_{\rm min}=165$ MeV and $T_{\rm max}\approx 2T_{\rm min}$, i.e., $L=\ln 2$.
\end{itemize}

Despite these considerations, we acknowledge that fixing the values of $\mu$, $\sigma$, and especially $L$ can be subjective. In the future, one may consider relaxing this assumption and treat $\mu, \sigma$ and $L$ as hyper-parameters that can also vary in the statistical inference.
In appendix \ref{sec:app:varyL}, we have studied the impact of using different correlation lengths in a toy jet quenching model.

With this Gaussian random field approach,  the prior distribution of $F=\ln\hat{q}$ can be expressed as
\begin{eqnarray}
    P_0[F(x)]=\frac{\mathcal{N}}{\sqrt{\mathrm{det}(C)}}e^{-\frac{1}{2}\int dx dx' \delta F(x) C^{-1}(x, x') \delta F(x')},
    \label{eq:IF-prior}
\end{eqnarray}
with $\delta F(x)=F(x)-\mu(x)$ and $\mathcal{N}$ a normalization factor independent of $C(x,x')$. In principle, it does not exclude any possible $F(x)$, but the prior probability of a particular realization will be suppressed if its typical variation length scale is very different from $L$.

Using the transformations as mentioned above between $F(x)$ and $\hat{q}(T)$, we have plotted $N=100$ realizations of the $\hat{q}(T)$ from the information field prior in the left panel of Fig.~\ref{fig:design} (thin red lines). The horizontal blue line is the mean median of $\hat{q}$ at each temperature, and the blue bands are the 60\% and 95\% prior credible intervals. One may notice that some realizations are moderately non-monotonic above $T_c$, this is unavoidable due to the probabilistic nature of the random field prior.
In the right panel of Fig.~\ref{fig:design}, we illustrate why the information field prior can resolve the first problem of explicit parametrizations mentioned above.
Here, we condition the random functions to satisfy the constraint $8<\hat{q}/T^3<10$ at $T=0.176$ GeV and plot the 60\% and 95\% credible intervals of the conditioned prior. Note that the prior conditioned tightly at low temperatures recovers the unconditioned prior in the high-temperature region because values of the random field with input separation greater than the correlation length effectively decorrelates.
Therefore, even if data from peripheral or low-beam-energy collisions already constrained $\hat{q}$ near $T_c$, it would not impair the flexibility of an extended analysis that also incorporate data from central or high-energy collisions to probe $\hat{q}$ at high temperatures.

\subsection{Training model emulators with random function realizations}
An emulator is often used to speed up the model simulations of the experimental observables. The emulator-assisted Bayesian analyses have already been applied to the extraction of jet transport coefficient~\cite{JETSCAPE:2021ehl,Xu:2017obm}. We will modify such a workflow to adapt to the use of the information field prior. 

For a practical application, we only take values of each random realization at 20 equally spaced temperature points between $T=0.15$ GeV and $T=0.75$ GeV and construct $\hat{q}$ from a piece-wise linear interpolation, 
\begin{eqnarray}
\left[\frac{\hat{q}(T)}{T^3}\right]_i = &&\sum\limits_{j=1}^{20} \theta(T-T_j)\theta(T_{j+1}-T) \nonumber\\
&\times& \left( \left[\frac{\hat{q}}{T^3}\right]_{i,j+1} \Delta_j +\left[\frac{\hat{q}}{T^3}\right]_{i,j}  (1-\Delta_j) \right),\nonumber \\
\Delta_j = &&\frac{T-T_j}{T_{j+1}-T_j}.
\label{eq:qhat-Bayes}
\end{eqnarray}
The index $i$ labels the 100 realizations that will be fed into the medium-modified NLO calculations.
These 100 by 20 grid of values $\left[\hat{q}/T^3\right]_{i,j}$ will become the input parameter matrix. 

With these inputs, we compute the training observables, including the suppression of single inclusive hadron spectra $R_{AA}(p_{\rm T})$, $\gamma$-hadron $I_{AA}^{\gamma h}$ and dihadron correlation $I_{AA}^{hh}$ in Au+Au collisions at $\sqrt{s_{\rm NN}}=0.2$ TeV and Pb+Pb collisions at $\sqrt{s_{\rm NN}}=2.76$ TeV and 5.02 TeV with different centralities:
\begin{itemize}
    \item {$R_{AA}^{\pi^0}$ in 0-5\%, 0-10\%, 10-20\%, 20-30\%, 30-40\%, 40-50\% 
    Au+Au collisions at $\sqrt{s}=0.2$ TeV  \cite{PHENIX:2008saf,Adare:2012wg};}
     \item {$R_{AA}^{h^{\pm}}$ in 0-5\%, 5-10\%, 10-20\%, 20-30\%, 30-40\%, 40-50\%
     Pb+Pb collisions at $\sqrt{s}=2.76$ TeV  \cite{CMS:2012aa,Abelev:2012hxa,Aad:2015wga};}
     \item {$R_{AA}^{h^{\pm}}$ in 0-5\%, 5-10\%, 10-20\%, 10-30\%, 20-30\%, 30-50\%
     Pb+Pb collisions at $\sqrt{s}=5.02$ TeV \cite{Khachatryan:2016odn,Acharya:2018qsh};}
     \item {$I_{AA}^{\gamma h^{\pm}}$ in 0-10\% Au+Au collisions at $\sqrt{s}=0.2$ TeV  \cite{Abelev:2009gu,STAR:2016jdz};}
     \item {$I_{AA}^{\pi^0 h^{\pm}}$ in 0-10\% Au+Au collisions at $\sqrt{s}=0.2$ TeV  \cite{Abelev:2009gu,STAR:2016jdz};}
       \item { $I_{AA}^{h^{\pm} h^{\pm}}$ in 0-5\%, 0-10\% Pb+Pb collisions at $\sqrt{s}=2.76$ TeV \cite{Aamodt:2011vg,Adam:2016xbp,Conway:2013xaa}.}
\end{itemize}
The detailed NLO pQCD parton model calculations of these observables with the 100 training realizations of $\hat q/T^3$ are shown in Appendix~\ref{sec:app:prior-obs} together with the above experimental data.

The procedures of training a model emulator are:
\begin{itemize}
\item Obtain the observable matrix  $y_{ik}$ with index $i=1\cdots 100$ labeling the set of observables that are computed using the $i^{\rm th}$ realization  $\left[\frac{\hat{q}}{T^3}\right]_i$. The $k$ index represents each observable point including $R_{AA}$ and $I_{AA}$ in each system, centrality and at each $p_{\rm T}$ ($z_{\rm T}$) point. The pair $\left(\left[\hat q/T^3\right]_{ij}, y_{ik}\right)$ forms the training data.
\item Transform the high-dimensional observables into the ``feature'' space using the Principal Component Analysis (PCA). The idea is that in the PCA basis, one can reconstruct the full observable vector with sufficient accuracy using only a few features ($N_{\rm feature} \ll N_{\rm obs}$).
\item Training the mapping from $\hat{q}$ to each one of the $N_{\rm feature}$ principal components with Gaussian emulators \cite{Rasmussen2004}.
\item To predict observables at new input $\hat{q}$, one transforms the corresponding principal components predicted by the Gaussian emulators, and then performs the inverse PCA transformation back to the observables space.
\end{itemize}
We refer readers to Ref.~\cite{Bernhard:2018hnz} for the details of PCA and Gaussian Processes.

\subsection{The posterior distribution}
According to the Bayes theorem, the posterior distribution is the product of the prior distribution and the likelihood function.
In the information field formulation, the posterior is also a functional distribution. 
If we denote the NLO pQCD parton model calculation as $y = M[\hat{q}(T)] = M'[F(x)]$, with $y$ being the collection of observables, and assume a multi-variate Gaussian form of the likelihood function, then, the posterior distribution of $F$ is,
\begin{widetext}
\begin{eqnarray}
      P[F(x)] &\propto& P_0[F(x)] \times {\rm Likelihood} \nonumber \\
      &=& \frac{\mathcal{N}}{\sqrt{\mathrm{det}(C)}}\exp\left\{-\frac{1}{2}\int dx dx' \delta F(x) C^{-1}(x, x') \delta F(x') - \frac{1}{2}(M'[F]-y_{\rm exp})^T \Sigma^{-1}_{\rm error} (M'[F]-y_{\rm exp})\right\},
\end{eqnarray}
\end{widetext}
where $y_{\rm exp}$ is the collection of the experimental observables and $\Sigma_{\rm error}$ is the uncertainty covariance matrix combining experimental, theoretical, and emulator uncertainties \cite{Bernhard:2018hnz,Ke:2020clc}.
To marginalize the values of $F^*$ at a fixed input $x^*$, one has to perform a path integral
\begin{eqnarray}
p(F^*) = \int [\mathcal{D} F] \delta(F(x^*)-F^*)  P[F(x)].
\end{eqnarray}
But for our practical application, they are performed on the 20 equally-spacing temperature grids.

\subsection{Sensitivity of the observables to the temperature dependence of $\hat q(T)$}\label{sec:stat:sensitivity}
\begin{figure}
    \centering
    \includegraphics[width=\columnwidth]{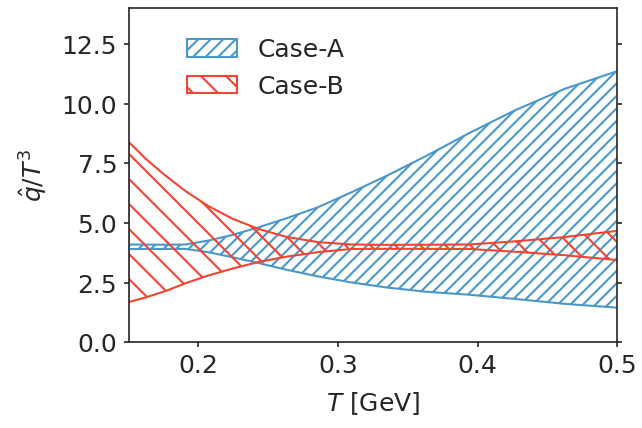}
    \caption{Two sets of random functions that vary $\hat{q}/T^3$ in different temperature regions.}
    \label{fig:Sensitivity-test-qhat}
\end{figure}

\begin{figure*}
    \centering
    \includegraphics[width=\textwidth]{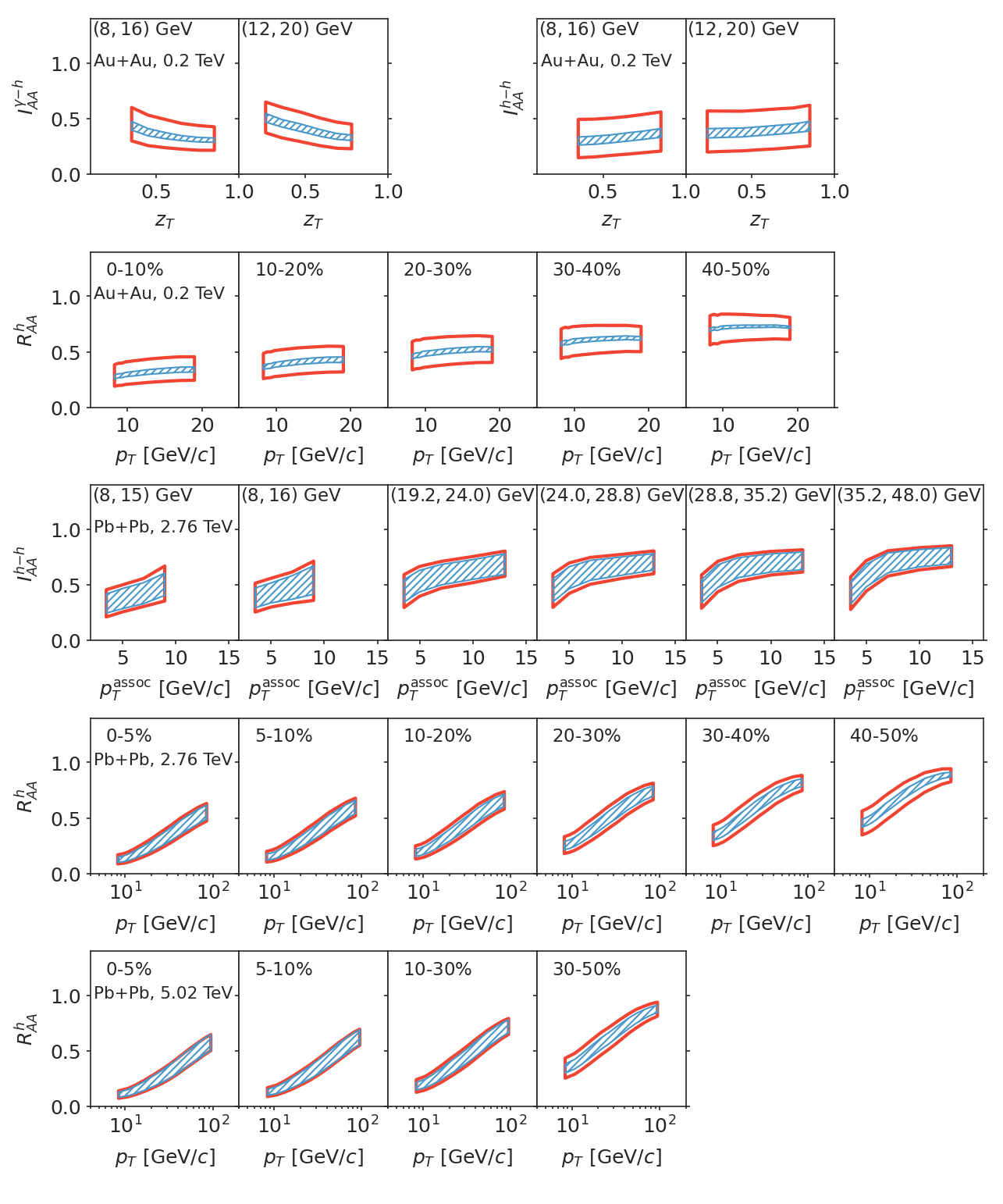}
    \caption{95\% variation around the median of the ensemble predictions using $\hat{q}/T^3$ sampled from Case-A (Red boxes) and B (blue shaded bands) of priors in Fig.~\ref{fig:Sensitivity-test-qhat}, corresponding to changes of $\hat{q}/T^3$ in low and high-temperature regions, respectively.
    }
    \label{fig:Sensitivity-test-obs}
\end{figure*}

One advantage of the information field formalism is that it is straightforward to analyze the sensitivity of the observables to the jet transport coefficient in different ranges of the temperature.  
To demonstrate this point, we create two sets of conditioned random functions of $\hat{q}$ as illustrated in Fig.~\ref{fig:Sensitivity-test-qhat}. 
In ``Case-A'', the random functions are required to take values between $\hat{q}/T^3 = 4\pm 0.1$ for $0.15<T<0.2$ GeV. 
In ``Case-B'', the condition $\hat{q}/T^3 = 4\pm 0.1$ is imposed to a different temperature region $0.3<T<0.4$ GeV. 
Case-A effectively allows  $\hat{q}$ to vary only in the high-temperature region, while ``Case-B'' can test the sensitivity of the observables to $\hat q$ in the low-temperature region. 

Using the emulator trained on the model calculations, we can make ensemble predictions using random functions from the two cases. 
The 95\% range of variations in the resulting observables is plotted in Fig.~\ref{fig:Sensitivity-test-obs}.
If a wider band of variation in observables is found for case-$A$ (case-$B$), the dataset is more sensitive to high(low)-temperature $\hat{q}$.

Examining the data at the RHIC energy (the first two rows), one can see that both single-hadron and trigger-hadron correlation data show a strong sensitivity to $\hat{q}/T^3$ at low temperature and become rather insensitive to its values at high temperature. Because the highest temperature reached is about 2.5$T_c$ at the RHIC energy, so the data should not constrain the jet transport coefficient at high temperatures, and this is exactly the purpose of this information-field construction.

For observables at the LHC energies (from the third to the last row) in Fig.~\ref{fig:Sensitivity-test-obs}, 
we find a similar level of sensitivity to the $\hat{q}$ variation in both Case-A and Case-B, except for the more peripheral collisions where the highest temperature reached is lower than in central collisions.  
 Because of the centrality dependence of the temperature range, the sensitivity to high-temperature $\hat{q}$ gradually decreases from central to peripheral collisions as seen from the narrowing of the case-A bands in $R_{AA}^h$ at $\sqrt{s} = 2.76$ TeV.
These are consistent with our construction: central collisions at higher beam energy are more sensitive to $\hat q$ in the high-temperature region.

In terms of percentage variation as we see in Fig.~\ref{fig:Sensitivity-test-obs}, the dihadron, and $\gamma$-hadron correlation data are more sensitive to the temperature dependence of $\hat q$ at both RHIC and LHC. This is consistent with early studies \cite{Zhang:2007ja,Zhang:2009rn} and shows the advantage of including trigger-hadron correlation data that have a different geometric bias effect than single-hadron suppression. However, we cannot fully exploit this advantage yet in this study, because the current uncertainties of the experimental on trigger-hadron correlation suppression are much larger than that of single inclusive hadron suppression. 
This will become clear later when we discuss the numerical results, and the theoretical calculations of dihadron suppression show some tension with data for high $p_{\rm T}$ trigger at the LHC. 
This implies that an improved theory calculation combined with precision measurements of dihadron and $\gamma$-hadron correlations can provide more stringent constraints on the jet transport coefficient in the high-temperature phase of the quark-gluon plasma.

\section{\label{sec:discussion} Constraining the temperature dependence of $\hat q(T)$}

In this section, we will perform the Bayesian analysis of the temperature dependence of the jet transport coefficient while neglecting a possible momentum dependence of $\hat{q}/T^3$. 
We will estimate the momentum dependence in Sec.~\ref{sec:discussion:qhat_vs_p} and leave a full $p$ and $T$ dependent extraction using a two-dimensional information field in future works.

\begin{figure*}
    \centering
    \includegraphics[width=\textwidth]{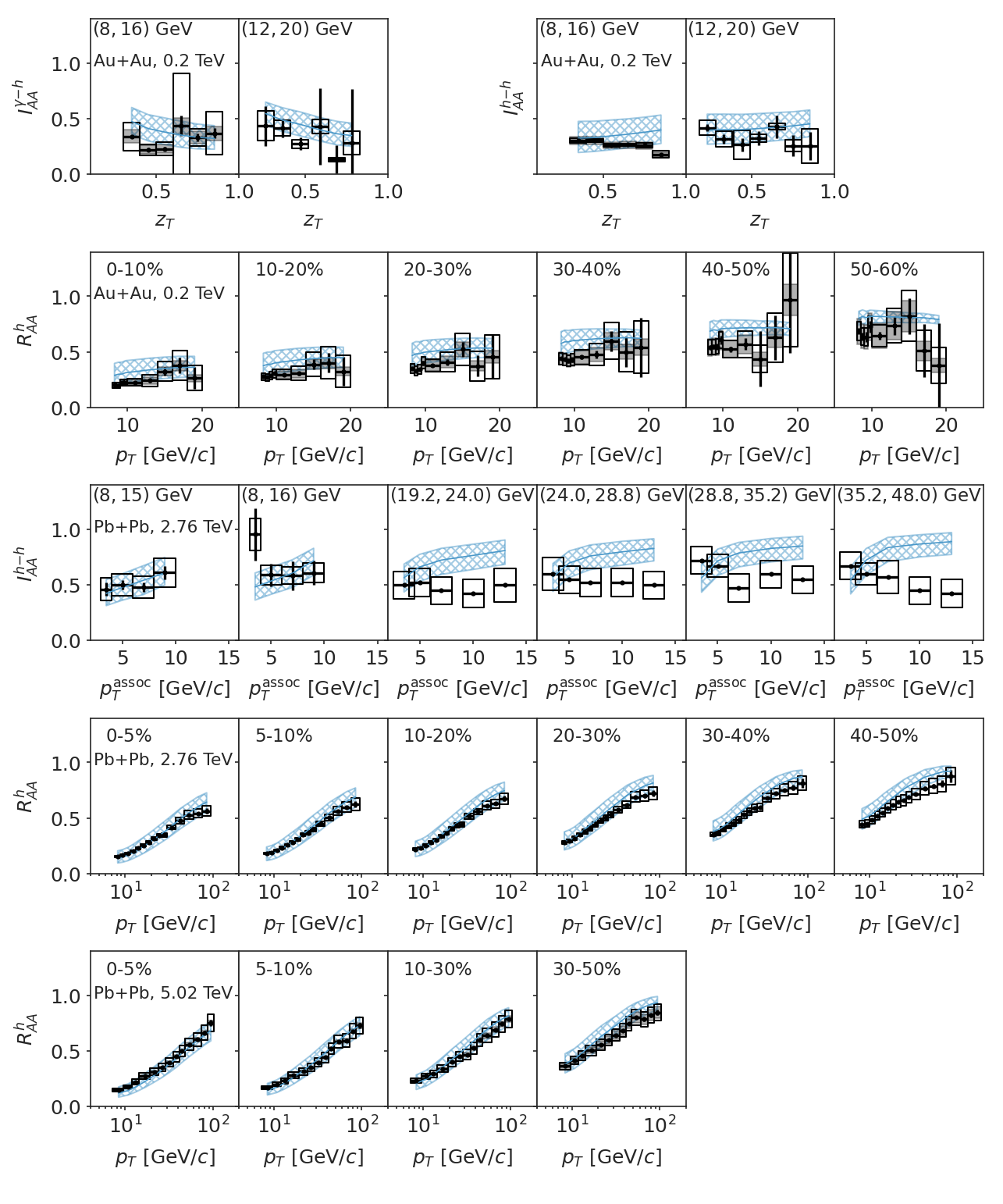}
    \caption{The posterior distributions of observables predicted by the emulator associated with the calibration as shown in Fig.~\ref{fig:qhat}. From top to bottom are $I_{AA}$ of $\gamma$-hadron and dihadron correlations in Au+Au collisions at 0.2 TeV, single hadron $R_{AA}$ in Au+Au collisions at 0.2 TeV, dihadron correlations in Pb+Pb collisions at 2.76 TeV, and the single hadron $R_{AA}$ for Pb+Pb collisions at 2.76 and 5.02 TeV. Different columns for $I_{AA}$ data correspond to different ranges of trigger particle $p_{\rm T}$, while different columns for $R_{AA}$ are for different centrality classes.}
    \label{fig:all-obs}
\end{figure*}

\subsection{Combined analysis using both $R_{AA}$ and $I_{AA}$}
\label{sec:discussion:main-qhat-results}
In Fig.~\ref{fig:all-obs}, we show the global descriptive power of the analysis after calibrating emulators of the NLO pQCD parton model calculations to experimental data on both the $R_{AA}$ of single inclusive hadron and $I_{AA}$ of dihadron and $\gamma$-hadron spectra. 
For $R_{AA}$, the centrality classes from 0 to 50\% are included, and we only include $R_{AA}$ data points with $p_{\rm T}^h>8$ GeV/$c$ and $I_{AA}$ data with $p_{\rm T}^{\rm assoc} > 5$ GeV/$c$ (or $z_{\rm T}\langle p_{\rm T}^{\rm trig}\rangle >5$ GeV/$c$).
Overall, the data are well reproduced by the 95\% credible interval around the median of the ensemble prediction of the model emulator. 
Some tensions are found between theory and experiments in $I_{AA}$ of dihadron in Pb+Pb collisions at $\sqrt{s} = 2.76$ TeV when $p_{\rm T}^{\rm trig}> 19.2$ GeV/$c$. 
This problem is already present in the prior calculations shown in Fig.~\ref{fig:IPbPb-had-had}, where $I_{AA}$ increases with $p_{\rm T}^{\rm assoc}$ for all $p_{\rm T}^{\rm trig}$ intervals .
However, the CMS data suggests $I_{AA}$ increases with $p_{\rm T}^{\rm assoc}$ at low $p_{\rm T}^{\rm trig}$ and decreases with $p_{\rm T}^{\rm assoc}$ at high $p_{\rm T}^{\rm trig}$. 

\begin{figure}
    \centering
    \includegraphics[width=\columnwidth]{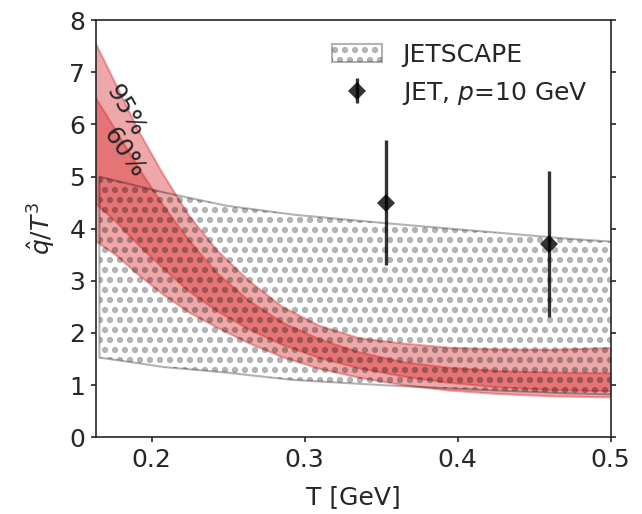}
    \caption{The final results of $\hat{q}/T^3$ using both single-hadron $R_{AA}$ and of $I_{AA}$ of hadron-hadron and $\gamma$-hadron correlation, assuming a momentum-independent $\hat{q}$. The 60\% and 95\% posterior probability density regions are shown in red. Symbols are $\hat{q}/T^3$ extracted by the JET Collaboration \cite{JET:2013cls}, and the shaded region is the result of JETSCAPE calibration using a multi-stage jet evolution model \cite{JETSCAPE:2021ehl}.}
    \label{fig:qhat}
\end{figure}

In Fig.~\ref{fig:qhat}, we plot the posterior bands of $\hat{q}/T^3$ as a function of temperature, with darker and lighter bands corresponding to 60\% and 95\% credible intervals around the median, respectively.
This result is compared to the extractions from the JET Collaboration \cite{JET:2013cls} and the more recent Bayesian analysis from the JETSCAPE Collaboration (95\% credible interval, the dot-hatched band) that calibrates a temperature, momentum, and virtuality dependent parametrization of $\hat{q}$ to single hadron data at both RHIC and LHC \cite{JETSCAPE:2021ehl}. 

Our result with the momentum-independent assumption is consistent with the JETSCAPE result at 95\% credible interval except for values at temperatures very close to $T_c$. Furthermore, this analysis exhibits a very strong temperature dependence of $\hat q/T^3$, decreasing with the temperature. In the following Subsection~\ref{sec:discussion:qhat_vs_T}, we will discuss how such a strong $T$ dependence results from the combination of central and peripheral data and measurements at different beam energies. In Appendix~\ref{sec:app:corr} one can find correlations of $\hat q/T^3$ in different regions of temperature which illustrate the power of combined constraints by data from different centralities and at different colliding energies. These correlations also show the relative independence of $\hat q$ in different regions of temperature as constrained by different set of data.
In Subsection.~\ref{sec:discussion:qhat_vs_p}, we will relax the momentum-independence assumption in the analysis, which will lead to larger uncertainty in the high-temperature region.

\subsection{Progressive constraints from collisions with different centrality and colliding energy }
\label{sec:discussion:qhat_vs_T}
As stated in the previous section, one of the advantages of using the information field method is that data sensitive to $\hat{q}$ in the low-temperature region does not constrain $\hat{q}$ at high temperatures. This allows us to include data progressively from peripheral and lower energy collisions to the central and high-energy collisions to demonstrate how different datasets contribute to the determination of the strong temperature dependence of $\hat{q}$ in Fig. \ref{fig:qhat}.

\begin{figure}
    \centering
    \includegraphics[width=\columnwidth]{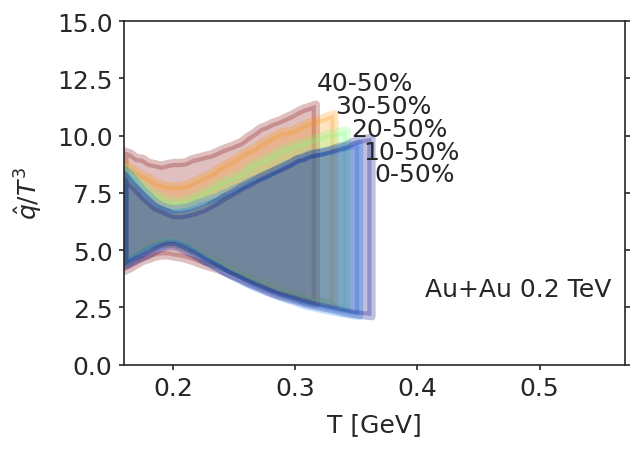}\\
    \vskip1em
    \includegraphics[width=\columnwidth]{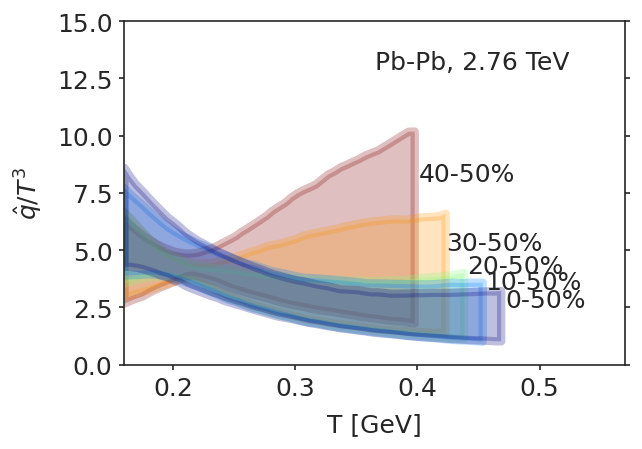}\\
    \vskip1em
    \includegraphics[width=\columnwidth]{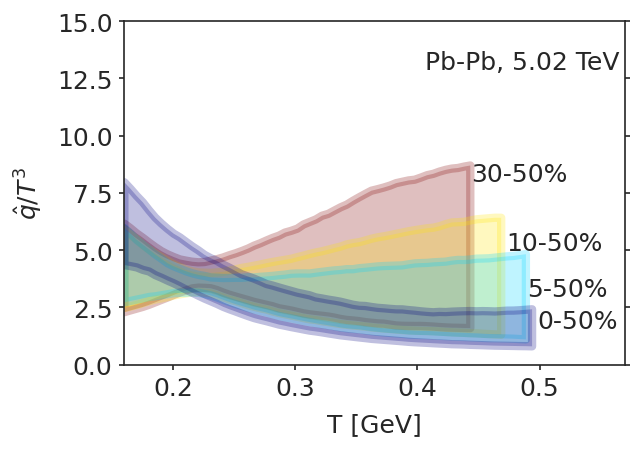}
    \caption{The 90\% credible intervals of $\hat{q}/T^3$ as a function of temperature constrained by $R_{AA}$ of single inclusive hadron suppression in different centrality class (indicated by the legend) of (upper) Au+Au collisions at $\sqrt{s}=0.2$ TeV, (middle) Pb+Pb collisions at $\sqrt{s}=2.76$ TeV and (lower) $\sqrt{s}=5.02$ TeV. The upper limit of the temperature range is given by the highest temperature reached in the hydrodynamics simulations within each centrality class.}
    \label{fig:qhat_cen}
\end{figure}

In the upper panel of Fig.~\ref{fig:qhat_cen}, we show the constrained $\hat{q}$ using $R_{AA}$ of $\pi^0$ in Au+Au collisions at $\sqrt{s}=0.2$ TeV, the middle and lower panels show results using charged hadron $R_{AA}$ in Pb+Pb collisions at $\sqrt{s}=2.76$ TeV and $5.02$ TeV. 
Different bands in each panel are the 90\% credible region of $\hat{q}/T^3$ using data from different combinations of centrality classes. 
From top to bottom legends, we start from calibrating to only the $R_{AA}$ in 40-50\% collisions and progressively including $R_{AA}$ from more central collisions. 
Eventually, the band labeled by the bottom legend contains all data from 0 to 50\% centrality in each colliding system. 
For each centrality combination, we only show the posterior of $\hat{q}(T)$ up to the highest temperature reached in the hydrodynamic simulation of the bulk matter in the corresponding centrality class. 

In the case of Au+Au collisions, $R_{AA}$ in peripheral collisions cannot constrain $\hat{q}$ very well. 
With the inclusion of data in more central collisions, the $\hat{q}/T^3$ posterior develops a ``neck'' around $T=0.2$ GeV.  Both the value of $\hat{q}/T^3$ below or above this temperature are less constrained.

The experimental uncertainty in Pb+Pb is smaller than those in Au+Au collisions. The data in 40-50\% peripheral Pb+Pb collisions can already constrain $\hat{q}/T^3$ around $T=0.2$ GeV pretty well with $\hat{q}/T^3 = 3 - 4.5$. As expected, data for 40-50\% collisions  cannot constrain $\hat{q}$ at higher temperatures, so the uncertainty band becomes large at $T>0.3$ GeV.
Including data from central collisions, the uncertainty band at high temperatures becomes smaller significantly. 
This clearly demonstrates that the centrality dependence of $R_{AA}$ contains the needed information to infer the temperature-dependent jet transport coefficient.
However, it should be noted that calibration using exclusive RHIC data in Au+Au collisions prefers higher values of $\hat{q}/T^3$ than that extracted from the LHC data. This problem was already noted in the pioneering work of the JET Collaboration \cite{JET:2013cls}.

\begin{figure}
    \centering
    \includegraphics[width=\columnwidth]{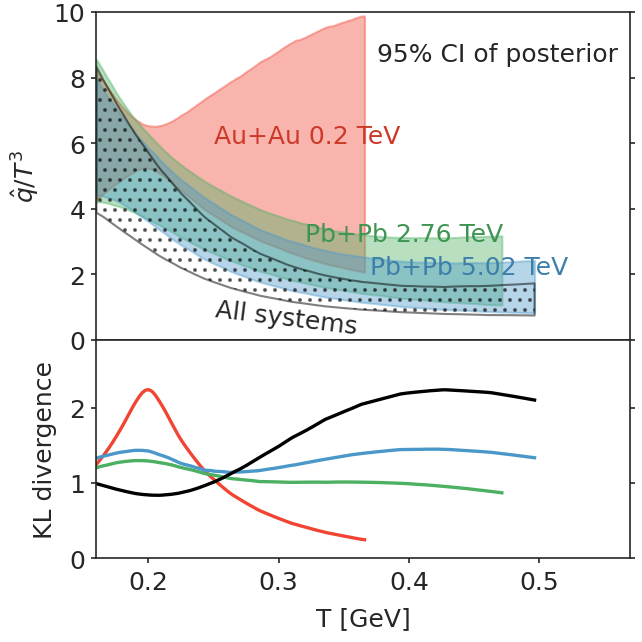}
    \caption{Upper panel: colliding system dependence of 95\% credible interval of $\hat{q}/T^3$ using single inclusive hadron suppression data. The red, green, and blue bands correspond to calibration using data from Au+Au collisions at $\sqrt{s}=0.2$ TeV, Pb+Pb collisions at $\sqrt{s}=2.76$ and $5.02$ TeV, respectively. Each calibration includes data from zero to $50\%$ centrality classes. The gray hatched band is the credible interval combing the data from all three colliding systems. Lower panel: the K-L divergence for each scenario with the same color coding as the upper panel.}
    \label{fig:qhat_sys}
\end{figure}

To look at this beam-energy dependence more systematically,  we compare in the upper panel of Fig.~\ref{fig:qhat_sys} the 95\% credible interval of $\hat{q}$ calibrated independently to Au+Au collisions at 0.2 TeV (red), Pb+Pb collisions at 2.76 TeV (green), and Pb+Pb collisions at 5.02 TeV (blue), with each analysis including single inclusive $R_{AA}$ from 0-50\% centrality classes.
In the lower panel of Fig.~\ref{fig:qhat_sys}, we also compute the information gain (with the same color coding as in the upper panel) quantified by the Kullback–Leibler (KL) divergence between the prior and posterior distribution of $\hat{q}$ at each temperature,
\begin{eqnarray}
D_{\rm KL}[P_1, P_2] = \int P_1(x) \ln\frac{P_1(x)}{P_2(x)} dx.
\end{eqnarray}
The KL divergence is a ``distance'' measuring the difference between two probability distributions (here, we take $P_1$ as the prior distribution and $P_2$ as the posterior distribution). 
A larger KL divergence indicates more information extracted from the data, and this method has already been used in Ref. \cite{JETSCAPE:2020shq}.

\begin{figure*}[ht!]
    \centering
    \includegraphics[width=.9\textwidth]{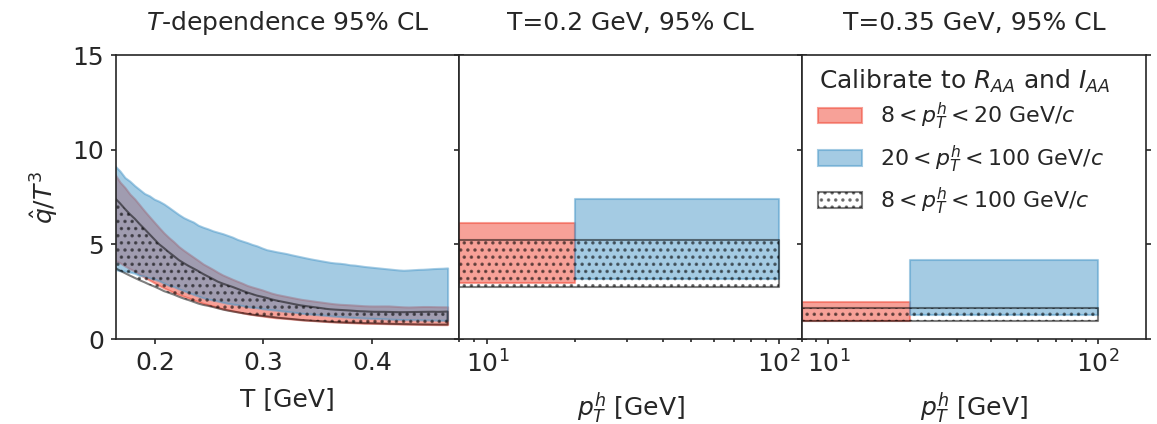}
    \caption{Dependence of $\hat{q}/T^3$ on hadron momentum range from calibration using both single-hadron $R_{AA}$ and $I_{AA}$ of hadron-hadron and $\gamma$-hadron correlation. The gray dotted band is the 95\% interval employing data from all three observables within $8 < p_{\rm T}^h < 100$ GeV/$c$, assuming there is no momentum dependence on $\hat{q}$. Red and blue bands are results separately calibrated to data within $8<p_{\rm T}^h<20$ GeV/$c$ and $20<p_{\rm T}^h<100$ GeV/$c$. The left panel compares the temperature dependence of $\hat q/T^3$ in these three momentum intervals, while the middle and right panels compare the hadron momentum dependence at two separate temperatures $T=0.2$ GeV and $T=0.35$ GeV, respectively.}
    \label{fig:qhat_pT}
\end{figure*}

Looking at either the credible interval or the peak of the KL divergence as a function of temperature, we conclude that RHIC data strongly prefer $\hat{q}/T^3 = 5-7$ around $T=0.2$ GeV.
Results using the LHC data show a rather uniform KL divergence, demonstrating their power to constrain $\hat{q}/T^3$ at both high and low $T$. Once RHIC and LHC data are combined, the information gain increases at high temperatures while decreasing at low temperatures. Such a decrease is the result of the tension between the posterior distributions obtained from RHIC and LHC data around $T=0.2$ GeV. One may argue that this is due to RHIC and LHC data having very different transverse momentum coverage and our analyses have neglected the momentum dependence of $\hat{q}$. 
However, we did not find a strong momentum dependence in Subsection \ref{sec:discussion:qhat_vs_p}, where we calibrated the model independently to high-$p_{\rm T}^h$, and low-$p_{\rm T}^h$ $R_{AA}$. This beam-energy dependence of the extraction may therefore reflect systematic problems that need to be improved in the theoretical framework, such as considering other factors that may systematically change with beam energy, e.g., thermalization time of the QGP and the quenching in the hadronic phase, etc.

Combining data from collisions at all three colliding energies into a single analysis (the shaded band), the Bayesian analysis will have to compromise the $\hat{q}$ accuracy to account for the tension between posterior distributions from data at RHIC and LHC colliding energies, resulting in a larger uncertainty in $\hat{q}$ in the low-temperature region.

\subsection{Momentum dependence of $\hat q$}
\label{sec:discussion:qhat_vs_p}

\begin{figure*}[ht!]
    \centering
    \includegraphics[width=.9\textwidth]{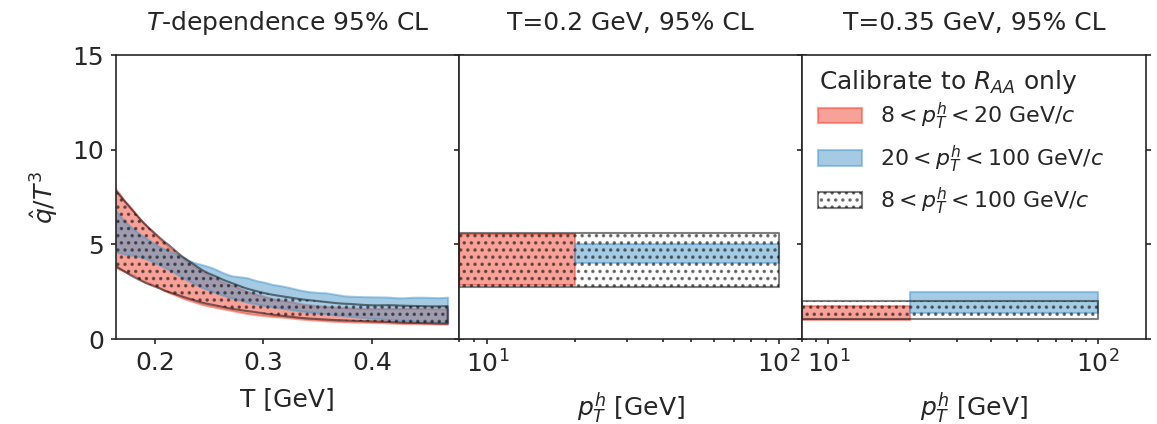}\\
    \vskip1em
    \includegraphics[width=.9\textwidth]{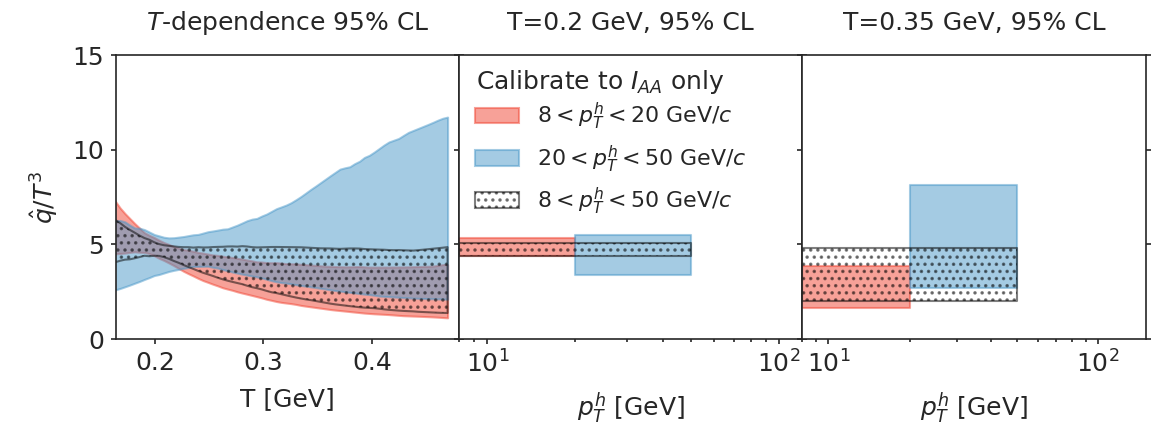}
    \caption{Hadron momentum dependence of $\hat{q}/T^3$ from calibration using exclusively $R_{AA}$ (upper) or $I_{AA}$ (lower). Red and blue bands are results calibrated separately to data within $8<p_{\rm T}^h<20$ GeV/$c$ and $20<p_{\rm T}^h<100$ GeV/$c$. The gray dotted bands are the 95\% interval employing data from data within $8 < p_{\rm T}^h < 100$ GeV/$c$, assuming no momentum dependence on $\hat{q}$. 
    The left panel compares the temperature dependence of $\hat q/T^3$ calibrated to three different momentum intervals, while the middle and right panels compare the hadron momentum dependence at two separate temperatures $T=0.2$ GeV and $T=0.35$ GeV, respectively.}
    \label{fig:qhat_RAA_IAA_pT}
\end{figure*}

As a first proof-of-principle application of the information field approach to Bayesian analyses in heavy-ion collisions, we have only applied it to parameterize the temperature dependence of $\hat{q}/T^3$. 
In the future, it is straightforward to extend this method to parameterize multi-variate functions, including, for example, the momentum dependence of $\hat q$. 
Nevertheless, even within this work, it is possible to relax the momentum-independent assumption of $\hat{q}$ by calibrating the model to data with different ranges of the hadron transverse momentum. To equate approximately the dependence on hadron and parton transverse momentum dependence of $\hat{q}$, we need to make two assumptions:
\begin{itemize}
    \item $\hat{q}$ changes slowly with parton momentum.
    \item The transverse momentum of hadron is of the same order as the parton momentum. This is true for inclusive hadron production since $R_{AA}$ is dominated by the modified fragmentation function at large $z$ due to the steep falling parton spectra. As for $I_{AA}$, this argument also applies to the trigger particles and we will exclude the region of small $z_{\rm T}$ for associated hadrons.
\end{itemize}

We divide the data into different groups based on the final hadron's momentum in $R_{AA}$ and the trigger momentum in $I_{AA}$ 
\begin{enumerate}
\item[I.] The low-momentum group: $8<p_{\rm T}^h, p_{\rm T}^{\rm trig}<20$ GeV/$c$.
\item[II.] The high-momentum group $20<p_{\rm T}^h, p_{\rm T}^{\rm trig}<100$ GeV/$c$.
\item[III.] The full region group: $8<p_{\rm T}^h, p_{\rm T}^{\rm trig}<100$ GeV/$c$.
\end{enumerate}

The 95\% credible interval of $\hat{q}(T)$ calibrated separately to data groups (I), (II), and (III) are shown as red, blue, and grey dotted bands in Fig.~\ref{fig:qhat_pT}. 
The left panel shows the temperature dependence, and the middle and right panels show $\hat{q}$ as a function of the momentum region for $T=0.2$ GeV and 0.35 GeV, respectively.

The results extracted from high-$p_{\rm T}$ data (group II) are comparable to those using low-$p_{\rm T}$ data (group I) at $T=0.2$ GeV, but are higher than group-I results at $T=0.35$ GeV.
Moreover, group II results also show significantly larger uncertainty, and we will see later in Sec.~\ref{sec:discussion:Raa_vs_IAA_qhat} that this is caused by the tension from the LHC $I_{AA}$ data at high $p_{\rm T}^{\rm trig}$.
The uncertainty band of the momentum-independent analysis (group III) is dominated by the low-$p_{\rm T}$ data group.
Despite this large uncertainty, our results suggest that a $\hat{q}$ that slowly increases with parton momentum.

\subsection{Calibrations with exclusively $R_{AA}$ or $I_{AA}$ data} \label{sec:discussion:Raa_vs_IAA_qhat}

We have seen in Fig.~\ref{fig:qhat_pT} that using high-$p_{\rm T}$ data (group II) results in a fairly large uncertainty. To better understand the cause, we perform two more analyses, with one calibrated exclusively to single inclusive hadron $R_{AA}$ and another exclusive to $\gamma$-hadron and dihadron $I_{AA}$.

In the upper panel of Fig.~\ref{fig:qhat_RAA_IAA_pT}, we show the 95\% credible intervals of $\hat{q}/T^3$ using exclusively $R_{AA}$ data, which have also been organized into different momentum regions. The outcome suggests negligible momentum dependence, with results from data groups (I) and (II) consistent with the uncertainty band of the momentum-independent analysis.

The situation is different when we calibrate exclusively to $I_{AA}$ data, shown in the lower panel of Fig.~\ref{fig:qhat_RAA_IAA_pT}. 
$I_{AA}$ data with $p_{\rm T}^{\rm trig}<20$ GeV/$c$ (dominated by RHIC data) result in a surprisingly good constraint on low-temperature $\hat{q}/T^3$ around  $T=0.2$ GeV.
However, the uncertainty from calibration using $p_{\rm T}^{\rm trig}>20$ GeV/$c$ data is extremely large.
The behavior is rooted in tension between the current model calculations and experimental data on $\pT^{\rm assoc}$ dependence of $I_{AA}$ for $p_{\rm T}^{\rm trig}>19.2$ GeV/$c$ in Pb+Pb collisions at 2.76 TeV.
It causes the same problem (larger uncertainty) that we have seen in the combined analysis of both $R_{AA}$ and $I_{AA}$ in Fig.~\ref{fig:qhat_pT} due to the tension in describing high-$p_{\rm T}^{\rm trig}$ $I_{AA}$ at the LHC.

Therefore, despite the expectation that $I_{AA}$ in central Pb+Pb collisions at the LHC energies are more sensitive to $\hat{q}$ in the high-temperature region, we do not gain much additional constrain power at LHC energy from the combined $R_{AA}$ and $I_{AA}$ analysis. This is caused by  (1) relatively large experimental uncertainty of $I_{AA}$ compared to $R_{AA}$ and (2) tension between model and data on $I_{AA}$ at large $p_{\rm T}^{\rm trig}$ at LHC. Nevertheless, in regions where the model provides a reasonable description (RHIC energy and low $p_{\rm T}^{\rm trig}$ data at the LHC), we see that the $I_{AA}$ analysis works as expected. Future progress in resolving the theory-experiment tension and more precise measurements of $\gamma$-hadron and dihadron correlation should lead to a more precise extration of $\hat q$, based on our sensitivity analysis in Sec.~\ref{sec:stat:sensitivity}.

\section{\label{sec:validation} Validations and predictions}
All the analyses presented in the previous sections are performed with model emulators. 
As a final step of this study, we will validate the Bayesian inference by comparing the NLO pQCD parton model calculations using the extracted jet transport coefficient $\hat{q}$ (combined analysis as described in section \ref{sec:discussion:main-qhat-results}) to the experimental data on the suppression of single inclusive hadron spectra $R_{AA}$, their elliptic anisotropy $v_2$ and the trigger-hadron correlations $I_{AA}$.

To go beyond emulator predictions as shown in Fig.~\ref{fig:all-obs},  one should in principle randomly sample $\hat{q}(T)$ from the functional posterior distribution and make ensemble predictions, which is highly computationally intensive.
Instead, we define a few representative parameter sets from the full posterior as originally proposed in Ref.~\cite{Ke:2020clc}. 
First, one transforms to the principal component space of the multidimensional posterior (in this case, a 20-dimensional space of the values of $\hat{q}$ at each of the 20 temperature points).
Then, one takes the first $N_{\rm p}$ principal direction that explains a large fraction of the total variance of the posterior. 
Along each principal direction $i=1, \cdots N_{\rm p}$, one defines the marginalized median value $V_i^{\rm med}$ and $5\%, 95\%$ quantile number $V_i^{\rm 5\%}$ and $V_i^{\rm 95\%}$ of this principal component. 
The ``central'' prediction $\hat{q}(T)$ is obtained by applying the inverse PCA transformation to
\begin{eqnarray}
&&\begin{bmatrix}
 V_1^{\rm med},  V_2^{\rm med}, \cdots,  V_{N_{\rm p}}^{\rm med}
\end{bmatrix}
\xrightarrow{\rm Inverse~PCA}\\\nonumber
&&\begin{bmatrix}
\hat{q}^{\rm med}(T_1), \hat{q}^{\rm med}(T_2), \cdots, \hat{q}^{\rm med}(T_{20})
\end{bmatrix}\equiv \hat{q}^{\rm med}(T)
\end{eqnarray}
The $\hat{q}(T)$ samples for error estimation are obtained in a similar way by varying each $V_i$ to the 5\% and 95\% quantile numbers,
\begin{eqnarray}
&&\begin{bmatrix}
 V_1^{\rm med},  \cdots V_j^{\rm 95\%} , \cdots,  V_{N_{\rm p}}^{\rm med}
\end{bmatrix}
\xrightarrow{\rm Inverse~PCA} \hat{q}^{j+}(T), \\
&&\begin{bmatrix}
 V_1^{\rm med},  \cdots V_j^{\rm 5\%} , \cdots,  V_{N_{\rm p}}^{\rm med}
\end{bmatrix}
\xrightarrow{\rm Inverse~PCA} \hat{q}^{j-}(T),
\end{eqnarray}
for $j=1, \cdots N_{\rm p}$. These constitute a total of $1+2N_{\rm p}$ representative parameters sets $\hat{q}^{\rm med}(T)$, $\hat{q}^{1\pm}(T)$, $\cdots$, $\hat{q}^{N_{\rm p}\pm}(T)$, and the envelope of the $1+2N_{\rm p} $calculations represents the calibration uncertainty of the theoretical model.
In this study, we use $N_{\rm p}=4$ and perform the full NLO pQCD parton model calculations of $R_{AA}^h$ and $I_{AA}$ at RHIC and LHC to verify the performance of the calibrated $\hat{q}(T)$. In Subsection.~\ref{sec:validation:OO}, these representative parameter sets are also applied to predict $R_{AA}$ and $v_2$ of the upcoming oxygen-oxygen collisions at the LHC.

\subsection{Model Validations}

\begin{figure*}[ht!]
\centering
\includegraphics[width=1.0\textwidth]{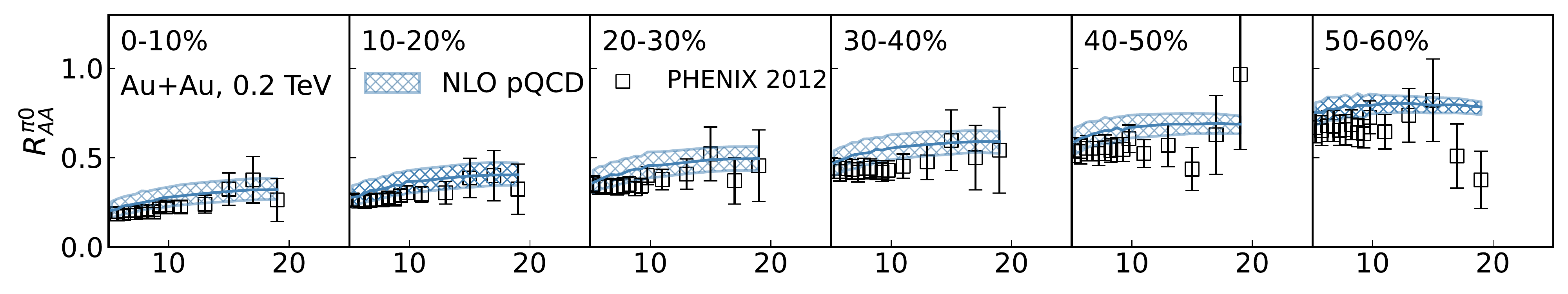}
\includegraphics[width=1.0\textwidth]{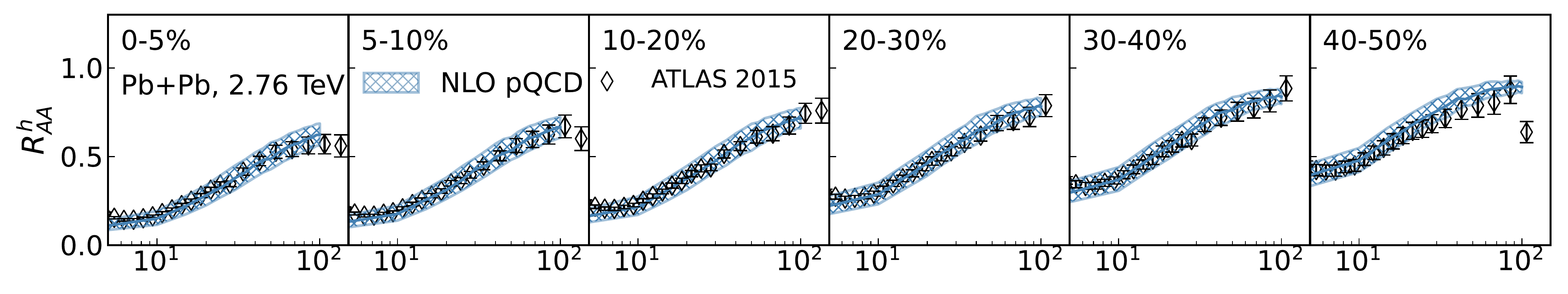}
\includegraphics[width=1.0\textwidth]{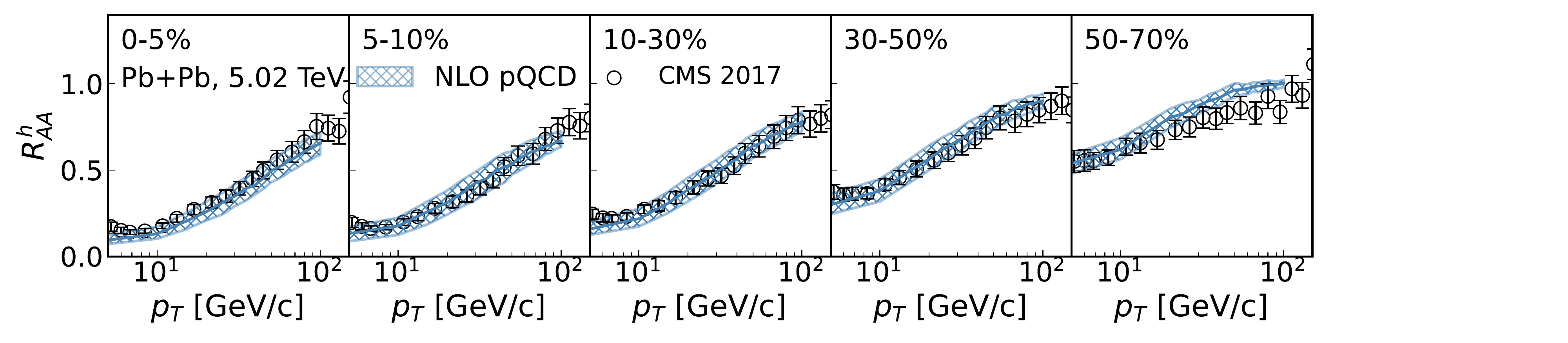}
\caption{Nuclear modification factor $R_{AA}$ as a function of $p_{\rm T}$ with $\hat{q}/T^3$ from Bayesian extraction for Au+Au collisions at $\sqrt{s}=0.2$ TeV (upper panel) and Pb+Pb collisions at $\sqrt{s}=2.76$ (middle panel) and 5.02 TeV (lower panel) within several centrality classes, as compared with experimental data \cite{PHENIX:2008saf,Adare:2012wg,CMS:2012aa,Aad:2015wga,Khachatryan:2016odn}. Steel blue curves are the NLO pQCD parton model results with the median value of $\hat{q}/T^3$, and the light blue bands are the estimated uncertainty.}
\label{fig:RAA-median}
\end{figure*}

Plotted in Fig.~\ref{fig:RAA-median} from top to bottom panels are the medium modification factors $R_{AA}$ of the single inclusive hadron spectra in $A+A$ collisions at $\sqrt{s}=0.2$, 2.76 and 5.02 TeV, respectively with several centrality classes. 
Steel blue solid lines are results with the median value of $\hat q/T^3$ and the hatched bands are the uncertainty from 90\% credible region of $\hat{q}/T^3$. 
One can see that the NLO parton model with parton energy loss and the jet transport coefficient $\hat{q}/T^3$ from the Bayesian extraction can describe the $R_{AA}$ of single inclusive hadrons well within the uncertainty of the experimental data in almost all centralities.

\begin{figure*}[ht!]
\centering
\includegraphics[width=0.7\textwidth]{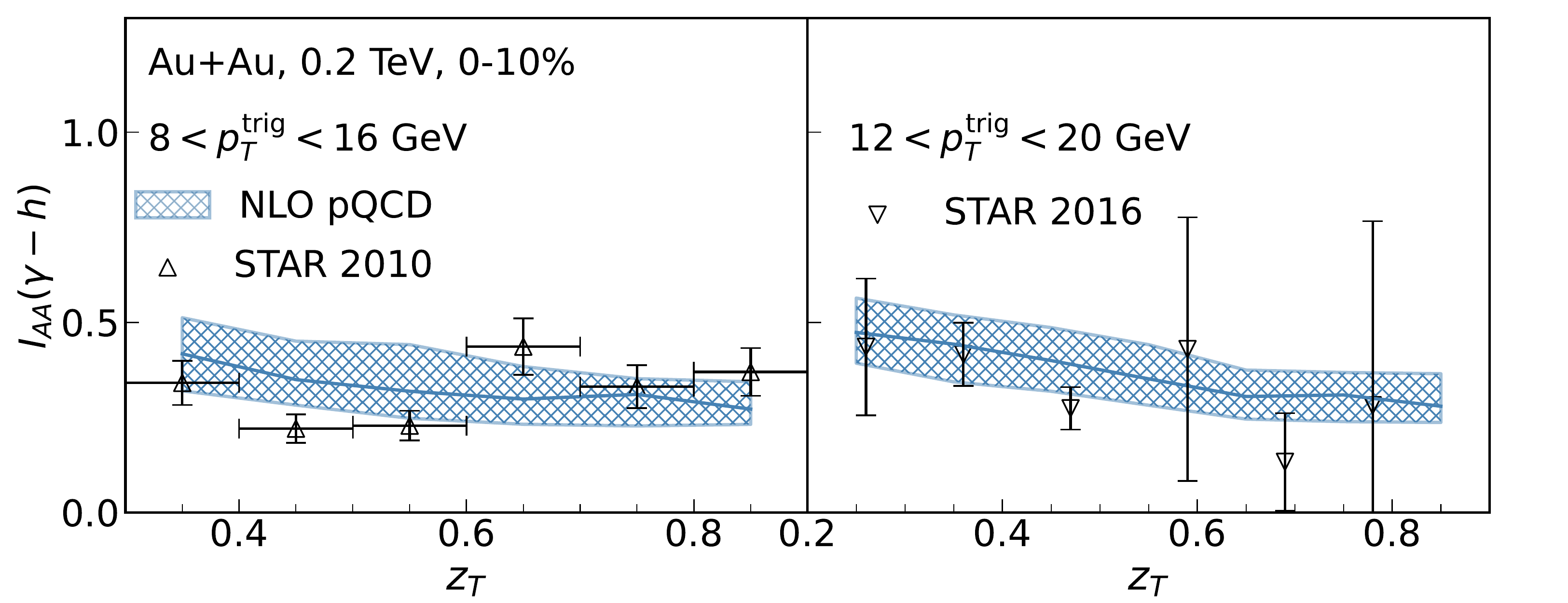}
\caption{$\gamma$-hadron suppression factor $I_{AA}$ as a function of $z_{\rm T}$ with $\hat{q}/T^3$ from Bayesian extraction for ($8<p_{\rm T}^{\rm trig}<16$ GeV$/c$, $3<p_{\rm T}^{\rm assoc}<16$ GeV$/c$) (left) and ($12<p_{\rm T}^{\rm trig}<20$ GeV$/c$, $1.2<p_{\rm T}^{\rm assoc}<p_{\rm T}^{\rm trig}$) (right) in 0-10\% central Au+Au collisions at $\sqrt{s}=0.2$ TeV, as compared with STAR data \cite{Abelev:2009gu,STAR:2016jdz}. Steel blue curves are the NLO pQCD parton model results with the median value of $\hat{q}/T^3$, and light blue bands are the estimated uncertainty.
}
\label{fig:IAuAu-gam-had-median}
\end{figure*}

\begin{figure*}[ht!]
\centering
\includegraphics[width=0.7\textwidth]{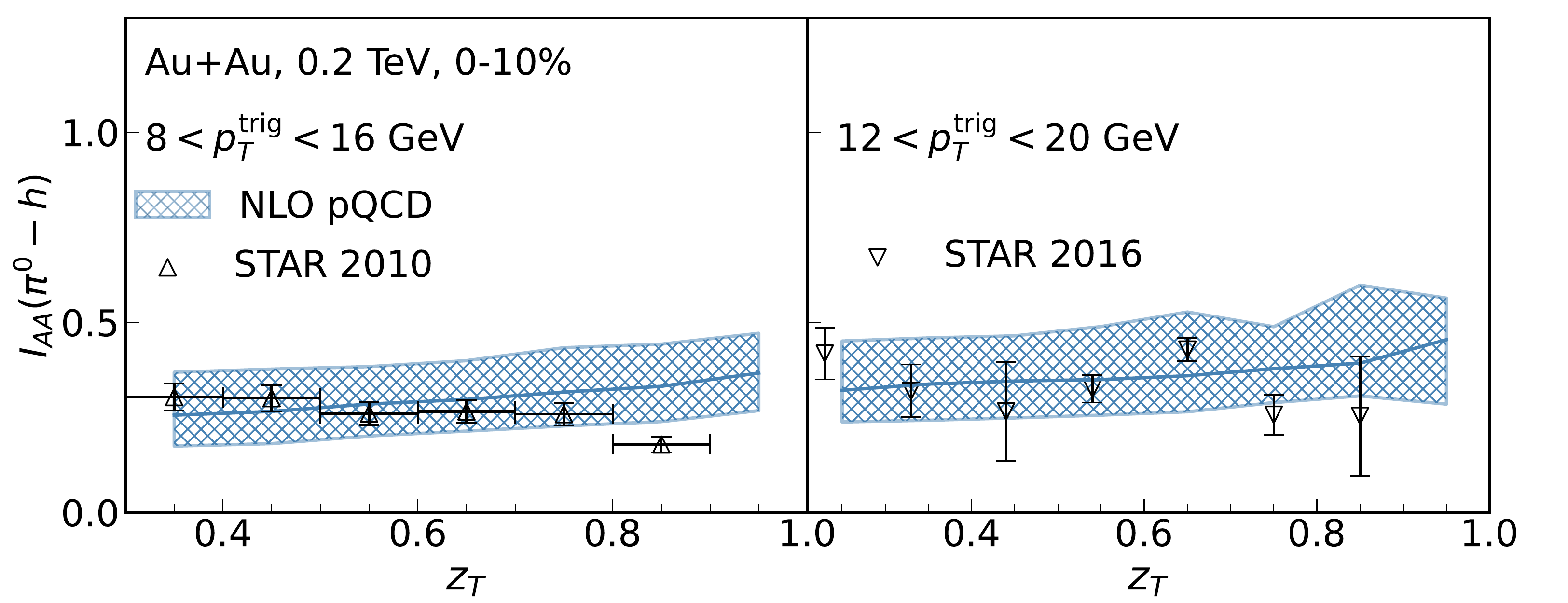}
\caption{$\pi^{0}$-hadron suppression factor $I_{AA}$ as a function of $z_{\rm T}$ with $\hat{q}/T^3$ from Bayesian extraction for ($8<p_{\rm T}^{\rm trig}<16$ GeV$/c$, $3<p_{\rm T}^{\rm assoc}<16$ GeV$/c$) (left) and  ($12<p_{\rm T}^{\rm trig}<20$ GeV$/c$, $1.2<p_{\rm T}^{\rm assoc}<p_{\rm T}^{\rm trig}$) (right) in 0-10\% central Au+Au collisions at $\sqrt{s}=0.2$ TeV, as compared with STAR data \cite{Abelev:2009gu,STAR:2016jdz}. Steel blue curves are the NLO pQCD parton model results with the median value of $\hat{q}/T^3$, and light blue bands are the estimated uncertainty.}
\label{fig:IAuAu-pi0-had-median}
\end{figure*}

The suppression of $\gamma$-hadron correlation $I_{AA}^{\gamma h}$ and $\pi^0$-hadron correlation $I_{AA}^{\pi^0 h}$ in 0-10\% central Au+Au collisions at $\sqrt{s}=0.2$ TeV  are shown in Figs.~\ref{fig:IAuAu-gam-had-median} and \ref{fig:IAuAu-pi0-had-median}, respectively, as compared to the experimental data. 
The left and right panels have different ranges of $p_{\rm T}^{\rm trig}$ for the trigger as in the different data sets from the STAR collaboration \cite{Abelev:2009gu,STAR:2016jdz} that we compare to. 
It is worthwhile to point out that the model with the median values of $\hat q/T^3$ can describe the data on $I_{AA}$ well for both $\gamma$ and $\pi^0$ triggers in central Au+Au collisions at the RHIC energy.  It somewhat under-predicts the suppression of single inclusive hadron $R_{AA}$ at the RHIC energy, though the data points are still within the uncertainty bands from 90\% credible region of $\hat{q}/T^3$.


\begin{figure*}[ht!]
\centering
\includegraphics[width=1.0\textwidth]{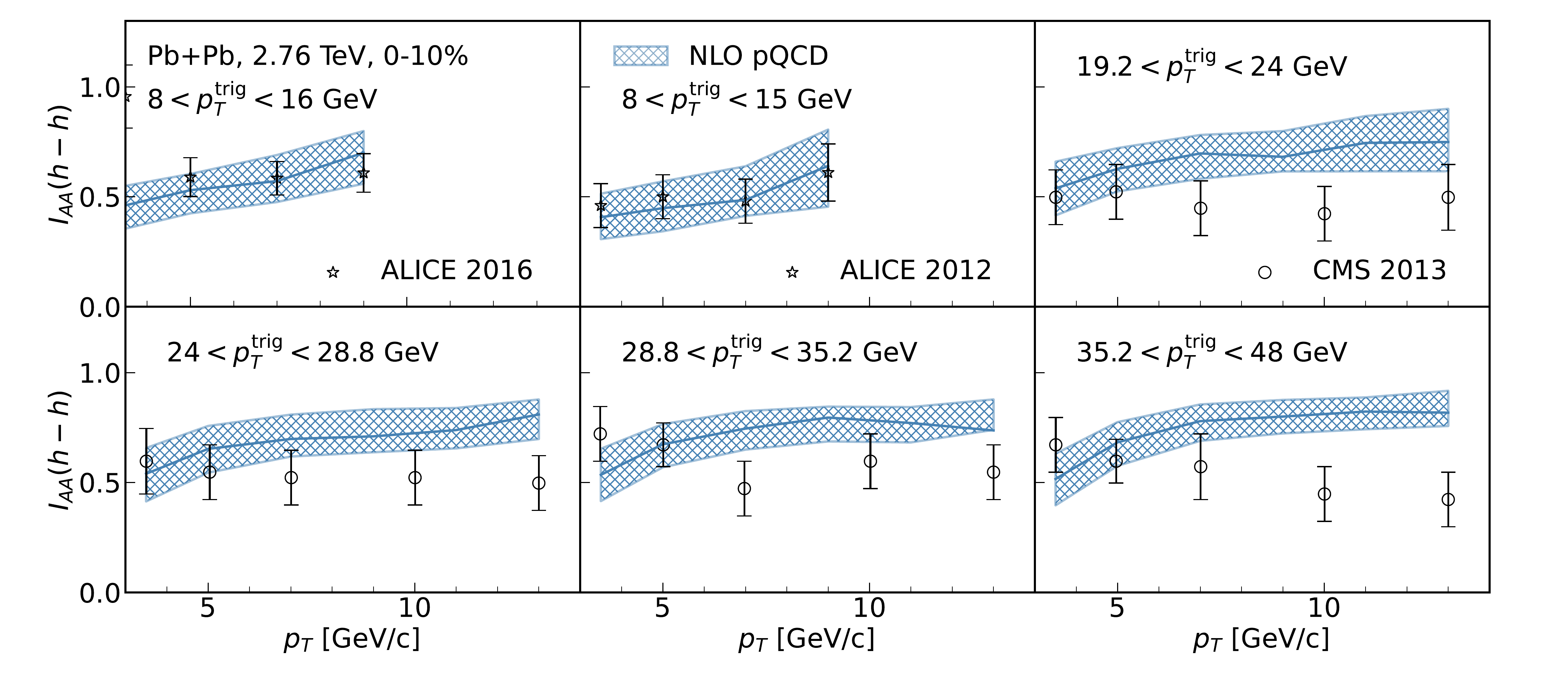}
\caption{Dihadron suppression factor $I_{AA}$ as a function of $p_{\rm T}^{\rm assoc}$ with $\hat{q}/T^3$ from Bayesian extraction for six different $p_{\rm T}^{\rm trig}$ ranges in 0-10\% central Pb+Pb collisions at $\sqrt{s}=2.76$ TeV, as compared with ALICE \cite{Aamodt:2011vg,Adam:2016xbp} and CMS \cite{Conway:2013xaa} experimental data. Steel blue curves are the NLO pQCD parton model results with the median value of $\hat{q}/T^3$, and light blue bands are the estimated uncertainty.}
\label{fig:IPbPb-had-had-median}
\end{figure*}

Shown in Fig.~\ref{fig:IPbPb-had-had-median} are the posterior distributions of
$I_{AA}^{hh}$ with six different $p_{\rm T}^{\rm trig}$ intervals in 0-10\% central Pb+Pb collisions at $\sqrt{s}=2.76$ TeV. The posterior predictions for $I_{AA}$ with lower $p_{\rm T}^{\rm trig}$ ranges can fit the experimental data well. 
The model with the median values of $\hat q/T^3$, however, under-estimates the dihadron suppression at higher $p_{\rm T}^{\rm trig}$.

\begin{figure*}[ht!]
\centering
\includegraphics[width=1.0\textwidth]{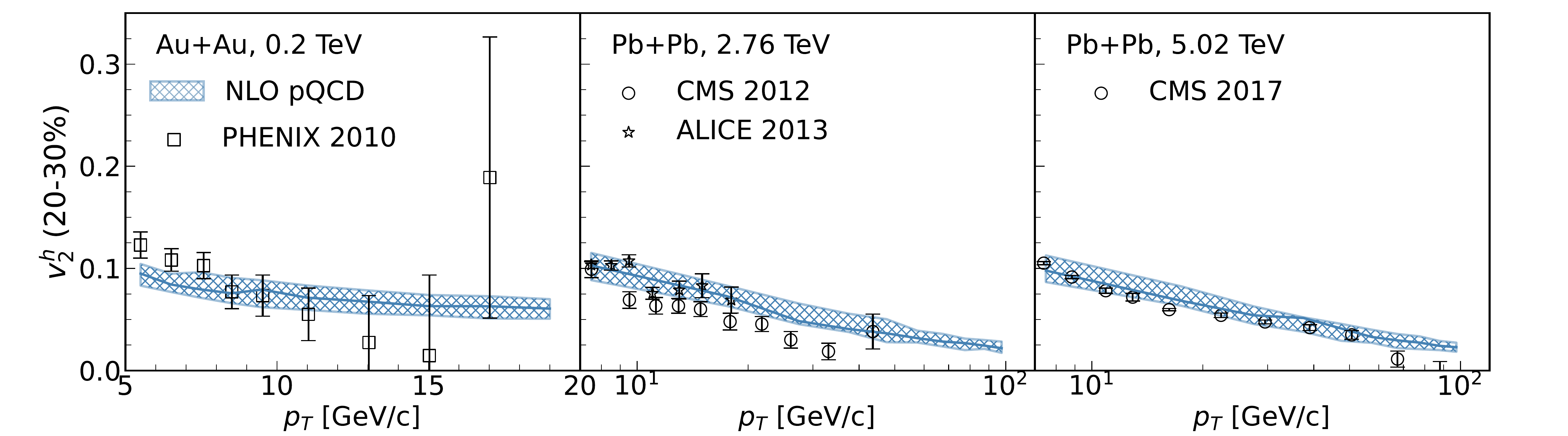}
\caption{The elliptic anisotropy $v_2$ for single inclusive hadrons with $\hat{q}/T^3$ from Bayesian extraction in 20-30\% Au+Au collisions at $\sqrt{s}=0.2$ TeV, Pb+Pb collisions at $\sqrt{s}=2.76$ TeV and 5.02 TeV, respectively. Steel blue curves are the NLO pQCD parton model results with the median value of $\hat{q}/T^3$, and light blue bands are the estimated uncertainty.}
\label{fig:Au-Pb-v2-20-30}
\end{figure*}

As another test of the NLO pQCD model with the Bayesian extracted temperature-dependent jet transport coefficient, we calculate and compare to experimental data on the momentum anisotropy $v_2$ of high $p_{\rm T}$ hadrons.
Results, as shown in Fig.~\ref{fig:Au-Pb-v2-20-30}, from left to right, for different collision systems: Au+Au collisions at $\sqrt{s}=0.2$ TeV, Pb+Pb collisions at $\sqrt{s}=2.76$ TeV and  $\sqrt{s}=5.02$ TeV, respectively, all in a typical 20-30\% centrality class, can describe the experimental data \cite{PHENIX:2010nlr,CMS:2012tqw,ALICE:2012vgf,CMS:2017xgk} well. We note again that the jet transport coefficient $\hat q/T^3$ extracted from our combined Bayesian analysis of the suppression of single inclusive spectra, dihadron, and $\gamma$-hadron correlations has a strong temperature dependence. 
This is part of the reason for the good description of the experimental data at different colliding energies. However, our extracted temperature dependence is far less dramatic near the critical temperature than some of the early models \cite{Cao:2017umt,Liao:2008dk,Xu:2014tda}. 
The NLO pQCD model will not be able to describe the experimental data on $v_2$ at lower and intermediate $p_{\rm T}$ where coalescence between jet and thermal partons becomes important \cite{Zhao:2021vmu}.

\begin{figure*}[ht!]
\centering
\includegraphics[width=1.0\textwidth]{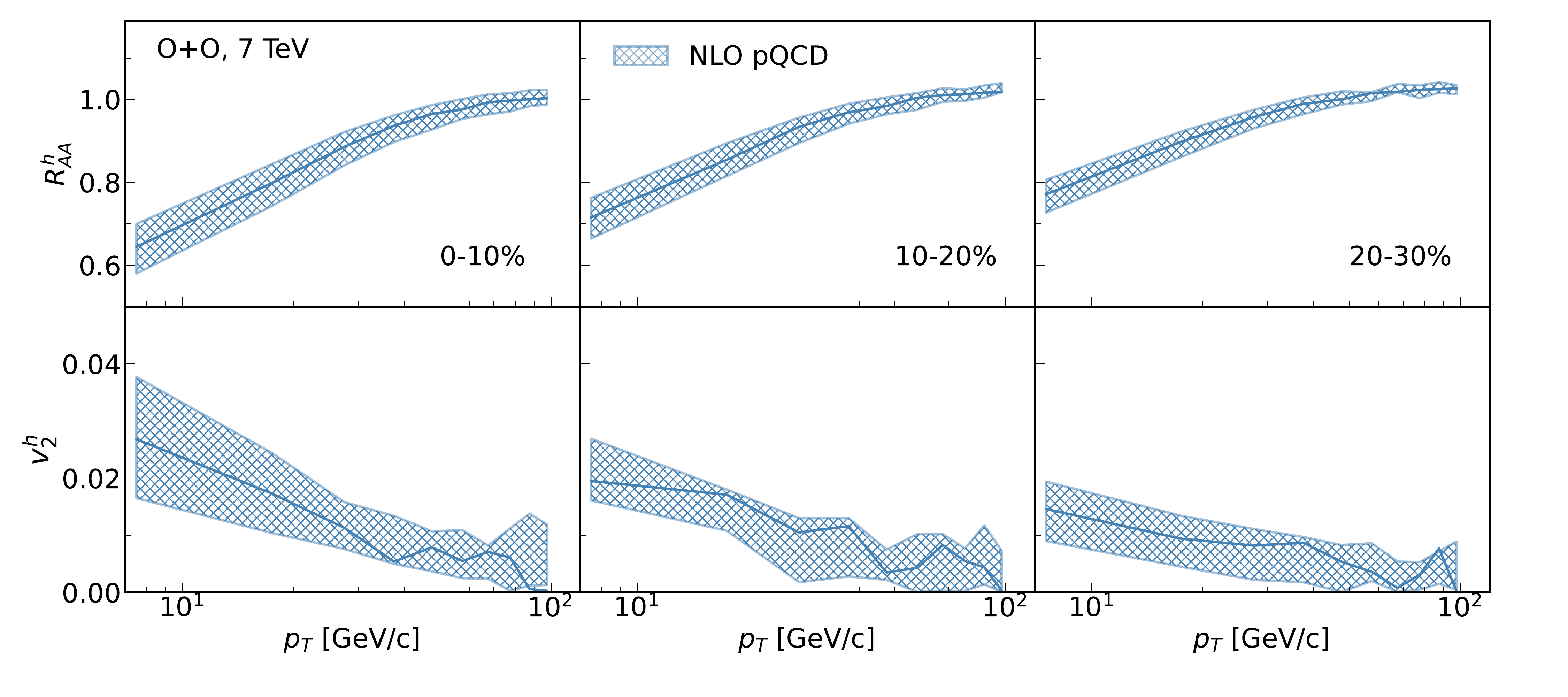}
\caption{Predictions for $R_{AA}$ (upper) and $v_2$ (lower) of single inclusive hadrons as functions of $p_{\rm T}$ with $\hat{q}/T^3$ from Bayesian extraction in $O+O$ collisions at $\sqrt{s}=7$ TeV within 0-10\%, 10-20\%, and 20-30\% centralities, respectively. Steel blue curves are the NLO pQCD parton model results with the median value of $\hat{q}/T^3$, and light blue bands are the estimated uncertainty.}
\label{fig:RAA-v2-OO}
\end{figure*}

\subsection{\label{sec:validation:OO} Prediction for $O+O$ collisions at $\sqrt{s}=7$ TeV}

Finally, we use the Bayesian extracted jet transport coefficient $\hat q/T^3$ to predict the single inclusive hadron suppression $R_{AA}$ and elliptic anisotropy $v_2$ in O+O collisions at $\sqrt{s}=7$ TeV. 
Such collisions are proposed as an intermediate colliding system between heavy-ion and proton-nucleus collisions to investigate the properties of dense matter in small systems \cite{Brewer:2021kiv,Brewer:2021tyv}.
Under the assumption that the QGP matter is also formed in $O+O$ collisions at $\sqrt{s}=7$ TeV, we use the CLVisc hydrodynamic model to describe the evolution of the bulk matter. 
The overall normalization factor for the initial bulk medium entropy density distribution from the Trento model~\cite{Moreland:2014oya} is fitted to reproduce the final total charged hadron multiplicity, which is estimated by extrapolation from an empirical parameterization of the beam energy dependence of the total charged hadrons multiplicity per participant pair \cite{ALICE:2016fbt}.  The scale factor for the TRENTo initial condition at $\sqrt{s}=7$ TeV is 205.

The predictions for the nuclear modification factor $R_{AA}$ (upper panel) and the elliptic flow coefficient $v_2$ (lower panel) for single inclusive hadrons are shown in Fig.~\ref{fig:RAA-v2-OO} in 0-10\%, 10-20\%, and 20-30\% centralities (from left to right). 
The steel blue solid lines correspond to the central results and the blue bands denote the uncertainty estimation. 
In the most 0-10\% central $O+O$ collisions, the suppression of single inclusive hadron spectra is about 30$\sim$40\% at $p_{\rm T} =7.5 \sim$ 10 GeV$/c$. 
The suppression mostly disappears when hadron's $p_{\rm T}$ reaches 60 GeV$/c$. In semi-central 20-30\% collisions, the single hadron spectra are suppressed by about 20\% at $p_{\rm T}\sim 10$ GeV$/c$.
As shown in lower panels, the elliptic anisotropy $v_2$ is only about 0.03 at $p_{\rm T}\sim7.5$ GeV$/c$ in 0-10\% $O+O$ collisions. It quickly disappears at high-$p_T$ and in more peripheral collisions.

\section{Impact of additional data sets}\label{sec:new-data}

\begin{figure}
    \centering
    \includegraphics[width=.5\textwidth]{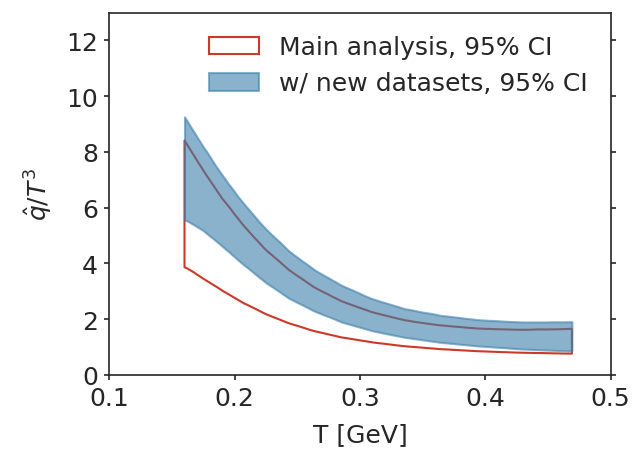}
    \caption{A comparison of the 95\% credible interval of $\hat{q}$ obtained in the main analysis (red open band) and that obtained by including $\gamma$-jet fragmentation functions in Pb+Pb collisions \cite{CMS:2018mqn} and hadron-hadron correlations in Au+Au collisions \cite{STAR:2006vcp}.}
    \label{fig:qhat_w_new_data}
\end{figure}

Besides the main results in Sec.~\ref{sec:discussion:main-qhat-results}, we found two additional data sets that can be included in the analysis. One is the 2006 STAR measurements of dihadron correlations in 0-5\% central and 20-40\% semi-central Au-Au collisions~\cite{STAR:2006vcp}. The other is the modification to the jet fragmentation functions in $\gamma$-jet events in Pb+Pb collisions at 5.02 TeV in 0-10\%, 10-30\%, and 30-50\% centrality classes as measured by the CMS experiment \cite{CMS:2018mqn}. Because there are no inclusive $\gamma$-hadron correlation measurements at LHC energy, these data on $\gamma$-jet fragmentation functions can partially serve the purpose. The details of these additional calculations are explained in Appendix \ref{sec:app:gam_jet}.
In Fig.~\ref{fig:qhat_w_new_data}, we compare the posterior distributions of $\hat q/T^3$ from the main and new analyses with the two additional data sets. The 95\% credible interval is consistent, while the new analyses slightly prefer a higher value of $\hat{q}/T^3$.

\section{Summary and outlook}\label{sec:summary}

In this work, we have carried out a first global Bayesian inference of the jet transport coefficient $\hat q$ using combined experimental data on suppression of single inclusive hadrons $R_{AA}$, dihadron and $\gamma$-hadron correlations $I_{AA}$ in heavy-ion collisions at both RHIC and LHC energies with all available range of centralities. The analyses use
the NLO pQCD parton model with a medium-modified fragmentation functions and parton energy given by the higher-twist approach.
This model provides a consistent description of single inclusive hadron, dihadron, and $\gamma$-hadron cross-sections in both p+p and A+A collisions. 

To take full advantage of the large amount of data at different colliding energies and centrality classes that probe the jet transport coefficient $\hat{q}/T^3$ in different temperature ranges, and to unambiguously test the constraining power of $R_{AA}$ of single inclusive hadrons and $I_{AA}$ of dihadron and $\gamma$-hadron correlations to $\hat{q}(T)/T^3$, we developed the Bayesian analysis with information field approach.
Compared to the use of an explicit parameterization of $\hat{q}/T^3$, the information field approach is shown to be free of complicated correlations in different temperature regions in the prior distributions.
Using the information-field approach, data that constrain $\hat{q}/T^3$ at low temperatures do not affect the prior distributions of $\hat{q}/T^3$ at high temperatures.
This makes the extraction of the temperature dependence of $\hat{q}/T^3$ more robust when combining RHIC and LHC data in a wide range of centralities that probe the quark-gluon plasma with different reach in temperature.
 
With this new Bayesian analysis framework, we demonstrate that the temperature dependence of $\hat{q}/T^3$ is progressively constrained by experimental data from peripheral to central collisions and from lower to higher beam energies. 
The combined analysis of $R_{AA}$ and $I_{AA}$ from all three colliding systems: Au+Au at 0.2 TeV, and Pb+Pb at 2.76 TeV and 5.02 TeV, including 0-50\% centrality classes, suggests $\hat{q}/T^3$ decreases with temperature from $5\pm 1$ near $T_c$ to $1.1\pm 0.3$ at $3T_c$. The uncertainty bands become larger if one relaxes the momentum-independent assumption and  calibrates $\hat{q}/T^3$ to experimental data in lower and higher $p_{\rm T}$ region separately.

In the sensitivity analysis, we found that $\gamma$-hadron and dihadron correlations are slightly more sensitive to the temperature variation of $\hat{q}/T^3$. 
However, restricted by the current level of experimental accuracy and tension between theory and data at the LHC energy, the inclusion of $I_{AA}$ data does not significantly improve the accuracy of the extracted $\hat{q}/T^3$ in the high-temperature region. We hope this will improve when more accurate data comes from future runs at LHC and RHIC.

As a validation check, we show the NLO pQCD parton model with parton energy loss and extracted temperature-dependent $\hat q/T^3$ can describe within the error bands the existing data on the suppression of single inclusive hadron spectra $R_{AA}$, dihadron and $\gamma$-hadron correlations $I_{AA}$. It is also shown to describe well the measured elliptic anisotropy $v_2$ of large $p_{\rm T}$ hadrons. We also give predictions for $R_{AA}$ and $v_2$ in $O+O$ collisions at $\sqrt{s}=7$ TeV. 

In the future, it is straightforward to apply the information field approach to parameterize higher dimensional unknown functions, such as the temperature $T$, momentum $p$, and the virtuality $Q$ dependence of $\hat{q}$ as postulated by the JETSCAPE Collaboration~\cite{JETSCAPE:2021ehl}. Compared to explicit parameterizations, the random field approach allows fully uncorrelated prior values separated by a large gap in input variables. These features should be very useful for other inverse problems of parameter extractions in addition to the jet transport coefficient.

\begin{acknowledgments}
We thank Chi Ding for providing the space-time profile of the bulk medium using the CLVisc hydrodynamic model. 
This work is supported in part by the National Science Foundation of China under Grant No. 11935007, No. 11861131009, No. 11890714 and No. 12075098, by the Guangdong Major Project of Basic and Applied Basic Research No. 2020B0301030008 and the Science and Technology Program of Guangzhou No. 2019050001, by the Director, Office of Energy Research, Office of High Energy and Nuclear Physics, Division of Nuclear Physics, of the U.S. Department of Energy under  Contract No. DE-AC02-05CH11231, by the US National Science Foundation under Grant No. OAC-2004571 within the X-SCAPE Collaboration.
WK is supported by the US Department of Energy through the Office of Nuclear Physics and the LDRD program at Los Alamos National Laboratory. Los Alamos National Laboratory is operated by Triad National Security, LLC, for the National Nuclear Security Administration of U.S. Department of Energy (Contract No. 89233218CNA000001). 
Computations are performed at Nuclear Science Computer Center at CCNU (NSC3) and the National Energy Research Scientific Computing Center (NERSC), a U.S. Department of Energy Office of Science User Facility operated under Contract No. DE-AC02- 05CH11231.
\end{acknowledgments}

\appendix

\section{Correlations of the posterior $\hat{q}(T)$}
\label{sec:app:corr}

\begin{figure}
    \centering
    \includegraphics[width=\columnwidth]{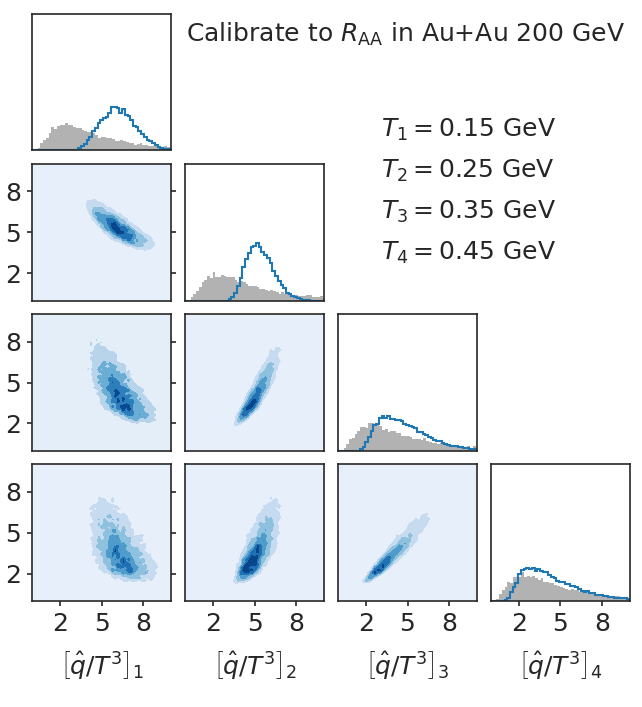}
    \caption{Correlation structures of the posterior of $\hat{q}(T)/T^3$ after calibration using $R_{AA}$ data from Au+Au collisions at $\sqrt{s}=200$ GeV.}
    \label{fig:qhat-corr-1}
\end{figure}

\begin{figure}
    \centering
    \includegraphics[width=\columnwidth]{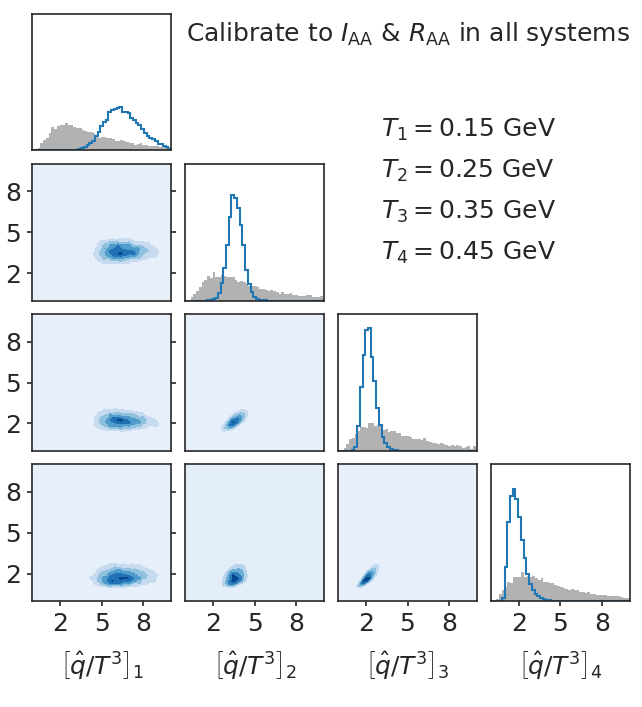}
    \caption{Correlation structures of the posterior of $\hat{q}(T)$ after calibration using both $R_{AA}$ and $I_{AA}$ data from Au+Au collisions at $\sqrt{s}=200$ GeV, Pb+Pb collisions at $\sqrt{s}=2.76$ TeV, and Pb+Pb collisions at $\sqrt{s}=5.02$ TeV.}
    \label{fig:qhat-corr-2}
\end{figure}

In this Appendix, we analyze the posterior of $\hat{q}(T)$ more closely by looking at the correlations of $\hat{q}$ at four different temperatures. These are shown in Figs.~\ref{fig:qhat-corr-1} and \ref{fig:qhat-corr-2} for results calibrated to $R_{AA}$ in Au+Au collisions at $\sqrt{s}=200$ GeV and to both $R_{AA}$ and $I_{AA}$ in all colliding systems, respectively. Diagonal plots, from top to bottom, show the prior (gray regions) and posterior (blue lines) distributions of $\hat{q}$ at $T=0.15, 0.25, 0.35$, and 0.45 GeV. Off-diagonal plots reflect the correlations among them.

In Fig.~\ref{fig:qhat-corr-1}, with information only from Au+Au collisions at $\sqrt{s}=200$ GeV, values of $\hat{q}$ at high temperature ($T_2, T_3, T_4$) are not well constrained and adjacent values are positively correlated as parameterized in the prior distribution. Values at the lowest temperature ($T_1$), however, are anti-correlated with high-temperature values. This is because data from a single colliding system can only constrain an effective temperature-averaged value of $\hat{q}$, with a high degree of flexibility in changing $\hat{q}$ at high and low temperatures in an anti-correlated manner.

When data from all colliding systems are included, as shown in Fig.~\ref{fig:qhat-corr-2}, not only the uncertainty marginalized at each temperature decreases, but the correlation structures also change. Especially, the low-temperature values of $\hat{q}(T_1)$ are almost uncorrelated with values at higher temperatures. This is the benefit of including more colliding systems that are sensitive to $\hat{q}$ in different temperature regions.

\label{sec:app:qhat-corr}

\section{Impact of varying the correlation length in a simple jet-quenching model}
\label{sec:app:varyL}

In this Appendix, we discuss the impact of changing the correlation length parameter in the information field prior on the posterior. Ideally, one should perform this test with the full model calculation. However, this will at least quadruple the amount of computation. Given that the model is already computationally intensive, we demonstrate it with a simple jet-quenching model.

The energy loss $\langle\Delta E\rangle$ in Eq.~\ref{eq:deltaE} is the only quantity in the current model that depends on $\hat{q}(T)$.
In the soft gluon limit, it approximately follows,
\begin{eqnarray}
\Delta E \approx \frac{3\pi  \alpha_s }{2}  \int_{\tau_0}^{\tau_{\rm max}} \hat{q}(\tau)\ln\left(\frac{2E}{\tau \mu_D^2}\right)  \tau d\tau + \cdots
\label{eq:toy-eloss}
\end{eqnarray}
where the ellipsis denotes terms that do not contribute to the $\ln E$ enhancement. We take $\alpha_s=0.3$, $\tau_0=0.6$ fm/$c$, $T^3 \tau = {\rm constant}$ for 1D expansion, and $\tau_{\rm max} = \min\{6.6 \, {\rm fm}/c, \tau(T=0.16\, {\rm GeV})\}$. Taking a hadron spectrum $d\sigma/dp_T \propto p_{\rm T}^{-8}$, the nuclear modification factor in this simple model is $R_{AA} = (1+\Delta E/E)^{-8}$. Then, we can compute $R_{AA}$ for three different ``beam energies'' using different initial temperatures $T_0=0.3, 0.4, 0.5$ GeV.

This is certainly an extremely simplified toy model compared to the full calculation. However, it captures the main features of the $\hat{q}$ extraction using the information field prior:
\begin{itemize}
\item[1.] Calculations are only sensitive to a temperature/time-integrated value of $\hat{q}(T)$ as shown in Eqs.~\ref{eq:toy-eloss} and \ref{eq:deltaE}.
\item[2.] Experimental measurements provide access to different temperature regions, with $0.3<T_0<0.5$ GeV.
\end{itemize}

\begin{figure}
\includegraphics[width=\columnwidth]{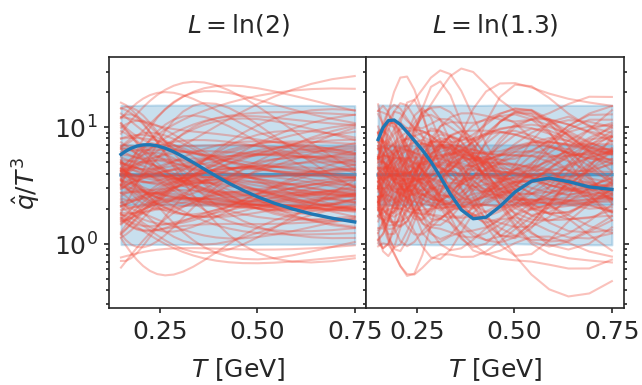}
\caption{Random field priors (bands) with long (left) and short correlation lengths (right). From each distribution, a particular realization of $\hat{q}(T)$ is chosen as the ``truth'' (solid lines) for the validation test.}
\label{fig:toy-truth-qhat}
\end{figure}

Now, we design the following test to illustrate the impact of changing the correlation length in the prior.
\begin{itemize}
\item[1.] Prepare two random field priors for $\hat{q}(T)/T^3$ with the same mean and variance as in the main analysis, but with different correlation length: $L=\ln(2)\approx 0.69$ and $\ln(1.3) \approx 0.26$. We will refer to these two cases as L(long)-set and S(short)-set hereafter.
\item[2.] Assume two truth distributions from the L-set and S-set, as shown in Fig.~\ref{fig:toy-truth-qhat}, such that the L-set and S-set prior samples are very different functions, with the latter containing more detailed, short-length-scale information.
\item[3.] Compute pseudo-data, i.e., $R_{AA}(T_0=0.3 {\rm~GeV})$, $R_{AA}(T_0=0.4 {\rm~GeV})$, and $R_{AA}(T_0=0.5 {\rm~GeV})$ according to each of the ``truth'' $\hat{q}(T)$ in the L-set and S-set. 5\% of uncorrelated uncertainty is added to each set of pseudo-data on $R_{AA}$.
\item[4.] Perform four independent Bayesian analyses 
\begin{itemize}
    \item[LL:] Use the L-set prior to infer the ``truth'' generated from the L-set.
    \item[SL:] Use the S-set prior to infer the ``truth'' generated from the L-set.
    \item[LS:] Use the L-set prior to infer the ``truth'' generated from the S-set.
    \item[SS:] Use the S-set prior to infer the ``truth'' generated from the S-set.
\end{itemize}
\end{itemize}
The purpose of the above procedures is to investigate the possibility that the truth function $\hat{q}(T)$ is not a high-probability realization of the prior. We expect that even though the truth may contain short-length-scale information, this information cannot be recovered in the analysis using neither L-set nor S-set prior due to the temperature-averaging nature of the observable $R_{AA}$.

\begin{figure}
\centering

\includegraphics[width=.5\columnwidth]{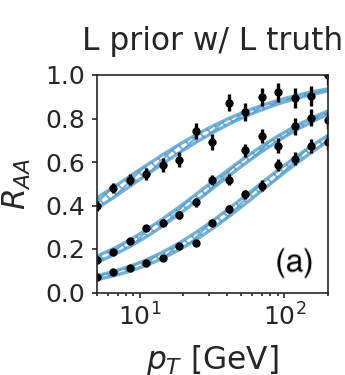}\includegraphics[width=.5\columnwidth]{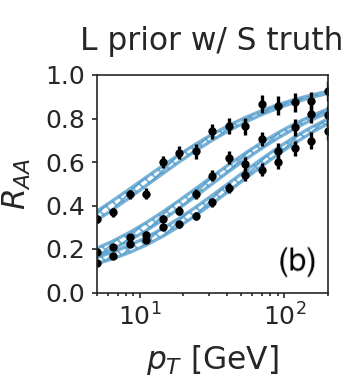}\\
\includegraphics[width=.5\columnwidth]{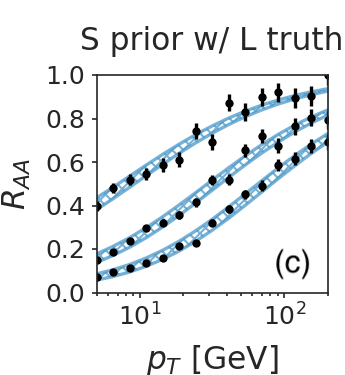}\includegraphics[width=.5\columnwidth]{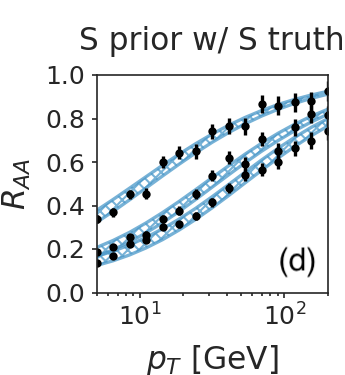}
\caption{Calculations from the toy jet-quenching model calibrated to the pseudo-data, which are generated for three different initial temperatures according to the corresponding ``true" $\hat q$ in each case. The four cases correspond to analyses using L-set or S-set prior and if the truth is generated from the L-set or S-set random function.}
\label{fig:toy-fit-Raa}
\end{figure}

In Fig.~\ref{fig:toy-fit-Raa}, we show the power of the calibrated toy models to describe the pseudo-data generated for three different initial temperatures with the corresponding ``true" $\hat q$. Each panel corresponds to one of the four scenarios (LL, SL, LS, SS) listed above. Models with either type of prior can fit the pseudo-data calculated with the ``true'' $\hat{q}$ sampled from the same or different type of the random fields. In particular, inference with the random field prior with a fairly large correlation length is able to describe the data calculated with a short correlation length. This is due to the information loss in the $R_{AA}$ observable that is not sensitive to the fast variation of $\hat{q}$ as a function of temperature as one would have expected.

\begin{figure}
\centering

\includegraphics[width=.5\columnwidth]{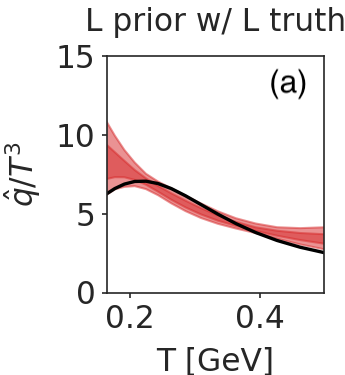}\includegraphics[width=.5\columnwidth]{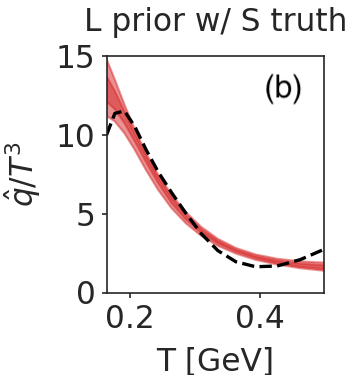}\\
\includegraphics[width=.5\columnwidth]{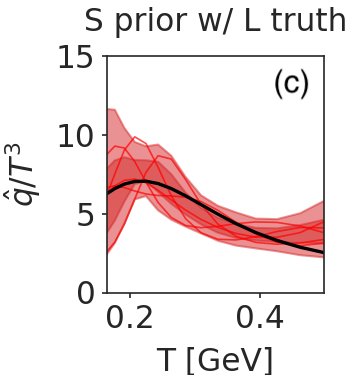}\includegraphics[width=.5\columnwidth]{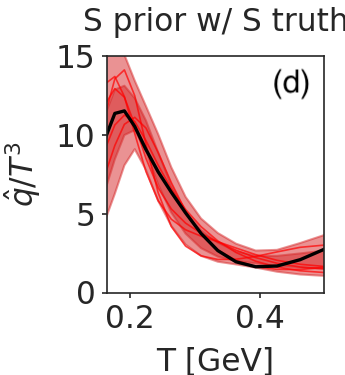}
\caption{The 60\% (dark red bands) and 95\% (light red bands) credible interval of the posterior distribution as compared to the ``truth'' (black lines). The four cases correspond to analyses using L-set or S-set prior with the truth generated from the L-set or S-set random function.}
\label{fig:toy-qhat-extraction}
\end{figure}

Looking at the posterior of $\hat{q}(T)$ as compared to the truth in Fig. \ref{fig:toy-qhat-extraction}, results from the LL scenario work as expected: the truth distribution is reasonably captured in the high-likelihood region of the posterior. In the SL scenario, the truth is well captured in the high-likelihood region, but the uncertainty bands get wider as compared to the LL scenario. This is because those short-range variations in the short-correlation length prior can never be constrained using the data provided.

The LS scenario is potentially problematic. The prior with a larger correlation length strongly disfavors a ``truth'' with an ``oscillating'' behavior. Nevertheless, the averaged trend of the truth is correctly reflected by the posterior while underestimating the uncertainty band.
Interestingly, the SS scenario also cannot recover the ``oscillating'' feature in the ``true'' $\hat{q}$. Again, this is due to the limitation of the observable that does not contain the short-length-scale information. 

We summarize two key observations from this practice regarding the correlation length scale in the prior, which also led to the use of $L=\ln 2$ in our main analysis:
\begin{itemize}
\item Features at the shortest length scale that can be extracted are limited by the differential power of the observable. For example, Au+Au collisions in the most peripheral centrality at $\sqrt{s}=200$ GeV probes the effective $\hat{q}$ from $T_c$ to $2 T_c$, which limits the temperature-differential power of the analysis. 
\item Using a prior with a correlation length much shorter than the resolution of the observable leads to highly-oscillating modes that cannot be constrained by data. Though the median of the extraction is not affected much, these modes increase the uncertainty bands of the posterior. Moreover, one of the main prior beliefs of $\hat{q}(T)$ is that it is monotonic above $T_c$ as there is no other scale in QCD at higher temperatures.
\end{itemize}

\section{Two additional observables added to the analysis}
\label{sec:app:gam_jet}

After completing our main analysis, we found two additional data sets that can be included in our analysis,
\begin{itemize}
     \item {$I_{AA}^{h^{\pm}h^{\pm}}$ in 0-5\% and 20-40\% Au+Au collisions at $\sqrt{s}=0.2$ TeV \cite{STAR:2006vcp};}
     \item {$I_{AA}^{\gamma h^\pm}$ from $\gamma$-triggered charged hadron fragmentation functions in 0-10\%, 10-30\%, 30-50\% Pb+Pb collisions at $\sqrt{s}=5..02$ TeV \cite{CMS:2018mqn}.}
\end{itemize}

In this Appendix, we show the $p+p$ baselines of these two observables and the nuclear modification factors with the 100 sets of the prior random $\hat{q}_i/T^3(i=1,100)$ functions obtained by the Bayesian analysis.

\begin{figure*}
\begin{center}
\includegraphics[width=0.33\textwidth]{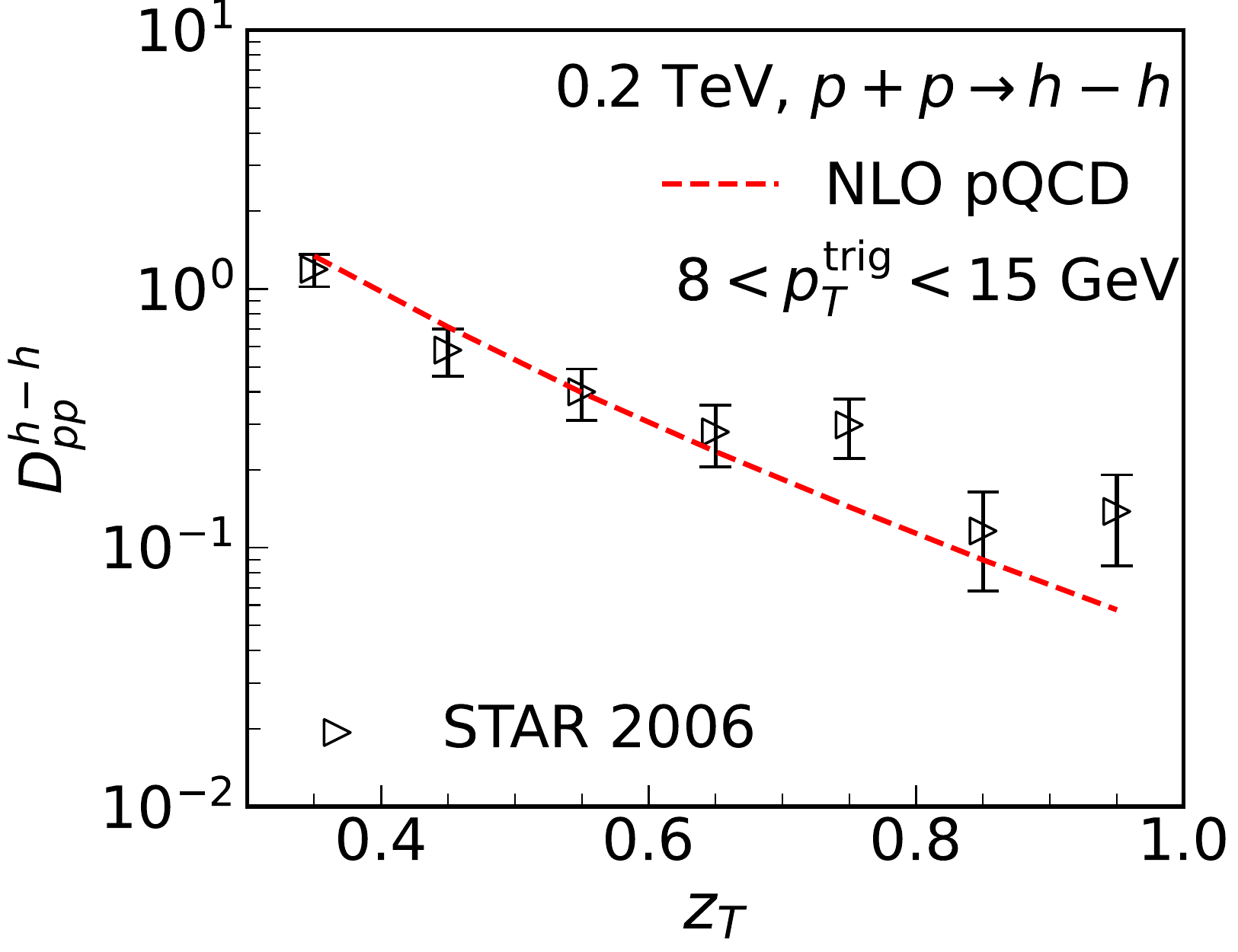}
\hspace{-3mm}
\includegraphics[width=0.66\textwidth]{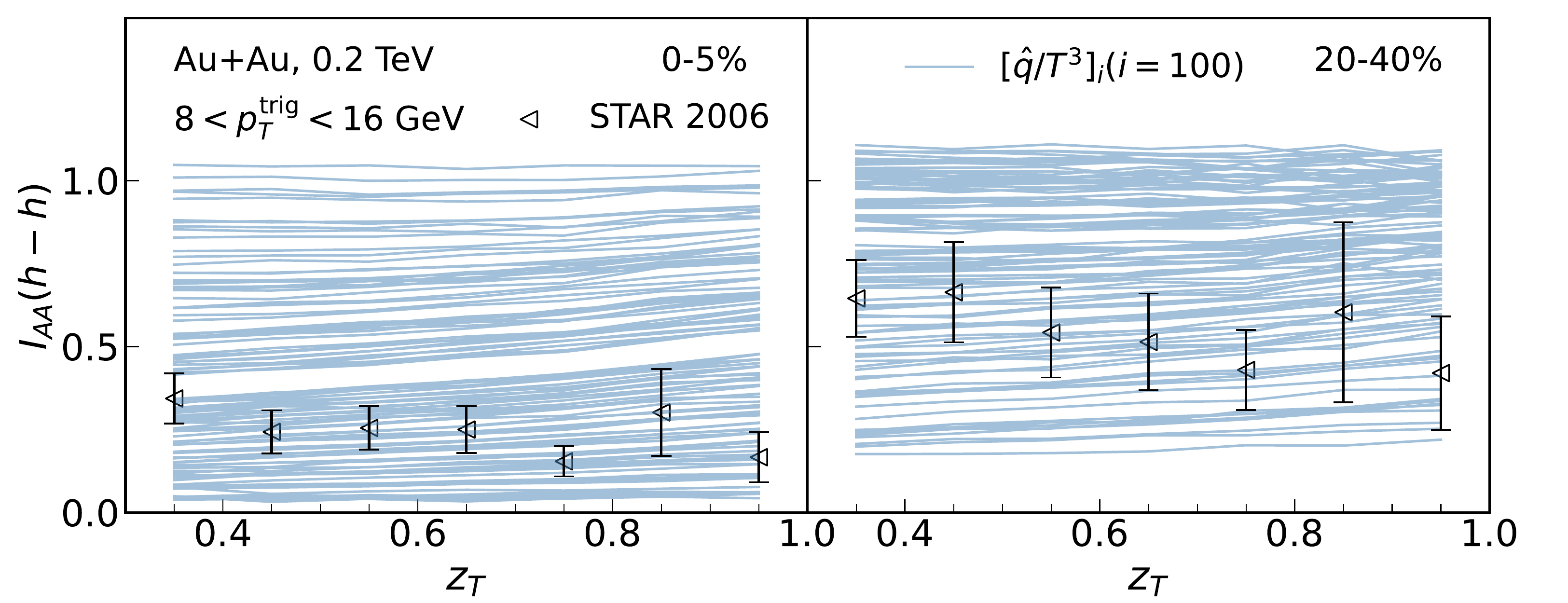}
\end{center}
\vspace{-5mm}
\caption{Charged-hadron-triggered FF's in $p+p$ collisions (left) and the dihadron suppression factors $I_{AA}$ in 0-5\% and 20-40\% Au+Au collisions (middle and right) at $\sqrt{s_{\rm NN}}=0.2$ TeV as a function of $z_{\rm T}$, for $8<p_{\rm T}^{\rm trig}<15$ GeV$/c$, 2.5 GeV$/c<p_{\rm T}^{\rm assoc}<p_{\rm T}^{\rm trig}$ as compared to STAR experimental data \cite{STAR:2006vcp}.}
\label{fig:IAA_STAR2006}
\end{figure*}

Shown in Fig. \ref{fig:IAA_STAR2006} are triggered FF's from charged dihadron correlations with $8<p_{\rm T}^{\rm trig}<15$ GeV$/c$, 2.5 GeV$/c<p_{\rm T}^{\rm assoc}<p_{\rm T}^{\rm trig}$ in $p+p$ collisions (left) and the dihadron suppression factors $I_{AA}$ in 0-5\% and 20-40\% Au+Au collisions (middle and right) at $\sqrt{s_{\rm NN}}=0.2$ TeV as a function of $z_{\rm T}$. These are similar to Fig. \ref{fig:Dpp-dihad}  and Fig. \ref{fig:IAuAu-pi0-had} except for the trigger $p_{\rm T}^{\rm tirg}$ range.

\begin{widetext}

\begin{figure}
\begin{center}
\includegraphics[width=0.35\textwidth]{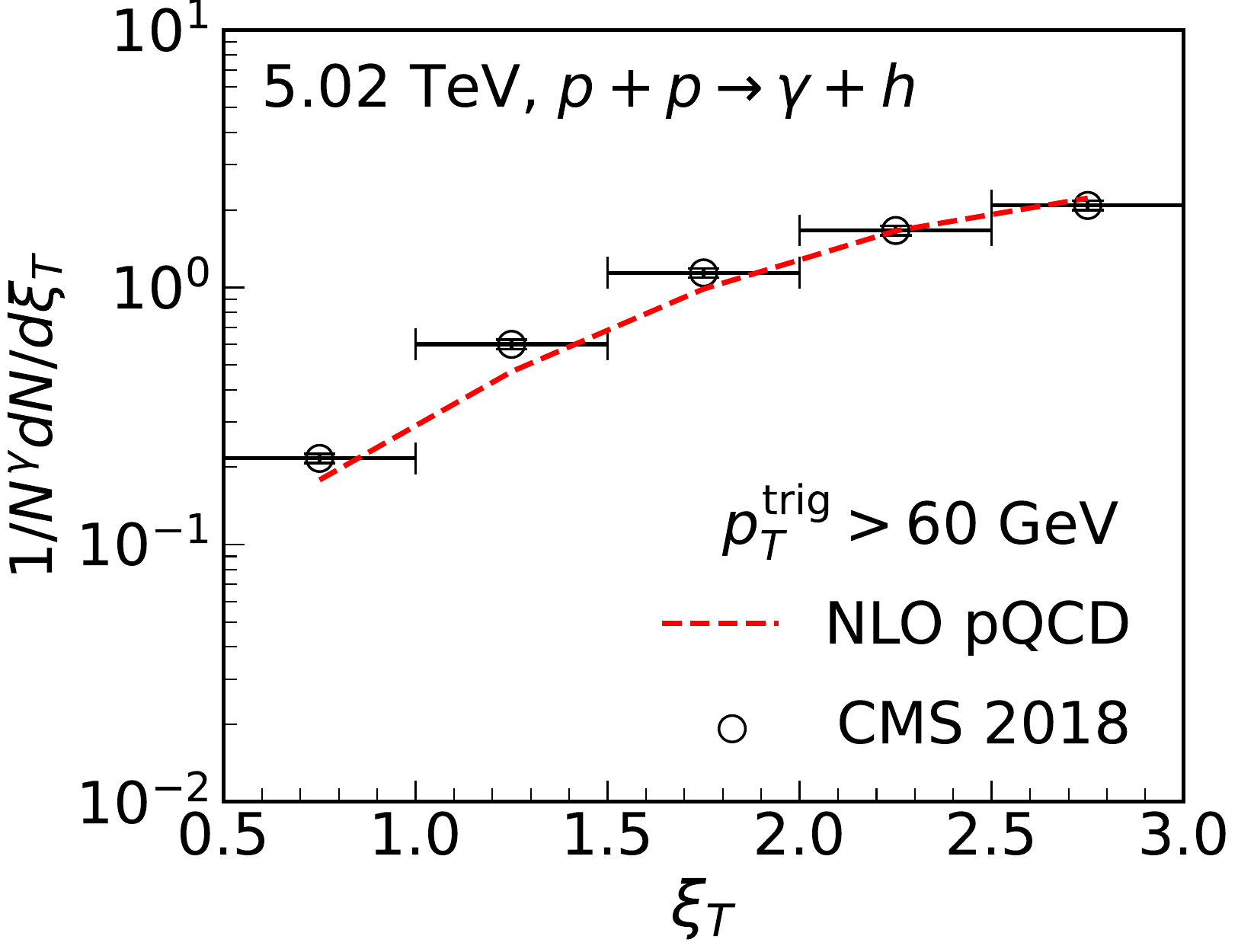}
\includegraphics[width=1.0\textwidth]{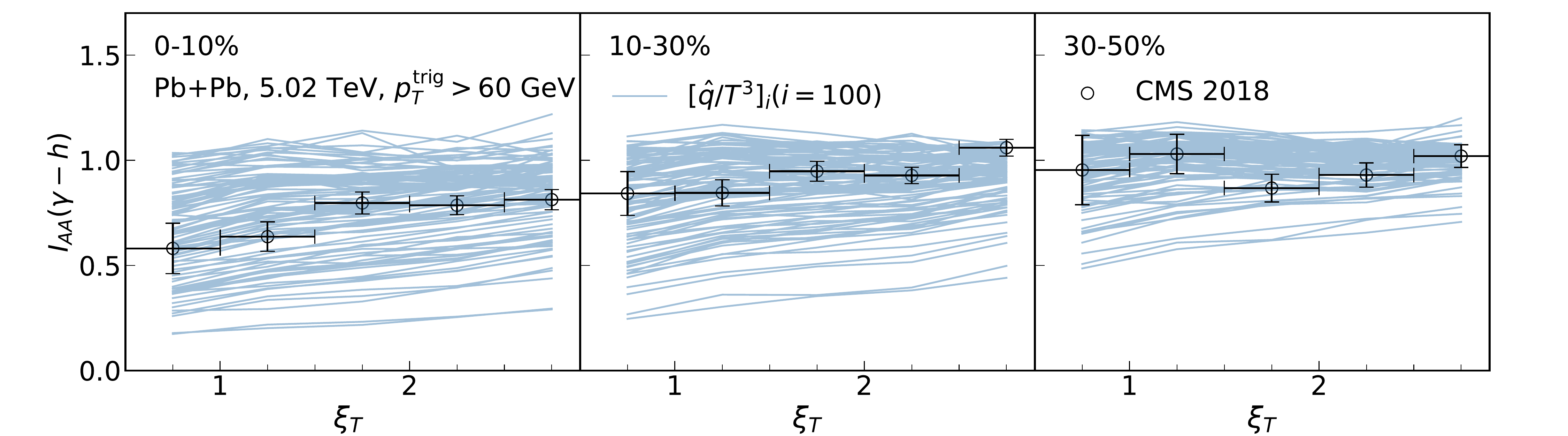}
\end{center}
\vspace{-5mm}
\caption{$\gamma$-triggered FF's in $p+p$ collisions (upper) and the modification factor $I_{AA}$ in 0-10\%, 10-30\%, and 30-50\% $Pb+Pb$ collisions at $\sqrt{s_{\rm NN}}=5.02$ TeV (lower) as a function of $\xi_{\rm T}=\ln [-|\vec{p}_{\rm T}^{\gamma}|^2  / \vec{p}_{\rm T}^{\gamma} \cdot \vec{p}_{\rm T}^{\rm assoc} ]$, for $p_{\rm T}^{\rm trig}>60$ GeV$/c$, 3 GeV$/c<p_{\rm T}^{\rm assoc}<p_{\rm T}^{\rm trig}$ as compared to CMS experimental data \cite{CMS:2018mqn}.}
\label{fig:IAA_CMS2018}
\end{figure}
\end{widetext}

Shown in Fig.~\ref{fig:IAA_CMS2018} are $\gamma$-triggered FF's in $p+p$ collisions (upper) and the modification factor $I_{AA}$ in Pb+Pb collisions at $\sqrt{s_{\rm NN}}=5.02$ TeV (lower). The experimental data are converted from measurements of $\gamma$-jet fragmentation functions \cite{CMS:2018mqn} as a function of $\xi_{\rm T}=\ln [-|\vec{p}_{\rm T}^{\gamma}|^2  / \vec{p}_{\rm T}^{\gamma} \cdot \vec{p}_{\rm T}^{\rm assoc} ]$ 
with a jet cone-size $R=0.3$,  $p_{trk}>1$ GeV$/c$, $p_{\rm T}^\gamma>60$ GeV$/c$, $p_{\rm T}^{\rm jet}>30$ GeV/$c$,  $\Delta{\phi}_{{\rm jet}\gamma}>7/8\pi$, $|\eta_{\rm jet}|<1.6$, $|\eta_\gamma|<1.44$. The selected charged-particle tracks are normalized by the total number of $\gamma$-jet pairs. In our NLO pQCD model calculation of $\gamma$-hadron correlation per triggered photon, we consider $|\Delta\phi_{\gamma h}|>7/8\pi-R$ and  3 GeV$/c<p_{\rm T}^{\rm assoc}<p_{\rm T}^{\gamma}$ which corresponds to the ${z_{\rm T}}_{min}=0.05$. The NLO pQCD result for $p+p$ collisions is consistent with the experimental data.


\section{Training observables of the main analysis}
\label{sec:app:prior-obs}

In this Appendix, we present the numerical results for suppression factors of single inclusive hadrons $R_{AA}$, $\gamma$-hadron and dihadron correlations $I_{AA}$ in Au+Au collisions at $\sqrt{s}=0.2$ TeV, Pb+Pb collisions at $\sqrt{s}=2.76$ and 5.2 TeV with different centralities. These observables are each evaluated within the NLO parton model with parton energy loss and 100 sets of the temperature-dependent jet transport coefficient $\hat{q}_i/T^3(i=1,100)$ as prior random functions in the Bayesian inference analysis. These results, together with an emulator for interpolation, are used for the Bayesian inference of the jet transport coefficient. 

\begin{figure*}[htb!]
\begin{center}
\includegraphics[width=1.0\textwidth]{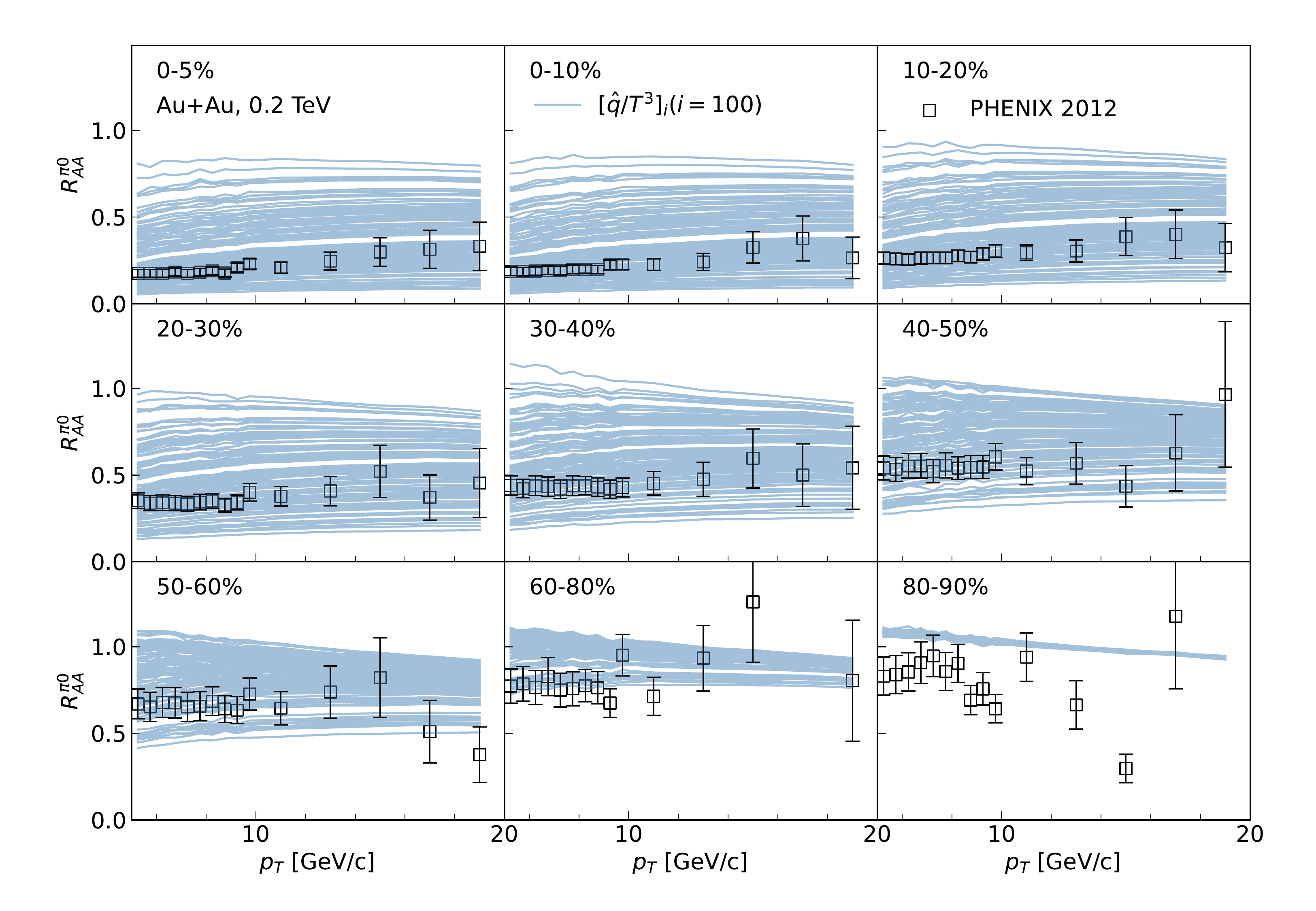}
\end{center}
\vspace{-9mm}
\caption{Nuclear modification factor $R_{AA}$ as a function of $p_{\rm T}$ in Au+Au collisions at $\sqrt{s}=0.2$ TeV in 0-5\%, 0-10\%, 10-20\%, 20-30\%, 30-40\%, 40-50\%, 50-60\%, 60-80\%, 80-90\% centrality bins, as compared to PHENIX \cite{PHENIX:2008saf,Adare:2012wg,Acharya:2018qsh} experimental data. Each plot contains 100 sets of $R_{AA}$ with $\hat{q}_i/T^3(i=1,100)$ priors provided by the Bayesian inference.}
\label{fig:RAA-200}
\end{figure*}

\begin{figure*}[h!]
\begin{center}
\includegraphics[width=1.0\textwidth]{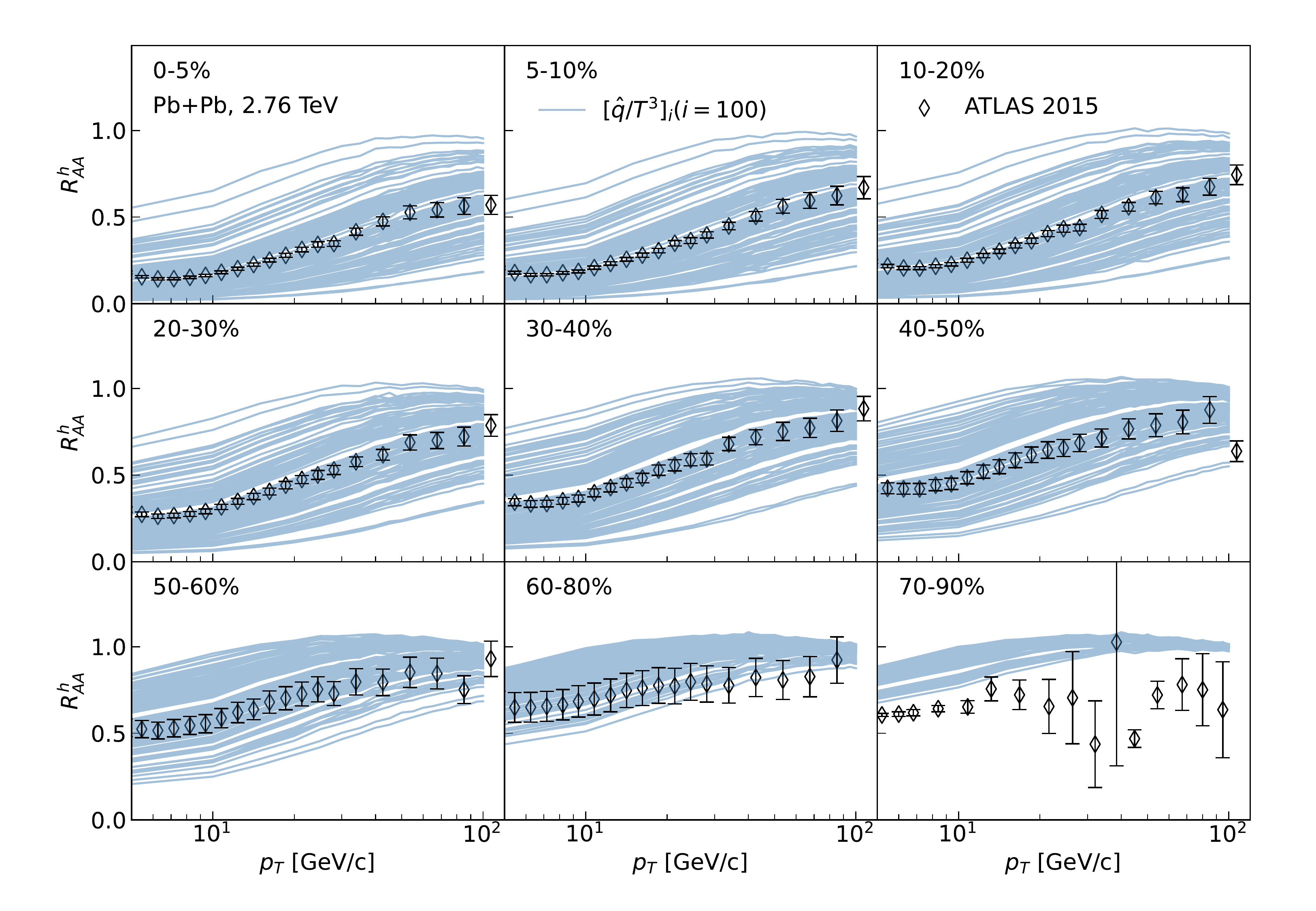}
\end{center}
\vspace{-9mm}
\caption{Nuclear modification factor $R_{AA}$ as a function of $p_{\rm T}$ in Pb+Pb collisions at $\sqrt{s}=2.76$ TeV in 0-5\%, 5-10\%, 10-20\%, 20-30\%, 30-40\%, 40-50\%, 50-60\%, 60-80\%, 70-90\% centrality bins, as compared to experimental data \cite{CMS:2012aa,Abelev:2012hxa,Aad:2015wga}.}
\label{fig:RAA-2760}
\end{figure*}

\begin{figure*}[h!]
\begin{center}
\includegraphics[width=1.0\textwidth]{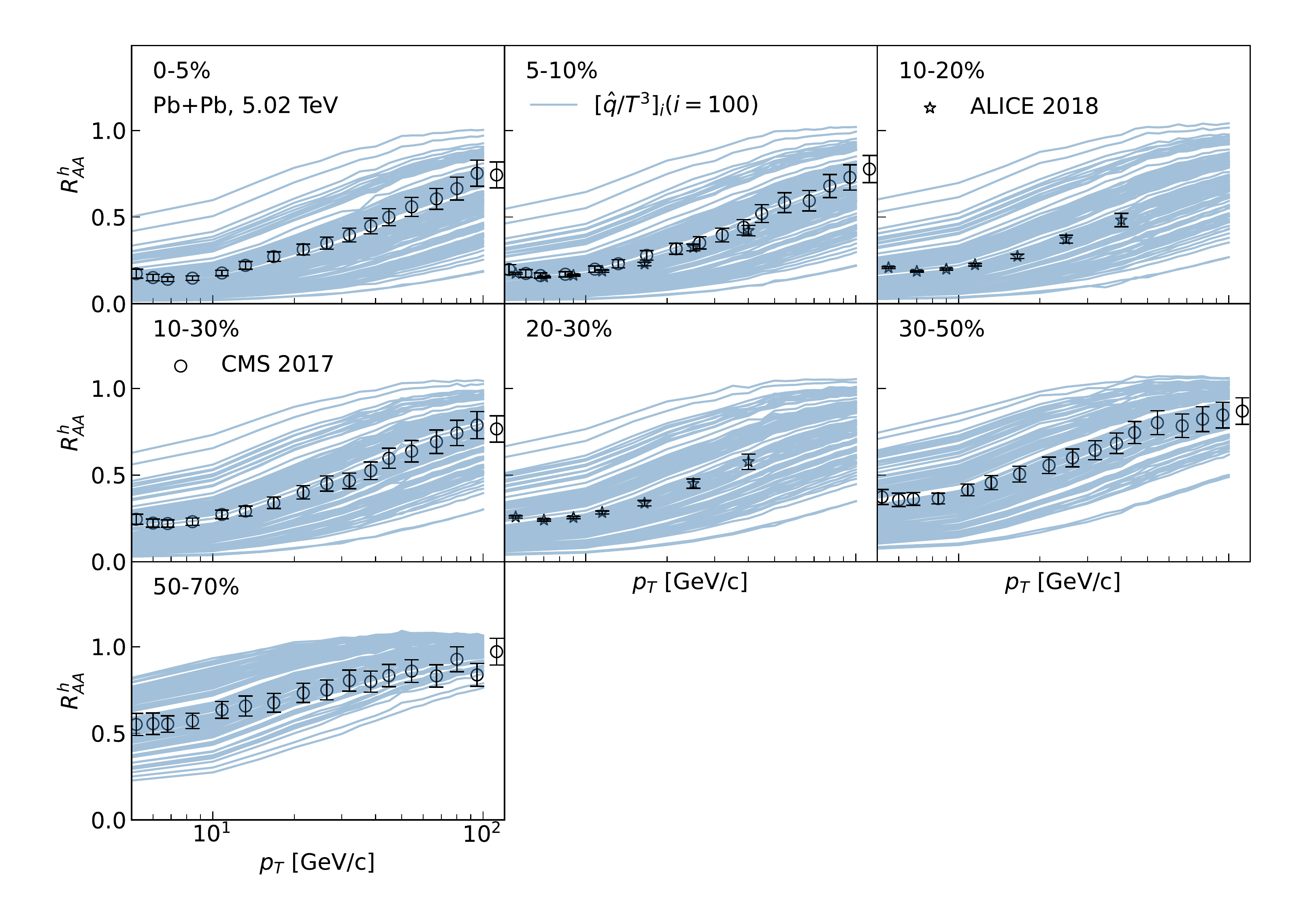}
\end{center}
\vspace{-9mm}
\caption{Nuclear modification factor $R_{AA}$ as a function of $p_{\rm T}$ in Pb+Pb collisions at $\sqrt{s}=5.02$ TeV in 0-5\%, 5-10\%, 10-20\%, 10-30\%, 20-30\%, 30-50\%, 50-70\% centrality bins, as compared to ALICE \cite{Acharya:2018qsh} and CMS \cite{Khachatryan:2016odn} experimental data.}
\label{fig:RAA-5020}
\end{figure*}

\begin{figure*}[h!]
\begin{center}
\includegraphics[width=0.7\textwidth]{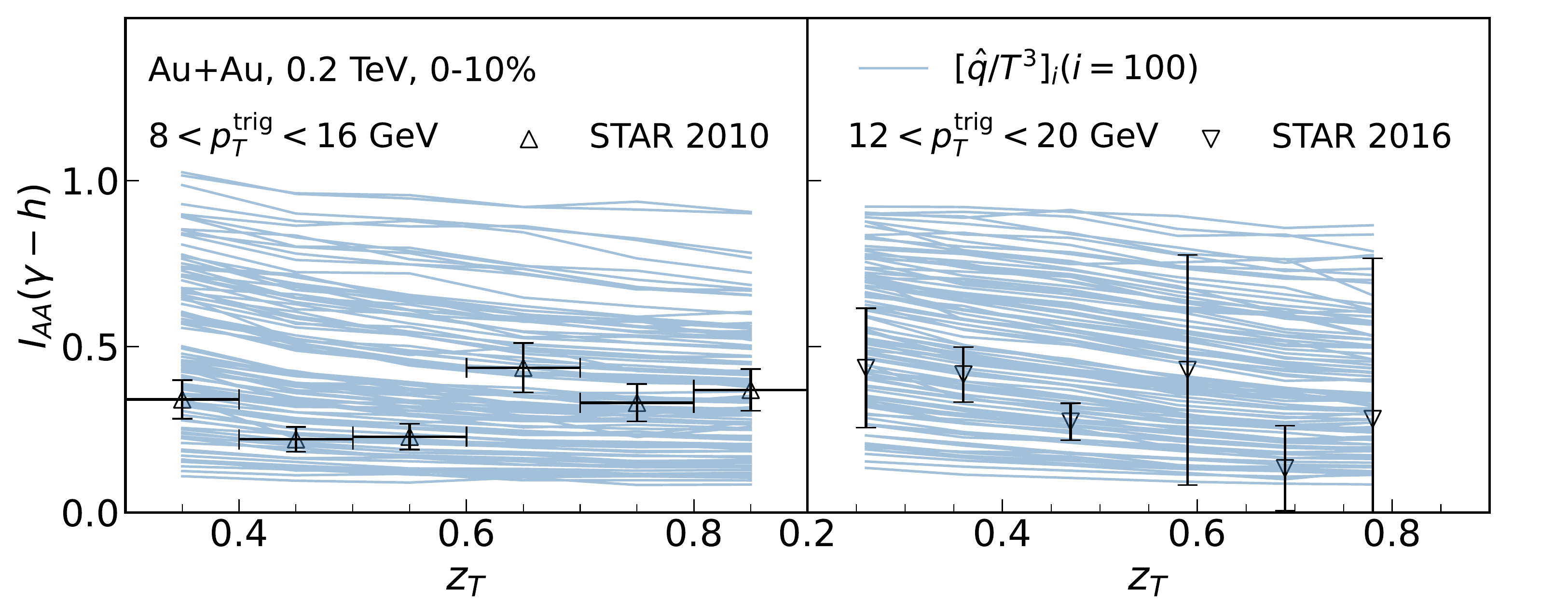}
\end{center}
\vspace{-5mm}
\caption{$\gamma$-hadron suppression factor $I_{AA}$ as a function of $z_{\rm T}$ with ($8<p_{\rm T}^{\rm trig}<16$ GeV$/c$, $3<p_{\rm T}^{\rm assoc}<16$ GeV$/c$) (left) and ($12<p_{\rm T}^{\rm trig}<20$ GeV$/c$, $1.2<p_{\rm T}^{\rm assoc}<p_{\rm T}^{\rm trig}$) (right) in Au+Au collisions at $\sqrt{s}=0.2$ TeV within 0-10\% centrality, as compared to STAR experimental data \cite{Abelev:2009gu,STAR:2016jdz}.}
\label{fig:IAuAu-gam-had}
\end{figure*}

\begin{figure*}[h!]
\begin{center}
\includegraphics[width=0.7\textwidth]{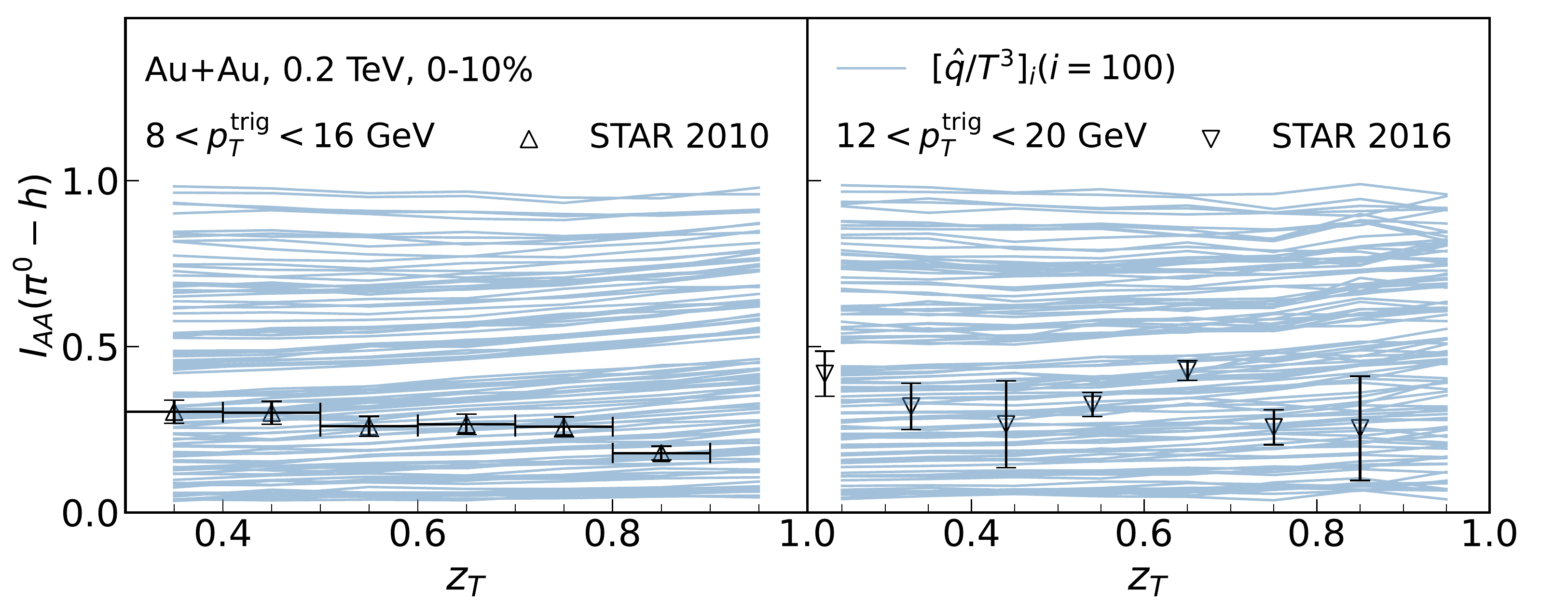}
\end{center}
\vspace{-5mm}
\caption{$\pi^{0}$-hadron suppression factor $I_{AA}$ as a function of $z_{\rm T}$ with ($8<p_{\rm T}^{\rm trig}<16$ GeV$/c$, $3<p_{\rm T}^{\rm assoc}<16$ GeV$/c$) (left) and ($12<p_{\rm T}^{\rm trig}<20$ GeV$/c$, $1.2<p_{\rm T}^{\rm assoc}<p_{\rm T}^{\rm trig}$) (right) in Au+Au collisions at $\sqrt{s}=0.2$ TeV within 0-10\% centrality, as compared to STAR experimental data \cite{Abelev:2009gu,STAR:2016jdz}.}
\label{fig:IAuAu-pi0-had}
\end{figure*}

\begin{figure*}[h!]
\begin{center}
\includegraphics[width=1.0\textwidth]{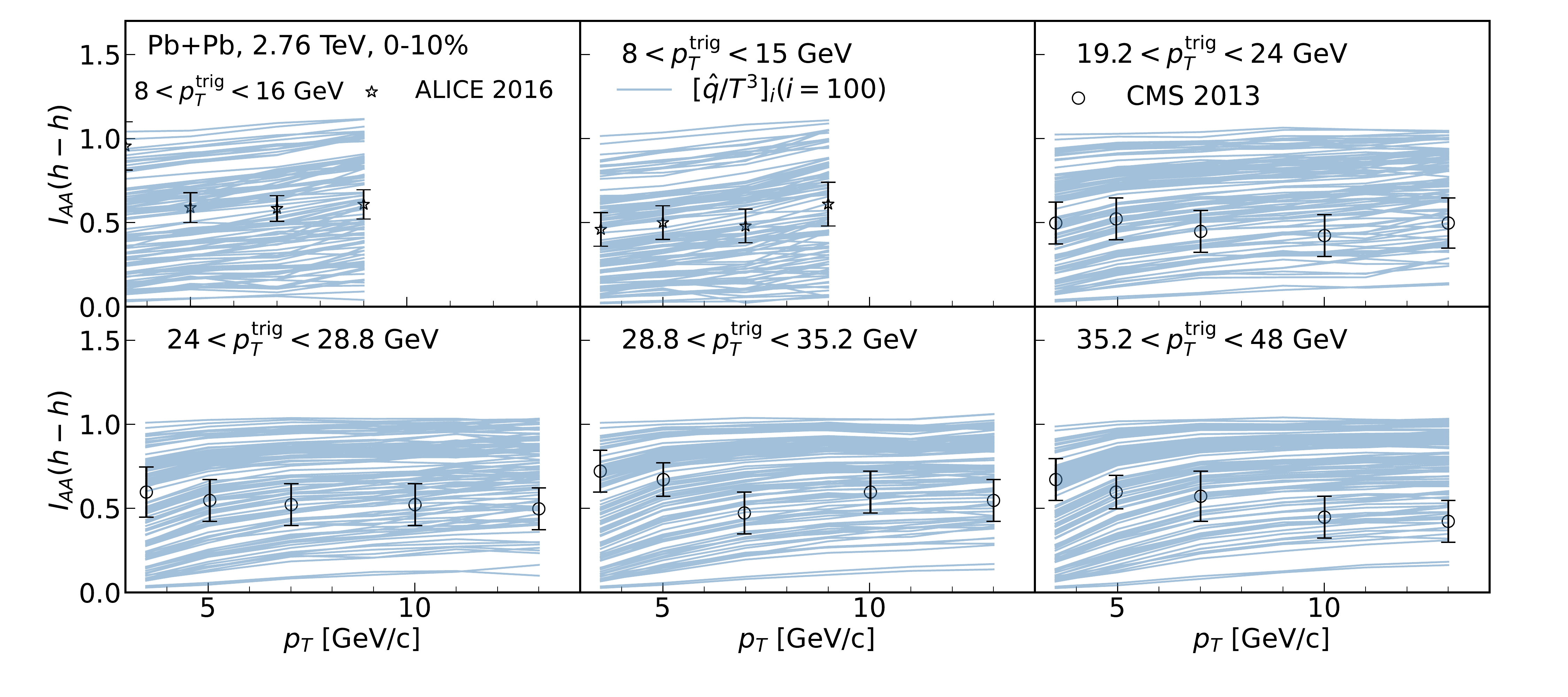}
\end{center}
\vspace{-5mm}
\caption{Dihadron suppression factor $I_{AA}$ as a function of $p_{\rm T}^{\rm assoc}$ with six different $p_{\rm T}^{\rm trig}$ ranges in Pb + Pb collisions at $\sqrt{s}=2.76$ TeV within 0-10\% centrality, as compared to ALICE \cite{Aamodt:2011vg,Adam:2016xbp} and CMS \cite{Conway:2013xaa} experimental data.}
\label{fig:IPbPb-had-had}
\end{figure*}

Figs. \ref{fig:RAA-200}, \ref{fig:RAA-2760} and \ref{fig:RAA-5020} are the suppression factors $R_{AA}$ of single inclusive hadrons in Au+Au collisions at $\sqrt{s}=0.2$ TeV in 9 different centrality bins, in Pb+Pb collisions at $\sqrt{s}=2.76$ TeV in 9 centrality bins and 5.02 TeV in 7 centrality bins, respectively, as compared with the experimental data \cite{PHENIX:2008saf,Adare:2012wg,CMS:2012aa,Abelev:2012hxa,Aad:2015wga,Khachatryan:2016odn,Acharya:2018qsh}.
Fig. \ref{fig:IAuAu-gam-had} shows the suppression factors $I_{AA}^{\gamma h}$ for $\gamma$-triggered hadron spectra as a function of $z_{\rm T}$ in 0-10\% central Au+Au collisions at $\sqrt{s}=0.2$ TeV. The difference between the left and right panels are the ranges of trigger and hadron transverse momentum, ($8<p_{\rm T}^{\rm trig}<16$ GeV$/c$, $3<p_{\rm T}^{\rm assoc}<16$ GeV$/c$) in the left panel and ($12<p_{\rm T}^{\rm trig}<20$ GeV$/c$, $1.2<p_{\rm T}^{\rm assoc}<p_{\rm T}^{\rm trig}$) in the right panel.
Figs.~\ref{fig:IAuAu-pi0-had} and \ref{fig:IPbPb-had-had} show the medium modification factors $I_{AA}^{\pi^0 h}$ for dihadron correlations as a function of $z_{\rm T}$ in Au+Au collisions at $\sqrt{s}=0.2$ TeV with two $p_{\rm T}^{\rm trig}$ ranges and $I_{AA}^{h h}$ as a function of $p_{\rm T}^{\rm assoc}$ in Pb+Pb collisions at $\sqrt{s}=2.76$ TeV with six $p_{\rm T}^{\rm trig}$ ranges, respectively.


\bibliography{prc}

\end{document}